\documentclass[a4paper,10pt]{article}

\usepackage[utf8]{inputenc}

\usepackage{authblk}

\usepackage{ifthen,ifpdf}

\ifpdf
  \usepackage{hyperref}
  \hypersetup{colorlinks=false,allcolors=blue}
  \usepackage{hypcap}
  \pdfpagewidth=\paperwidth
  \pdfpageheight=\paperheight
\fi

\usepackage[utf8]{inputenc}
\usepackage[T1]{fontenc}

\usepackage{amsmath, amsthm, amssymb}
\usepackage{mathtools}
\usepackage{enumerate}
\usepackage{afterpage}
\usepackage{tabularx}
\usepackage{dsfont}

\usepackage{graphicx}

\usepackage{algorithm}
\usepackage{algpseudocode}
\usepackage{caption}
\usepackage[labelformat=simple]{subcaption}

\usepackage{color}
\usepackage{units}
\usepackage{array, multirow}
\usepackage{booktabs}

\newcolumntype{M}[1]{>{\vspace{3pt}\raggedleft\arraybackslash}m{#1}}

\usepackage{siunitx}
\usepackage{tikz}
\usepackage{pgfplots}
\usepackage{pgfplotstable}
\usepgfplotslibrary{units}
\pgfplotsset{compat=1.9}
\pgfplotstableset{col sep=comma}

\usepackage{cite}

\usepackage{anysize} 
\marginsize{2.5cm}{2.5cm}{2cm}{2cm}
\usepackage{stackengine}
\usepackage{wrapfig}

\theoremstyle{plain}

\theoremstyle{remark}

\numberwithin{equation}{section}

\newcommand{\I}{\mathds{I}}

\newcommand{\fd}{\sty{ d}}
\newcommand{\fsigma}{\mbox{\boldmath $\sigma$}}
\newcommand{\fn}{\sty{ n}}
\newcommand{\fe}{\sty{ e}}
\newcommand{\ffC}{\styy{ C}}

\newcommand{\feps}{\mbox{\boldmath $\varepsilon $}}

\newcommand{\sign}[1]{{\rm sgn}\left( #1 \right)}

\newcommand{\bfeps}{\bar{\feps}}

\newcommand{\sty}[1]{\mbox{\boldmath $#1$}}

\newcommand{\styy}[1]{{\mathbb{#1}}}
\newcommand{\fsym}[1]{{\rm sym }( #1 )}

\newcommand{\abs}[1]{|#1|}

\newcommand{\la}{\langle}
\newcommand{\ra}{\rangle}

\newcommand{\Geff}{{\bar{G}}}
\newcommand{\ffK}{{\styy{ K}}}

\newcommand{\fQ}{{\sty{ Q}}}
\newcommand{\fmu}{{\sty{ \mu}}}
\newcommand{\fzero}{\sty{ 0}}

\newcommand{\ms}{\hspace{0.78em}}

\title{FFT-based investigation of the shear stress distribution in face-centered cubic polycrystals} 

\author{Flavia Gehrig}
\author{Daniel Wicht}
\author{Maximilian Krause}
\author{Thomas Böhlke}

\affil{Karlsruhe Institute of Technology (KIT), Institute of Engineering Mechanics\\ correspondence to: \texttt{thomas.boehlke@kit.edu}}

\date{\today}

\begin{document}

\maketitle 

\begin{abstract}
\noindent 

The onset of nonlinear effects in metals, such as plasticity and damage, is strongly influenced by the heterogeneous stress distribution at the grain level.
This work is devoted to studying the local stress distribution in fcc polycrystals using FFT-based solvers. In particular, we focus on the distribution of shear stresses resolved in the slip systems as the critical driving force for plastic deformations. Specific grain orientations with respect to load direction are investigated in the linear elastic regime and at incipient plastic deformations based on a large ensemble of microstructures.
The elastic anisotropy of the single crystal is found to have a crucial influence on the scatter of the stress distribution, whereas the Young's modulus in the respective crystal direction governs the mean stress in the grain. 
It is further demonstrated that, for higher anisotropy, the shear stresses deviate from the normal distribution and are better approximated by a log-normal fit. 
Comparing the full-field simulations to the Maximum Entropy Method (MEM), reveals that the MEM  provides an excellent prediction up to the second statistical moment in the linear elastic range.
In a study on the spatial distribution of shear stresses, the grain boundary is identified as a region of pronounced stress fluctuations and as the starting point of yielding during the elastic-plastic transition.

\end{abstract}

\section{Introduction}
\label{sec:intro}

Due to their polycrystalline structure, metals display anisotropic behavior at the grain level inducing a heterogeneous local stress distribution. This distribution has a major influence on nonlinear effects and failure mechanism, such as micro-plastic deformations, e.g. grain boundary pop-ins \cite{javaid2020local}, or incipient damage, e.g. fatigue fracture \cite{blochwitz1996analysis}. Therefore, investigating the local stress distribution in polycrystalline materials has been the subject to numerous experimental, analytical and numerical studies.

In experiments, techniques such as neutron diffraction \cite{clausen1999lattice,pang2000generation}, x-ray diffraction \cite{tamura2003scanning,lienert2004investigating}, electron back-scattering diffraction (EBSD) \cite{blochwitz1996analysis, raabe2001micromechanical} and digital image correlation (DIC) \cite{tasan2014strain,raabe2001micromechanical} were used for investigating the mechanical fields on the microscale. For instance, in their recent article Berger et al. \cite{berger2022experimental} use the latter two techniques for a detailed examination of the spatial strain distribution in $\alpha$-FE polycrystals. As experiments are associated with considerable cost and effort, analytical and numerical approaches appear attractive for predicting the stress distribution, owing to their higher flexibility. However, in this context, experimental data is still crucial for validating analytical \cite{clausen1999lattice,raabe2001micromechanical} and numerical results \cite{lienert2004investigating,demir2021investigation}. Therefore, according to Berger et al.\cite{berger2022experimental}, the dialogue between experiments and simulations should be sought.

Early analytical homogenization approaches assumed a homogeneous strain  \cite{voigt1928lehrbuch,taylor1938plastic,bishop1951cxxviii,bishop1951xlvi} or stress field \cite{Reus1929,sachs1928zietschrift}, enabling a (crude) estimate for the localization of the conjugate stress and strain field, respectively. 
More sophisticated models, using a statistical description of the microstructure, were developed to arrive at more accurate predictions. For instance, the self consistent approach for linear \cite{hershey1954elasticity,kroner1958berechnung} and nonlinear material behavior \cite{hill1965self,hutchinson1976bounds,molinari1987self,lebensohn1993self} has been exploited for predicting the first and second statistical moment of the stress field in polycrystals \cite{LEBENSOHN20045347,BRENNER20093018}.
A different analytical technique is the Maximum Entropy Method (MEM) pioneered by Kreher and Pompe \cite{kreher:89}, see Krause and Böhlke \cite{krause:20} for a review and an in-depth study on the capabilities of the method.

In numerical studies, full-field simulations are carried out on discretized microstructures. As solvers, commercial finite element software \cite{hashimoto1983role,kumar1996micro,BARBE2001537,sauzay2007cubic,WONG20101658} and dedicated FFT-based micromechanics codes \cite{LEBENSOHN20045347,BRENNER20093018,LAVERGNE2013387,castelnau2006effect} are most commonly used.
In particular, the computational efficiency of FFT-based methods \cite{moulinec1994fast,moulinec1998numerical} and improvements in solver technology \cite{michel2001computational,gelebart2013non,QuasiNewton2019} have enabled the study of larger samples \cite{BRENNER20093018,kasemer2020numerical}.
Due to the flexibility of the full-field approach, various studies have focused on different aspects of the stress localization in polycrystals. For instance, it was shown that the neighborhood effect - comprising the impact of the neighboring grains' stiffness, their relative position with respect to loading axis and their crystallographic orientation \cite{BRETIN201936} - leads to a large scatter of the mean stress in a grain \cite{sauzay2007cubic,GUILHEM20101748,BRETIN201936,castelnau2020multiscale,gelebart2021grain}. This influence is suspected to be greater than the scatter resulting from the grain's own orientation \cite{sauzay2007cubic}. However, a grain's orientation has a major impact on the stress distribution in the plastified regime \cite{WONG20101658,bieler2009role}, whereas, in the elastic regime and during the elastic-plastic transition, the single crystal anisotropy is more crucial \cite{sauzay2007cubic,kumar1996micro,WONG20101658,lebensohn2012elasto}. In addition to the grain's orientation, the hardening behavior tends to influence the stress distribution in fully plastified polycrystals \cite{kasemer2020numerical}. A detailed study on the impact of phenomenological and physical hardening laws can be found in Demir et al. \cite{demir2021investigation}.
At the grain boundaries, on average, larger disorientations \cite{kuhbach2020quantification}, a higher number of active slip systems and an increased sum of plastic slips \cite{cailletaud2003some} can be found. The resulting local stress and strain concentrations are suspected to be the precursor of crack initiation and growth \cite{cailletaud2003some,GUILHEM20101748} although Bieler et al. \cite{bieler2009role} state that the first microcracks do not occur at the exact location of the stress or strain peaks but in close proximity. Therefore, the spatial distribution of the stresses has attracted increased research interest \cite{cailletaud2003some,rollett2010stress,GONZALEZ201449,hure2016intergranular,el2020full,kuhbach2020quantification}. 
\\
\\
The present work is devoted to investigating the local shear stress distribution using FFT-based solvers in face-centered cubic (fcc) polycrystals \cite{lebensohn2012elasto,sauzay2007cubic,lim2019investigating}. In particular, we study the stress distribution with respect to grain orientations, focusing on the $\la 100\ra$,$\la 110\ra$ and $\la 111\ra$ grains. 
For the elastic regime, we follow the strategy of Brenner et al. \cite{BRENNER20093018} and Sauzay \cite{sauzay2007cubic}, studying the characteristics of the stress distribution based on a large ensemble. In particular, we focus on the impact of elastic anisotropy and the specific shape of the stress distribution. The results are supplemented by spatial information, taking the distance with respect to grain boundaries into account \cite{rollett2010stress,castelnau2020multiscale}.
Furthermore, we compare our results to predictions of the Maximum Entropy Method (MEM), which has proven to be a powerful analytical tool for polycrystals \cite{krause:20}.
In the plastic regime, studies are often limited to a single realization \cite{rollett2010stress,GONZALEZ201449} or a small ensemble (e.g. 30 realizations \cite{vincent2011stress}), due to the associated computational cost. Thus, an analysis based on a large ensemble is of interest to obtain representative results.
Building upon the elastic results, we investigate the onset of plastification. More precisely, we study the characteristics of the stress distribution during the elastic-plastic transition and correlate the grain boundary distance to incipient plasticity.
\\
\\
The outline is as follows.
After a short introduction of the methods in Sec.~\ref{sec:methods}, a preliminary study follows in Sec.~\ref{sec:resolution_realization} in order to determine the required resolution and ensemble size. 
In the first part of the actual study, the influence of anisotropy on the stress distribution in the linear elastic range is examined  (see Sec.~\ref{sec:lin_elast_anisotropy}).
An analysis of the specific shape of the stress distribution is given in Sec.~\ref{sec:normality}, followed by a comparison to results of the Maximum Entropy Method (MEM) in Sec.~\ref{sec:compare_to_mem} and an investigation of the stress distribution with respect to grain boundary distance in Sec.~\ref{sec:lin_elast_space}. Secondly, the stress distribution is examined during the elastic-plastic transition in Sec.~\ref{sec:onset}, taking a closer look at the location of the plastified regions in Sec.~\ref{sec:spatial_plastification}.

\section{Methods}
\label{sec:methods}
\subsection{Crystal plasticity}
\label{sec:crystal_plasticity}
For the present study, we rely on a small-strain single-crystal elasto-viscoplasticity model. The elastic behavior is governed by Hooke's law
\begin{equation}
\fsigma = \ffC [\feps_\mathrm{e}] \quad \text{with} \quad \feps = \feps_\mathrm{e} + \feps_\mathrm{p} ,
\label{eq:Hooke}
\end{equation}
where the total strain $\feps$ is decomposed additively into an elastic part $\feps_\mathrm{e}$ and a plastic part $\feps_\mathrm{p}$. As we restrict to crystals with cubic crystal symmetry, the stiffness tensor $\ffC$ is specified by three elastic constants (${C_{11}=C_{1111}}$, ${C_{12}=C_{1122}}$, ${C_{44}=C_{1212}}$) and the orientations of the lattice vectors. The Zener parameter 
\begin{equation}
	A=\frac{2 C_{44}}{C_{11}-C_{12}}
	\label{eq:zener}
\end{equation}
serves as a measure of the elastic anisotropy and is equal to one in the isotropic case.

In single-crystal plasticity, a flow rule of the form
\begin{equation}
\dot{\feps}_\mathrm{p} = \sum_{\alpha = 1}^{12} \ \dot{\gamma}_\alpha \  \fsym{\fd_\alpha \otimes \fn_\alpha}
\label{eq:Verzerrungsrate}
\end{equation}
is typically assumed, i.e., the plastic strain arises as a linear combination of shears on crystallographic slip systems described by the Schmid tensors $\fd_\alpha \otimes \fn_\alpha$. The plastic slip in a system is activated when the shear stress, $\tau_\alpha = \fsigma : (\fd_\alpha \otimes \fn_\alpha)$ reaches a critical resolved yield stress $\tau^\mathrm{F}_\alpha$. In this context the operator~$:$ stands for a double tensor contraction creating a projection of the Cauchy stress tensor on the Schmid tensor. Thus, in elasto-viscoplastic formulations, $\dot{\gamma}_\alpha$ is generally a function of $\tau_\alpha$ and $\tau^\mathrm{F}_\alpha$.
For the present study, we use the flow rule proposed by Chaboche \cite{Lemaitre1990}
\begin{equation}
\dot{\gamma}_\alpha = \dot{\gamma}_{\mathrm{0}}\ \sign{\tau_\alpha} \ \left\la \frac{\abs{\tau_\alpha} - \tau^{\mathrm{F}}}{\tau^{\mathrm{D}}}\right\ra^{m},
\label{eq:Gleitrate}
\end{equation}
where we assume a common yield stress for all slip systems. The Macaulay brackets are defined by $\left\la \cdot \right\ra=\mathrm{max}(0,\cdot)$. Note that, for this particular model, the true elastic regime is bounded by $\tau^{\mathrm{F}}$. However, for large stress exponents $m$, significant plastic slip only occurs for stress values above $\tau_\textrm{crit} = \tau^\textrm{F} + \tau^\textrm{D}$.
Based on the accumulated plastic slip $\dot{\gamma}^{\mathrm{V}}$, a phenomenological linear-exponential approach 
\begin{equation}
\tau^{\mathrm{F}}=\tau_0+(\tau_{\infty}-\tau_0)\Bigl(1-\exp\Bigl(-\frac{\Theta_0-\Theta_{\infty}}{\tau_{\infty}-\tau_0} \ \gamma^{\mathrm{V}}\Bigr)\Bigr)+\tau_{\infty} \gamma^{\mathrm{V}} \quad\text{with}\quad \dot{\gamma}^{\mathrm{V}}=   \sum_{\alpha = 1}^{12} \ \abs{\dot{\gamma}_\alpha}
\label{eq:lin_exponentielle_Verfestigung}
\end{equation}
serves as the hardening law.
Irrespective of the particular formulations for \eqref{eq:Gleitrate} and \eqref{eq:lin_exponentielle_Verfestigung}, the shear stresses $\tau_\alpha$ constitute the driving force for plastic deformations on the single crystal level. Hence, we consider the maximum shear stress $\tau_{\mathrm{max}}$ as our primary quantity of interest when investigating the stress distribution. More precisely, we focus on the dimensionless shear stress
\begin{equation}
\tau_\mathrm{max}^*=\frac{\tau_{\mathrm{max}}}{\overline{\sigma}_{11}},
\label{eq:Tau_Stern}
\end{equation} 
restricting to uniaxial loading in $\boldmath{e}_1$ direction with associated macroscopic stress $\overline{\fsigma} =  \overline{\sigma}_{11}\, \sty{e}_1\otimes\sty{e}_1$. Note that $\tau_\mathrm{max}^*$ coincides with the Schmid factor under the assumption of a homogeneous stress-field. However, in practice, its value may exceed the maximum value of $0.5$ for the Schmid factor, due to local stress concentrations in the microstructure.

\subsection{Statistical evaluation}
\label{sec:statistics}

For the present study, we focus on the stress distribution in grains with certain characteristic orientations with respect to the macroscopic loading. More precisely, we consider grains with a $\la 100\ra$, $\la 110\ra$ and $\la 111\ra$ configuration, where the Miller index denotes the crystallographic direction parallel to the tensile axis. However, for a single grain, its particular size, shape and neighborhood \cite{sauzay2007cubic,LAVERGNE2013387,gelebart2021grain} may have a large impact on its stress distribution. To obtain representative results, i.e., the distribution for all grains of a certain orientation in the infinite volume limit, we follow the strategy of Brenner et al. and Lebensohn et al. \cite{BRENNER20093018,LEBENSOHN20045347}. In particular, we generate an ensemble of finite-sized polycrystalline microstructures, each containing a $\la 100\ra$, $\la 110\ra$ and $\la 111\ra$ grain embedded in a random neighborhood. Thus, by evaluating the stress distribution for a particular orientation over all grains in the ensemble, we mitigate the influence of the neighborhood effect on our results.
For characterizing the stress distribution, we consider the mean value and standard deviation
\begin{equation}
\la \tau_\mathrm{max}^* \ra= \frac{1}{N} \ \sum_{i = 1}^{N} \ (\tau_\mathrm{max}^*)_i \quad\text{and}\quad
SD(\tau_\mathrm{max}^*)= \sqrt{\frac{1}{N} \ \sum_{i = 1}^{N} \ ((\tau_\mathrm{max}^*)_i-\la \tau_\mathrm{max}^* \ra)^2}.
\label{eq:Mean_and_SD}
\end{equation} 
When discussing the shape of the stress distribution, see Sec.~\ref{sec:normality}, we further examine the standardized third and fourth statistical moments, i.e., the skewness and the excess kurtosis.
Firstly, the skewness, defined by
\begin{equation}
SK(\tau_\mathrm{max}^*) = \frac{\frac{1}{N} \ \sum_{\alpha = 1}^{N} \ ((\tau_\mathrm{max}^*)_\alpha-\la \tau_\mathrm{max}^* \ra)^3}{SD(\tau_\mathrm{max}^*)^3},
\label{eq:SK}
\end{equation}
serves as a measure for the asymmetry of a distribution, where positive values signify right-skewed distributions and negative values indicate left-skewed distributions \cite{d1990suggestion}. As a simple heuristic, a right-skewed distribution is associated with a longer tail on its right and vice versa. Some care is advised, as the textbook rule, where a long tail on the right (left) is indicated by the mean being farther out in the tail than the median, fails surprisingly often \cite{von2005mean}. However, for the large samples of the present study which lead to close to Gaussian shaped distributions, the aforementioned geometric intuition is expected to hold.
Secondly, the excess kurtosis
\begin{equation}
EK(\tau_\mathrm{max}^*) = \frac{\frac{1}{N} \ \sum_{i = 1}^{N} \ ((\tau_\mathrm{max}^*)_i-\la \tau_\mathrm{max}^* \ra)^4}{SD(\tau_\mathrm{max}^*)^4} - 3
\label{eq:KU}
\end{equation} 
quantifies the tail-heaviness, i.e., the occurrence of outliers, in comparison to the normal distribution and does not measure the peakedness \cite{westfall2014kurtosis}. Positive values of the excess kurtosis represent heavier tails than the normal distribution and vice versa \cite{d1990suggestion}. Both skewness and excess kurtosis are zero in case of a normal distribution and may therefore additionally serve as measures for the normality of a distribution. In this context, one approach is to interpret absolute values smaller than one as fairly close to normal \cite{huck2012reading}. Lei and Lomax \cite{lei2005effect} propose to assume normality only at absolute values of 0 and slight non-normality at a skewness between 0.3 and 0.4 and an excess kurtosis around 1.

\subsection{Computational setup and algorithms}
\begin{figure}[H]
	\centering
	
	\begin{subfigure}[c]{0.4\textwidth}
		\begin{tikzpicture}
		\node(fig)[anchor = south west, inner sep = 0] at (0,0)
		{\includegraphics[trim=170 210 130 280,clip,width=\textwidth]{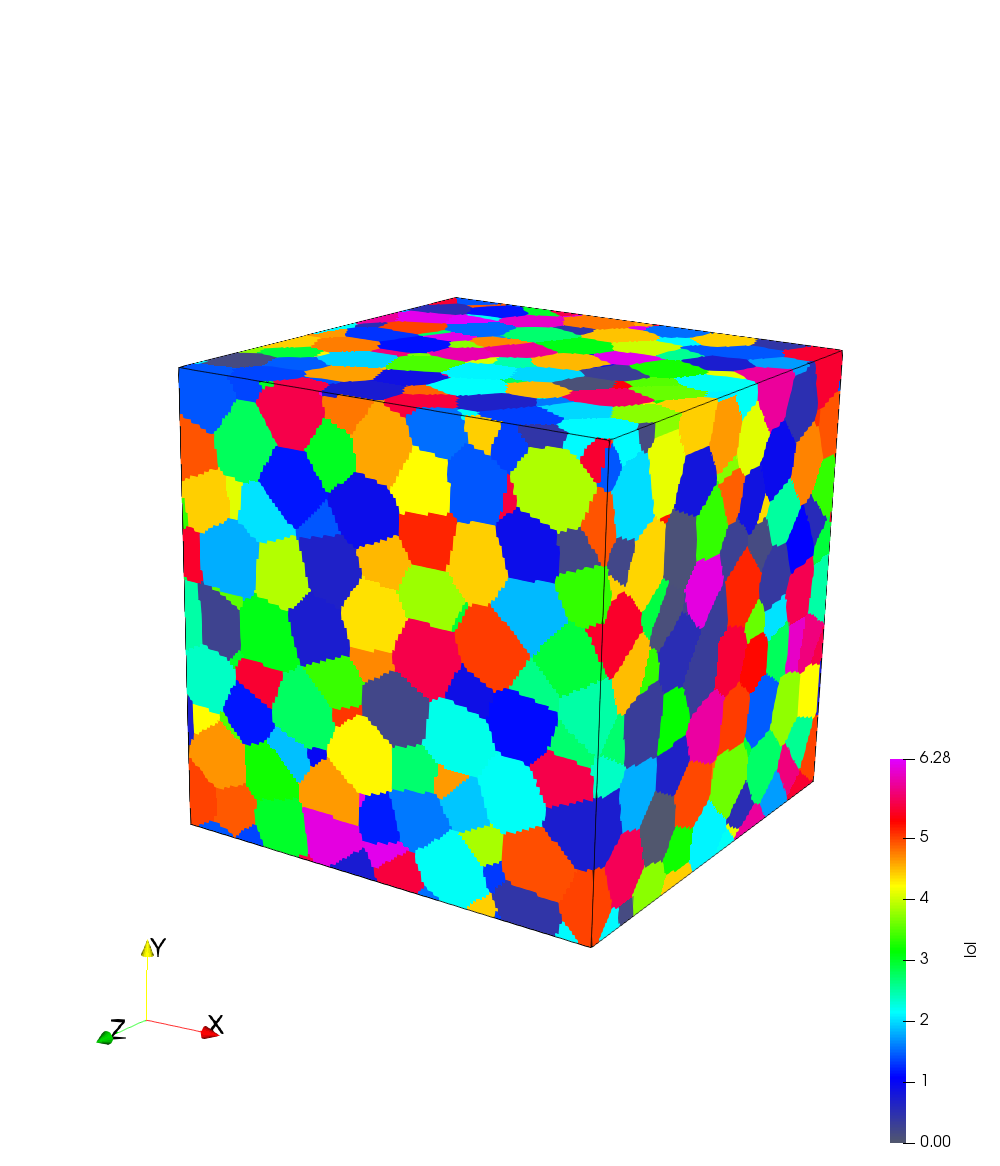}};
		\node(cos)[anchor = south west, inner sep = 0] at (5.2,0)
		{\includegraphics[trim=90 120 770 920,clip,width=0.2\textwidth]{figures/Mikrostruktur_bunt.png}};
		\end{tikzpicture}
	\end{subfigure}
	\begin{subfigure}[c]{0.4\textwidth}
		\begin{tikzpicture}
		\node(fig)[anchor = south west, inner sep = 0] at (0,0)
		{\includegraphics[trim=170 210 130 280,clip,width=\textwidth]{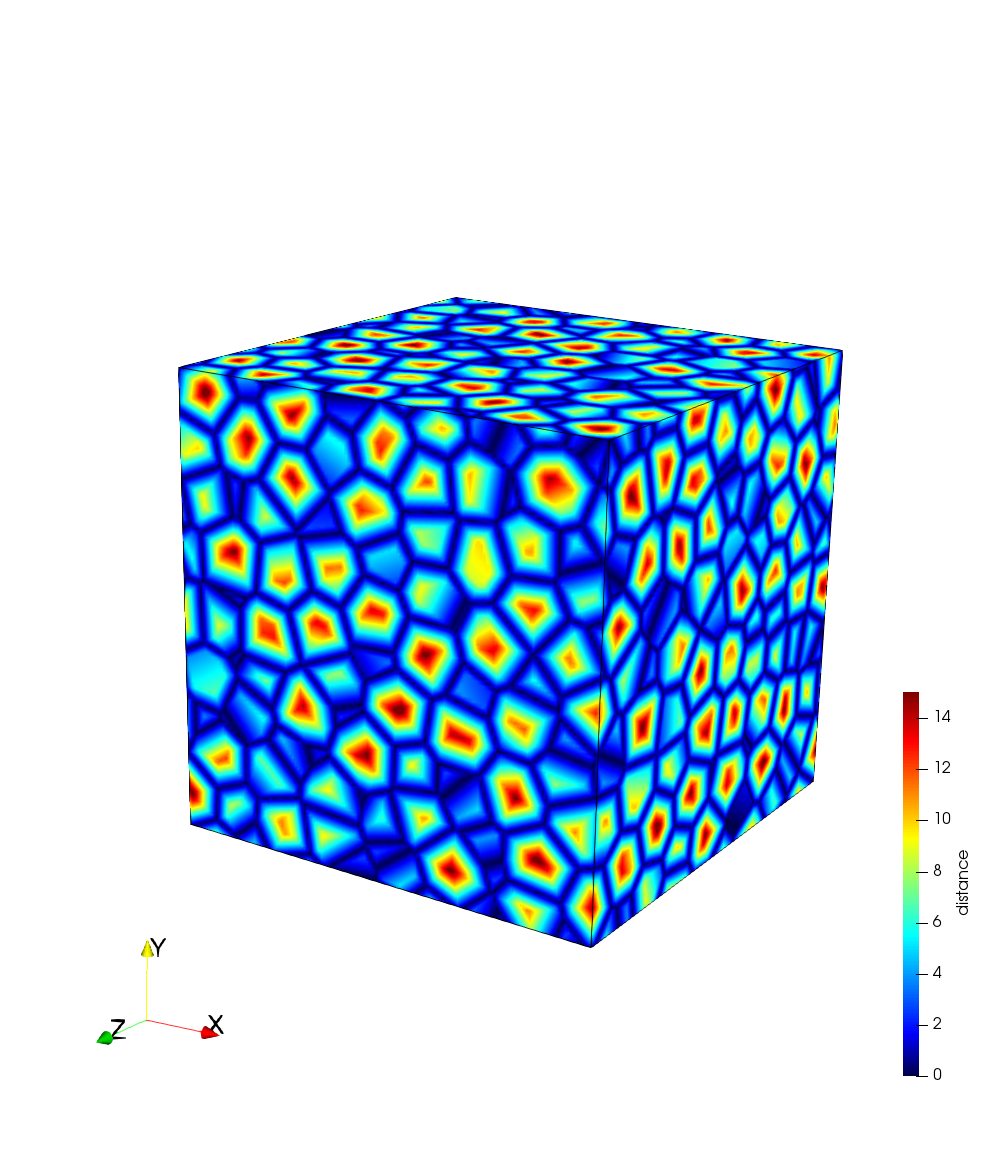}};
		\node(cos)[anchor = south west, inner sep = 0, label=below:$d$ in $\mu$m ] at (6.5,1)
		{\includegraphics[trim=910 80 45 600,clip,width=0.08\textwidth]{figures/Mikrostruktur_DN1_jet.png}};
		\end{tikzpicture}
	\end{subfigure}
	\caption{Microstructure (left, colored according to orientation) and associated distance map (right, colored according to  distance $d$ to the closest grain boundary)}
	\label{fig:DistanceMap}
\end{figure}

For generating the ensemble of polycrystalline microstructures, we rely on the algorithm for computing centroidal Laguerre tessellations by Kuhn et al. \cite{kuhn2020fast} based on the framework by Bourne et al. \cite{bourne2020laguerre}, see Fig.~\ref{fig:DistanceMap}(left). By prescribing the same volume fraction for all grains and enforcing a centroidal tessellation, the influence of grain size and shape on the resulting stress distributions is minimized. For all generated structures, the orientations are distributed uniformly and the grains closest to the chosen set of investigated orientations are set to $\la 100\ra$, $\la 110\ra$ and $\la 111\ra$, respectively. All microstructures consist of $8^3=512$ grains, so that each grain is embedded in a sufficiently large random neighborhood of at least 3 layers \cite{BRETIN201936}.
In a post-processing step, the spatial distribution of stresses is investigated in Sec.~\ref{sec:lin_elast_space}. In this context distance maps, see Fig.~\ref{fig:DistanceMap}(right), are used which assign to each voxel the Euclidean distance of the voxel center to the nearest grain boundary \cite{rollett2010stress,castelnau2020multiscale,kuhbach2020quantification}.

For computing the stress-fields, we use an in-house FFT-based micromechanics code written in Python 3.7 with Cython extensions. Computations in the linear elastic range are solved with the conjugate gradient method \cite{zeman2010accelerating}. For the crystal-plasticity computations, the Newton-CG method \cite{NewtonKrylov, gelebart2013non} in the stress-based framework \cite{wicht2020efficient} is used. To avoid ringing artifacts, typically observed for the classic discretization by trigonometric polynomials \cite{moulinec1998numerical}, we rely on the staggered grid discretization by Schneider et al.\cite{StaggeredGrid}.

In this study the Conjugated Gradient (CG) method \cite{zeman2010accelerating} is used for the linear elastic simulations and the Newton-CG method in the dual formulation \cite{wicht2020efficient} for the elasto-viscoplastic simulations. 

More details concerning FFT-based solvers can be found in Wicht, Schneider and Böhlke \cite{wicht2020efficient, QuasiNewton2019}.

 \subsection{The Maximum Entropy Method as an analytical benchmark}

 \label{sec:MEM}
 
 In addition to full-field simulations, an analytical approach known as the Maximum Entropy Method (MEM) \cite{kreher:89} is used to estimate the distribution of maximum shear stress for the elastic regime. For a detailed introduction of the method and a comprehensive study of its capabilities, we refer to Krause and Böhlke \cite{krause:20}. 
 The MEM estimates the probability distribution of local stresses and strains by the principle of maximum entropy first developed by Jaynes \cite{jaynes:63}. It takes the probability distribution of phase properties and the effective properties as an input, then returns the first and second moments of local stresses and strains. Using this approach, higher statistical moments are zero.
 
 The MEM solution given here is based on Krause and Böhlke \cite{krause:20} box 12, which is meant for general thermoelastic polycrystals. As the linear elastic case is considered here, the formulas are adapted with $\bar{\feps}^\mathrm{I} = \bar{\feps}$, $\bar{\feps}^\mathrm{II} = \fzero$. For any given single crystal orientation $\fQ$, the formulas yield one six-dimensional gaussian probability distribution of strains with the strain means and covariances 
 \begin{align}
\la \feps \ra_\fQ &= \fQ \star \ffC_0^{-1}[\fmu_\varepsilon] + \fmu_\sigma, \\
\ffK_\fQ &= \frac{1}{2 d} \ffC^{-1}.
 \end{align}
 
The $\fmu$ and $d$ terms are defined in terms of Voigt and Reuss bounds. For statistical isotropy, these bounds are isotropic. For cubic $\ffC_0$ in specific, the compression modulus $K$ is homogeneous. The spherical part of the strain $\rm{sph}(\bfeps)$ then becomes homogeneous, and the formulas in box 12 of the aforementioned paper \cite{krause:20} are singular. A MEM solution can nonetheless be obtained if the deviatoric part $\rm{dev}(\bfeps)$ is solved separately in terms of the shear moduli of the Voigt and Reuss bounds, which are denoted as $G_+$ and $G_-$ respectively. The solution takes the form
\begin{align}
\fmu_\varepsilon &= 2 G_- \frac{G_+ - \Geff}{G_+ - G_-} \mathrm{dev}(\bfeps), \\
\fmu_{\sigma} &= \frac{\Geff - G_-}{G_+ - G_-} \rm{dev}(\bar{\feps}) + \rm{sph}(\bfeps), \\
d &= \frac{2}{3} \frac{(G_+ - \Geff)(\Geff- G_-)}{G_+ - G_-} \rm{dev}(\bfeps) \cdot \rm{dev}(\bfeps).
\end{align}

After combining the results, stress means and covariances are recovered through $\fsigma = \ffC[\feps]$, leading to
\begin{align}
\la \fsigma \ra_\fQ &= \frac{G_- (G_+-\Geff)}{G_+-G_-}\mathrm{dev}(\bfeps) + \frac{(\Geff - G_-)}{G_+ - G_-}\fQ \star \ffC_0[\mathrm{dev}(\bfeps)] + 3K\mathrm{sph}(\bfeps), \label{eq:MEM1}\\
\ffK_{\sigma\fQ} &= \frac{G_+ - G_-}{(G_+ - \Geff)(\Geff- G_-)} \frac{3}{\rm{dev}(\bfeps) \cdot \rm{dev}(\bfeps)} \fQ\star\ffC_0. \label{eq:MEM2}
\end{align}

 The MEM prediction for the resulting shear stress distribution for $\tau^*_\textrm{max}$ is obtained by a Monte-Carlo approach, sampling from the distribution of the Cauchy stress $\fsigma$ using \eqref{eq:MEM1} and \eqref{eq:MEM2}.
 
 \section{Influence of resolution and ensemble size}
 \label{sec:resolution_realization}
 \subsection{Resolution study for a single microstructure}
\label{sec:resolution_study}

\begin{figure}[htb]
	\begin{subfigure}[c]{\textwidth}
		\begin{subfigure}[c]{0.33\textwidth}
			\centering 
			\large $\la 100\ra$
			\includegraphics[width=\textwidth]{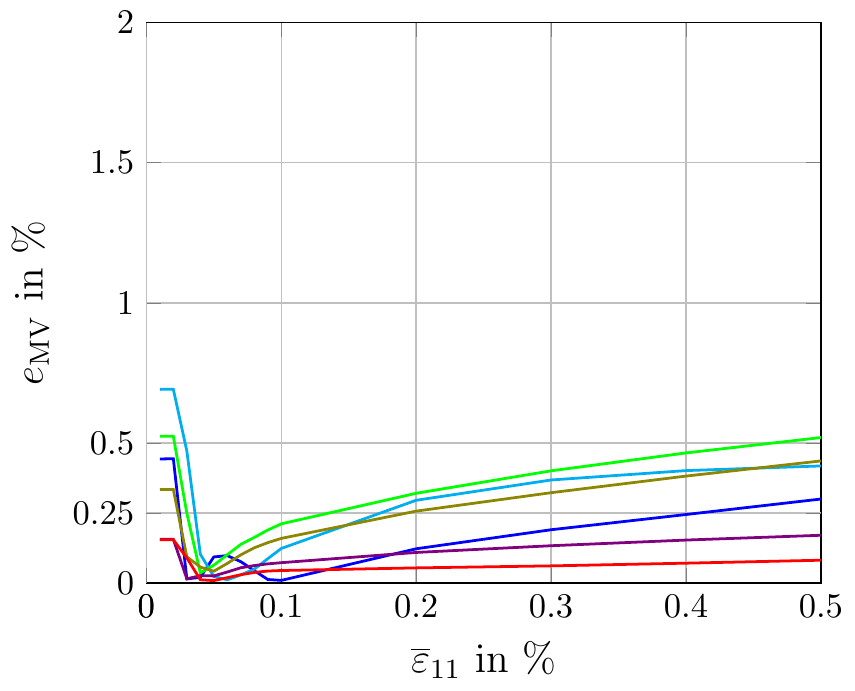}
		\end{subfigure}
		\begin{subfigure}[c]{0.33\textwidth}
			\centering 
			\large $\la 110\ra$
			\includegraphics[width=\textwidth]{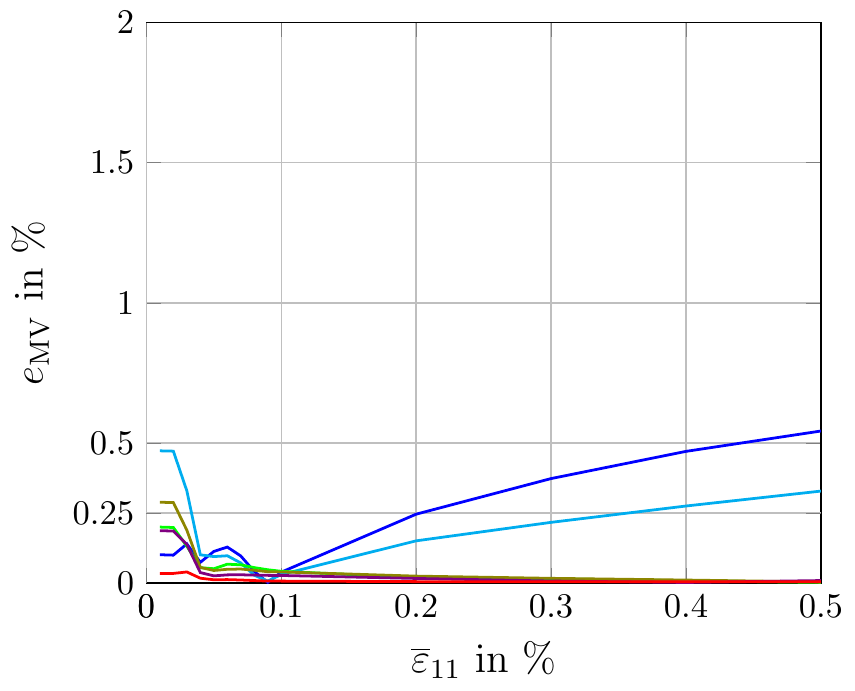}
		\end{subfigure}
		\begin{subfigure}[c]{0.33\textwidth}
			\centering 
			\large $\la 111\ra$
			\includegraphics[width=\textwidth]{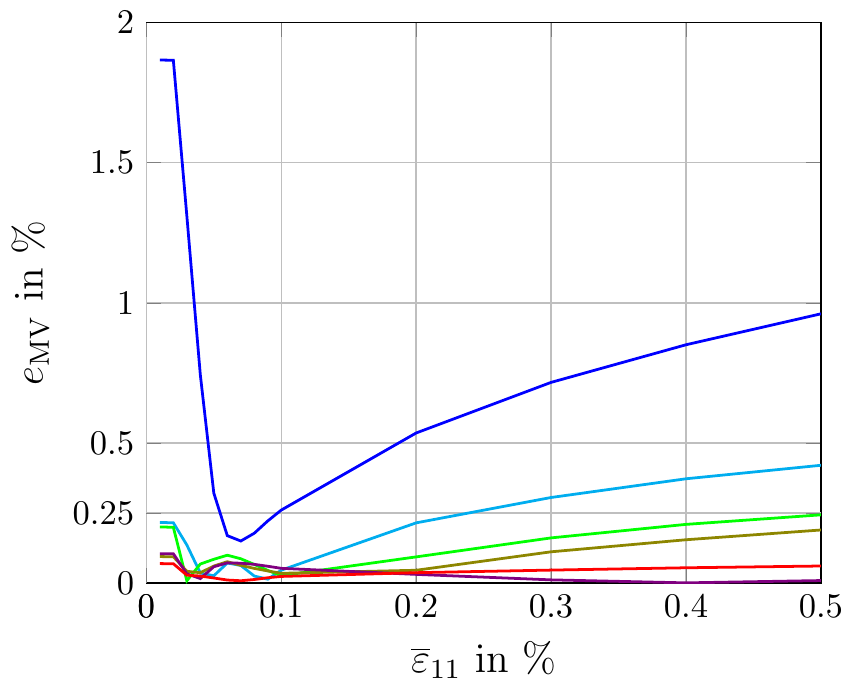}
		\end{subfigure}
		\subcaption{Relative error of mean value for uniaxial extension to $0.5\%$ strain}
		\vspace{5pt}
	\end{subfigure}
	\begin{subfigure}[c]{\textwidth}
		\begin{subfigure}[c]{0.33\textwidth}
			\centering 
			\includegraphics[width=\textwidth]{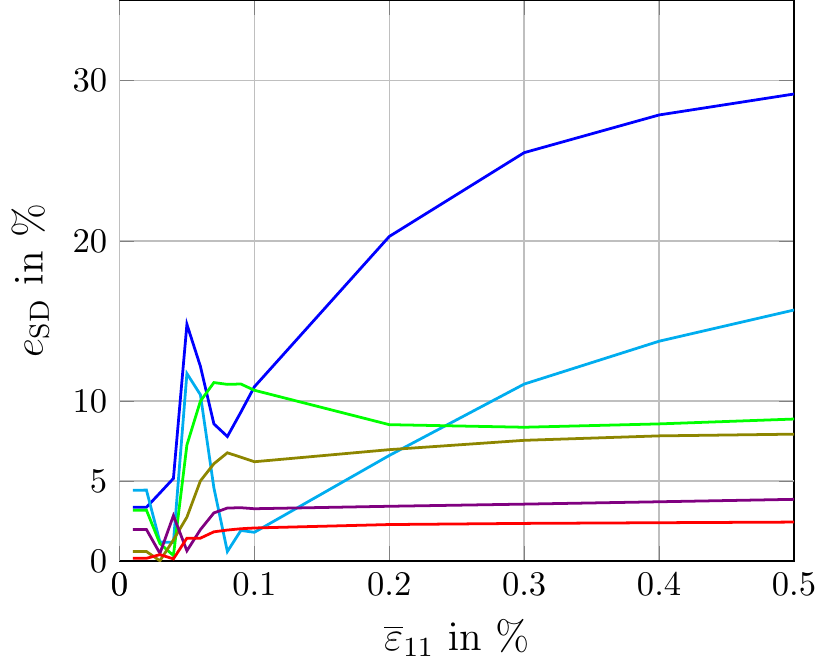}
		\end{subfigure}
		\begin{subfigure}[c]{0.33\textwidth}
			\centering 
			\includegraphics[width=\textwidth]{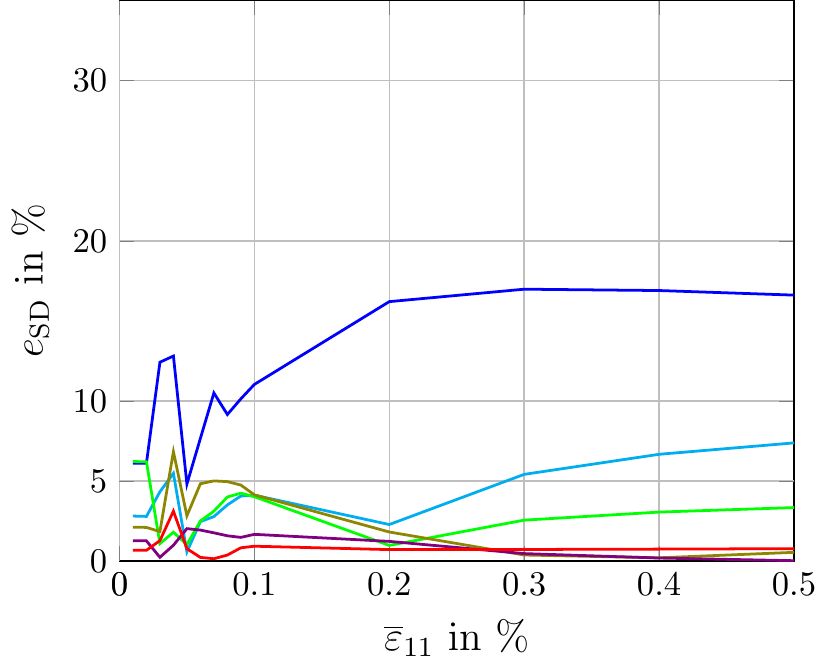}
		\end{subfigure}
		\begin{subfigure}[c]{0.33\textwidth}
			\centering 
			\includegraphics[width=\textwidth]{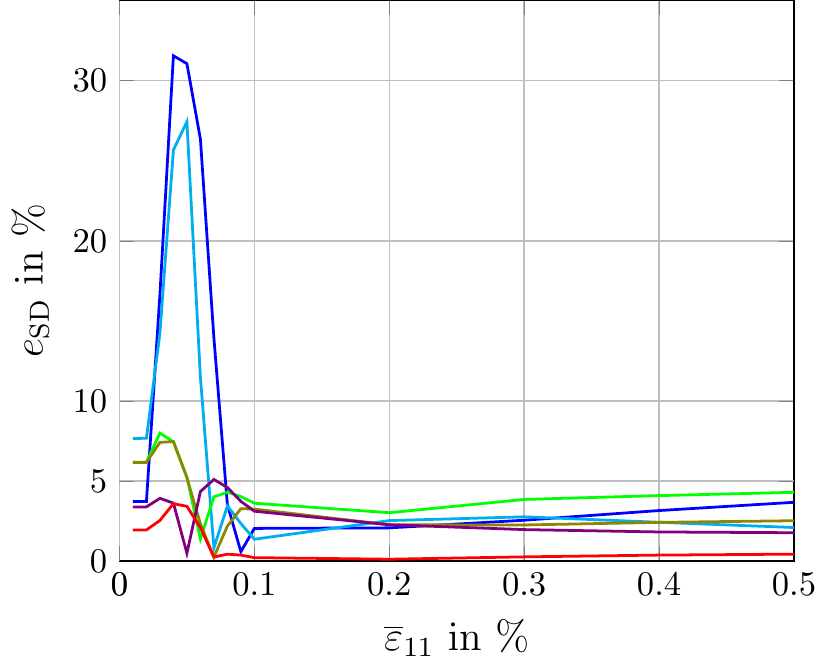}
		\end{subfigure}
		\subcaption{Relative error of standard deviation for uniaxial extension to $0.5\%$ strain}
		\vspace{5pt}
	\end{subfigure}
	\includegraphics[width=\textwidth]{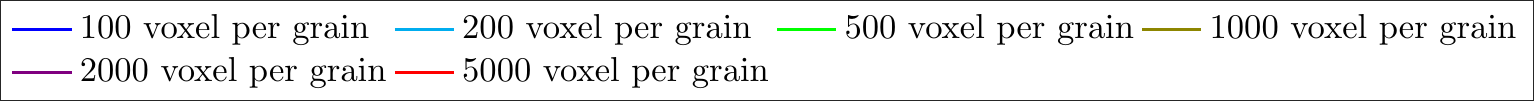}
	\caption{Resolution study for a single realization}
	\label{fig:resolution_ID1}
\end{figure}

\begin{table}[htb]
	\centering
	\begin{tabular}{lcc}
		\toprule
		Resolution & $e_\mathrm{MV}^\mathrm{max}$ in $\%$&$e_\mathrm{SD}^\mathrm{max}$ in $\%$\\
		\midrule
		$100$&$ 1.86$&$ 31.56$\\
		$200$& $0.69$&$ 27.44$\\
		$500$& $0.52$&$11.15$\\
		$1000$& $0.44$&$7.93$\\
		$2000$&$ 0.19$&$ 5.10$\\
		$5000$& $0.15$&$ 3.59$\\
		\bottomrule
	\end{tabular}
	\caption{Maximum relative error caused by resolution}
	\label{tab:max_error_resolution}
\end{table}

Computing the stress and strain field on a microstructure is associated with considerable computational effort, especially for more involved material laws such as the crystal plasticity model in Sec.~\ref{sec:crystal_plasticity}. Thus, before committing to computations on a large ensemble, we seek a suitable resolution which limits the computational cost but is still fine enough to permit a meaningful investigation of the stress field.
More precisely, for a single microstructure discretized with various resolutions, we track the mean value and standard deviation of $\tau^*_\textrm{max}$ in the designated $\la 100\ra$, $\la 110\ra$ and $\la 111\ra$ grains for a uniaxial extension to $0.5\%$ (see Sec.~\ref{sec:onset} for material parameters and boundary conditions).
Based on Lim et al. \cite{lim2019investigating}, a resolution of $10000$ voxels per grain is taken as reference. The relative errors in mean value and standard deviation for various resolutions are shown in Fig.~\ref{fig:resolution_ID1} and the maximum errors over the whole loading path are listed in Tab.~\ref{tab:max_error_resolution}. 
Notably, the accuracy for computing the mean of $\tau^*_\textrm{max}$ is hardly affected by the chosen resolution. Even for the coarsest choice of 100 voxels per grain, the relative error mostly stays well below $1\%$ and never exceeds $2\%$. However, much larger errors are observed for the standard deviation, indicating that the stress distribution cannot be properly resolved for very small voxel counts. Indeed, for an error which is reliably below $5\%$, more than 2000 voxels per grain are necessary. To be on the safe side, we use the conservative choice of 5000 voxels per grain for the remainder of this study, corresponding to an edge length of 137 voxels for a cubic volume element. Thus, in comparison to similar studies, such as Brenner et al. \cite{BRENNER20093018} (500 grains with 4200 voxels per grain) or Rollett \cite{rollett2010stress} (426 grains with 5000 voxels per grain), we arrive at a comparable resolution and volume element size.

 \subsection{Ensemble size}
\label{sec:realization_study}

\begin{figure}[H]
	\centering
	\begin{subfigure}[c]{0.4\textwidth}
		\begin{subfigure}[c]{\textwidth}
			\centering 
			\includegraphics[width=\textwidth]{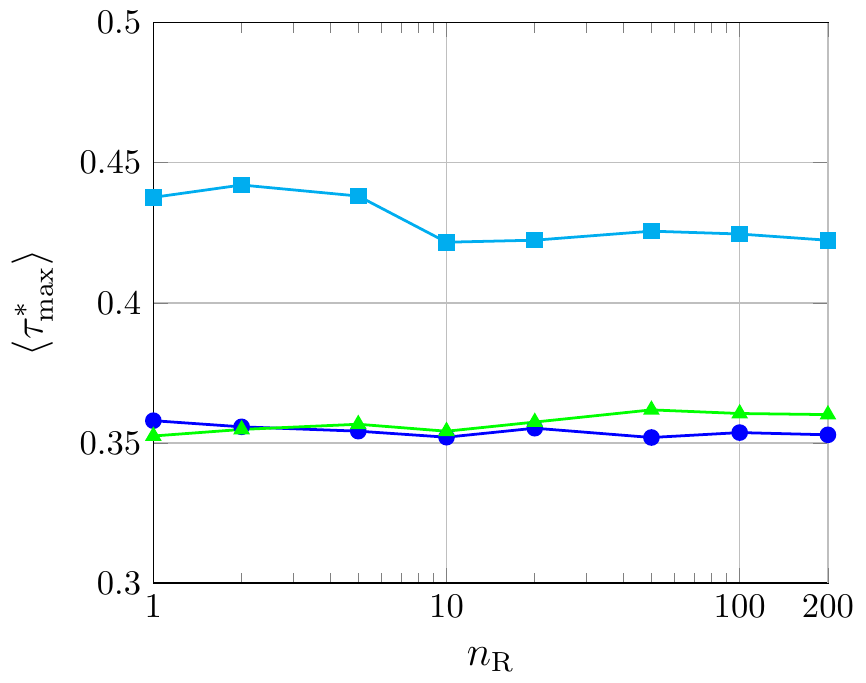}
		\end{subfigure}
		\subcaption{Mean value}
		\vspace{5pt}
	\end{subfigure}
	\begin{subfigure}[c]{0.4\textwidth}
		\begin{subfigure}[c]{\textwidth}
			\centering 
			\includegraphics[width=\textwidth]{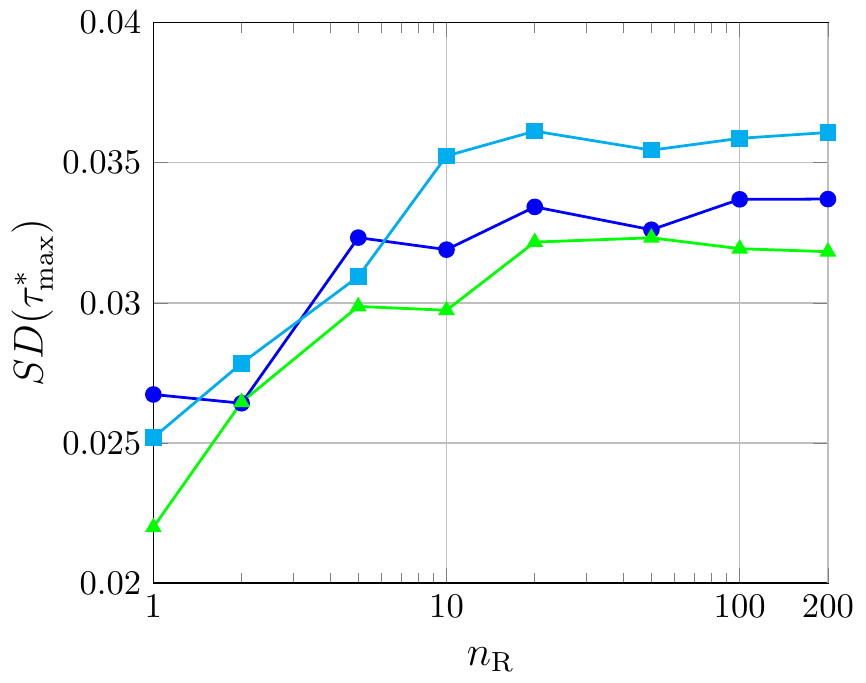}
		\end{subfigure}
		\subcaption{Standard deviation}
		\vspace{5pt}
	\end{subfigure}
	\includegraphics[width=0.5\textwidth]{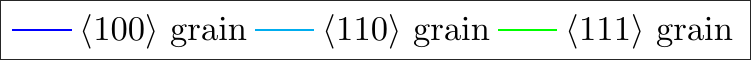}
	\caption{Absolute values in the realization study}
	\label{fig:realization_study_elastic_log}
\end{figure}

\begin{figure}[H]
	\begin{subfigure}[c]{0.4\textwidth}
		\begin{subfigure}[c]{\textwidth}
			\centering 
			\includegraphics[width=\textwidth]{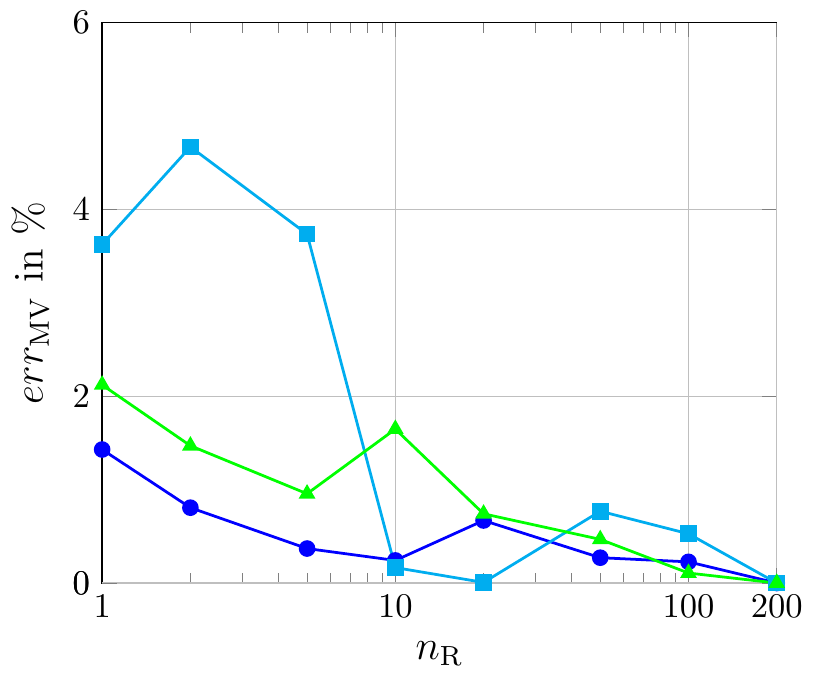}
		\end{subfigure}
		\subcaption{Relative error of mean value  \qquad \qquad}
		\vspace{5pt}
	\end{subfigure}
	\begin{subfigure}[c]{0.4\textwidth}
		\begin{subfigure}[c]{\textwidth}
			\centering 
			\includegraphics[width=\textwidth]{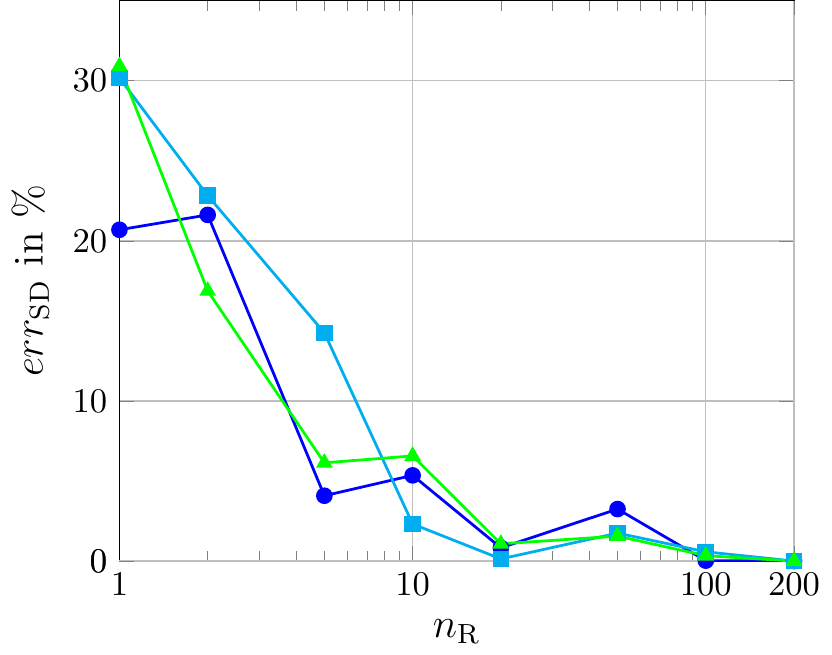}
		\end{subfigure}
		\subcaption{Relative error of standard deviation}
		\vspace{5pt}
	\end{subfigure}
	\centering
	\begin{subfigure}[c]{0.5\textwidth}
		\vspace{5pt}
		\includegraphics[width=\textwidth]{figures/Realization_convergence_error_mv_id_orient_legend.pdf}
	\end{subfigure}
	\caption{Relative error in the resolution study (reference 200 realizations)}
	\label{fig:realization_study_elastic_rel_error_log}
\end{figure}
As noted by Brenner et al. and Lebensohn et al. \cite{BRENNER20093018,LEBENSOHN20045347} the stress distribution in a single grain embedded in a specific neighborhood may differ substantially from the representative stress distribution obtained by considering a large ensemble.
To quantify the influence of the ensemble size, we plot the mean value and standard deviation of $\tau^*_\textrm{max}$ for an increasing number of realizations $n_\textrm{R}$ in Fig.~\ref{fig:realization_study_elastic_log}. For ease of exposition, we restrict to the elastic regime and use the stiffness tensor of copper, see Sec.~\ref{sec:lin_elast}.
Similar to the observations of Sec.~\ref{sec:resolution_study}, the average maximum shear stress $\langle\tau^*_\textrm{max}\rangle$ changes marginally with ensemble size. Taking a look at the relative error to our maximum ensemble size of 200, the error never exceeds $5\%$ and is consistently below $2\%$ for ensembles of 10 or more microstructures, see Fig.~\ref{fig:realization_study_elastic_rel_error_log}. Moving on to the standard deviation of $\tau^*_\textrm{max}$, the influence of ensemble size on stress distribution becomes more notable. For ensembles containing less than 10 realizations, the observed standard deviation is distinctly lower than for the reference of $n_\textrm{R} = 200$. Subsequently, the values for $SD(\tau^*_\textrm{max})$ stagnate, however, more than 50 realizations are necessary for the relative error to consistently drop below $5\%$. Thus, an ensemble size of at least 50 is recommended for representative results. Having already generated the large data set, we use the results for all 200 realizations throughout. We revisit the topic of ensemble size in Sec.~\ref{sec:normality}, when the shape of the shear stress distribution is discussed in greater detailed.

 \section{Distribution of shear stresses in the linear elastic range}
 \label{sec:lin_elast}
 
 \subsection{Influence of anisotropy}
  \label{sec:lin_elast_anisotropy}
 
\begin{figure}[htb]
	\begin{center}
		\begin{subfigure}{0.24\textwidth}
			\centering 
			\includegraphics[trim=30 0 0 0,clip,width=\textwidth]{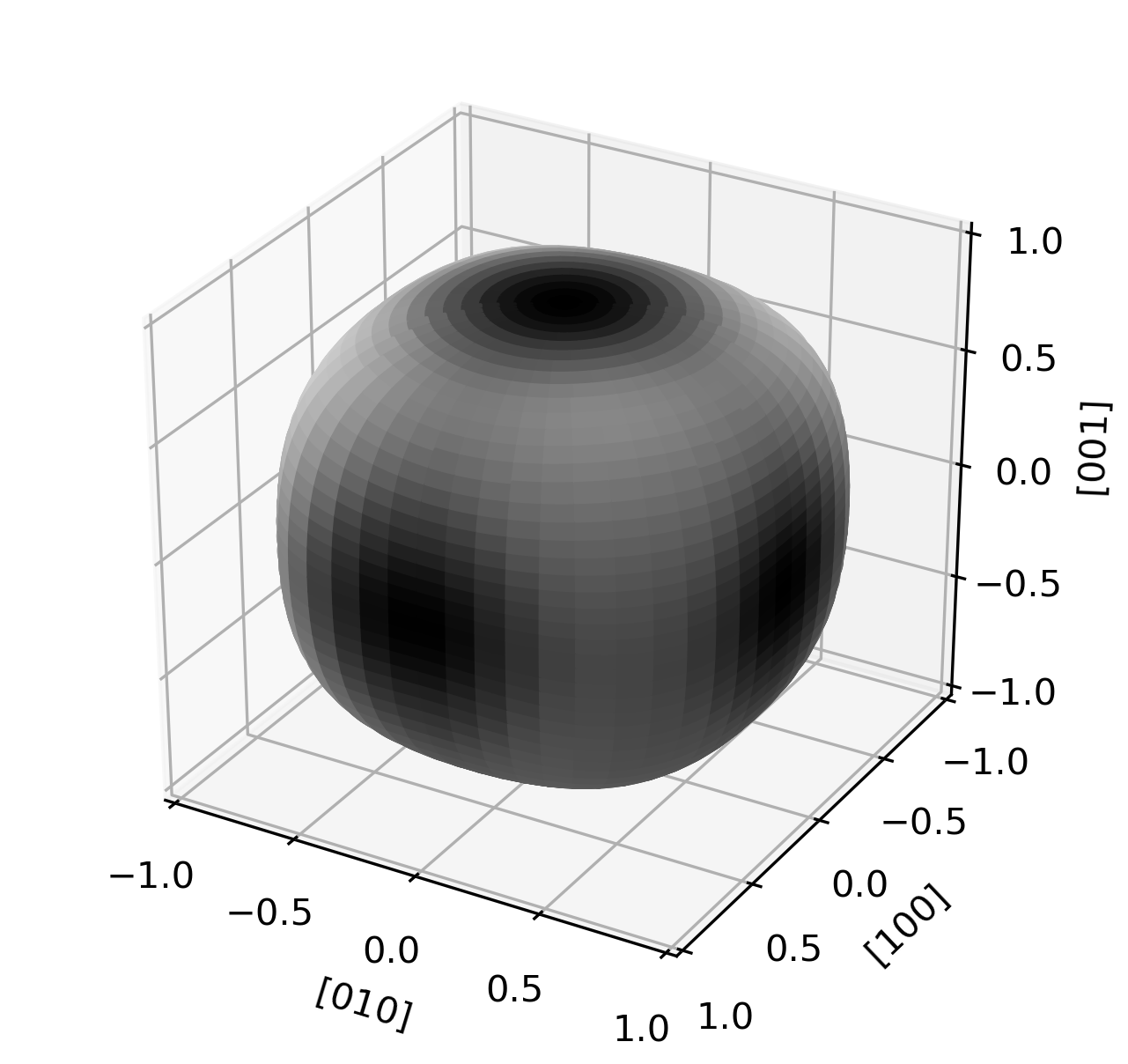}
			\subcaption{Aluminum}
		\end{subfigure}
		\begin{subfigure}{0.24\textwidth}
			\centering 
			\includegraphics[trim=30 0 0 0,clip,width=\textwidth]{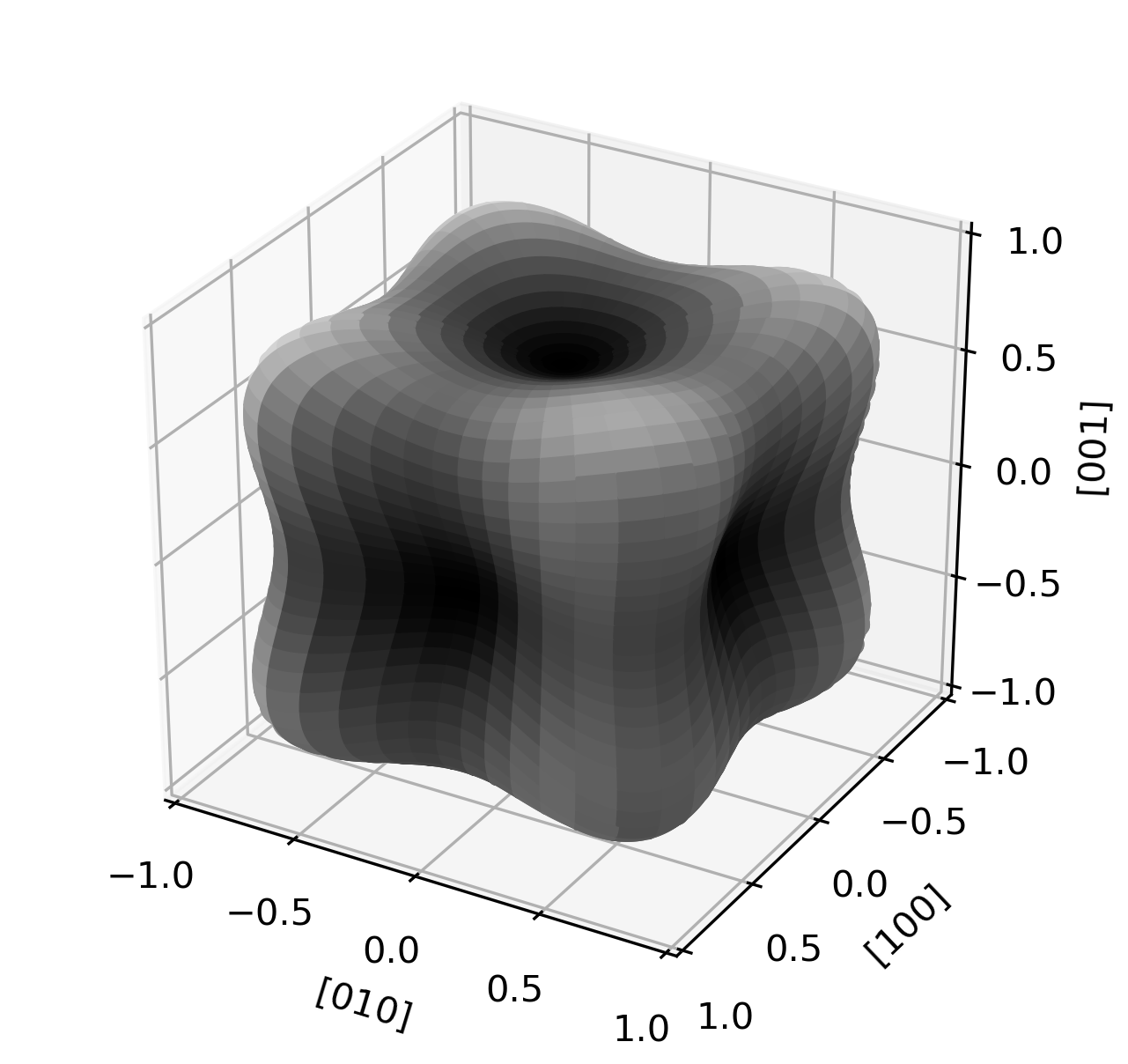}
			\subcaption{Copper}
		\end{subfigure}
		\begin{subfigure}{0.24\textwidth}
			\centering 
			\includegraphics[trim=30 0 0 0,clip,width=\textwidth]{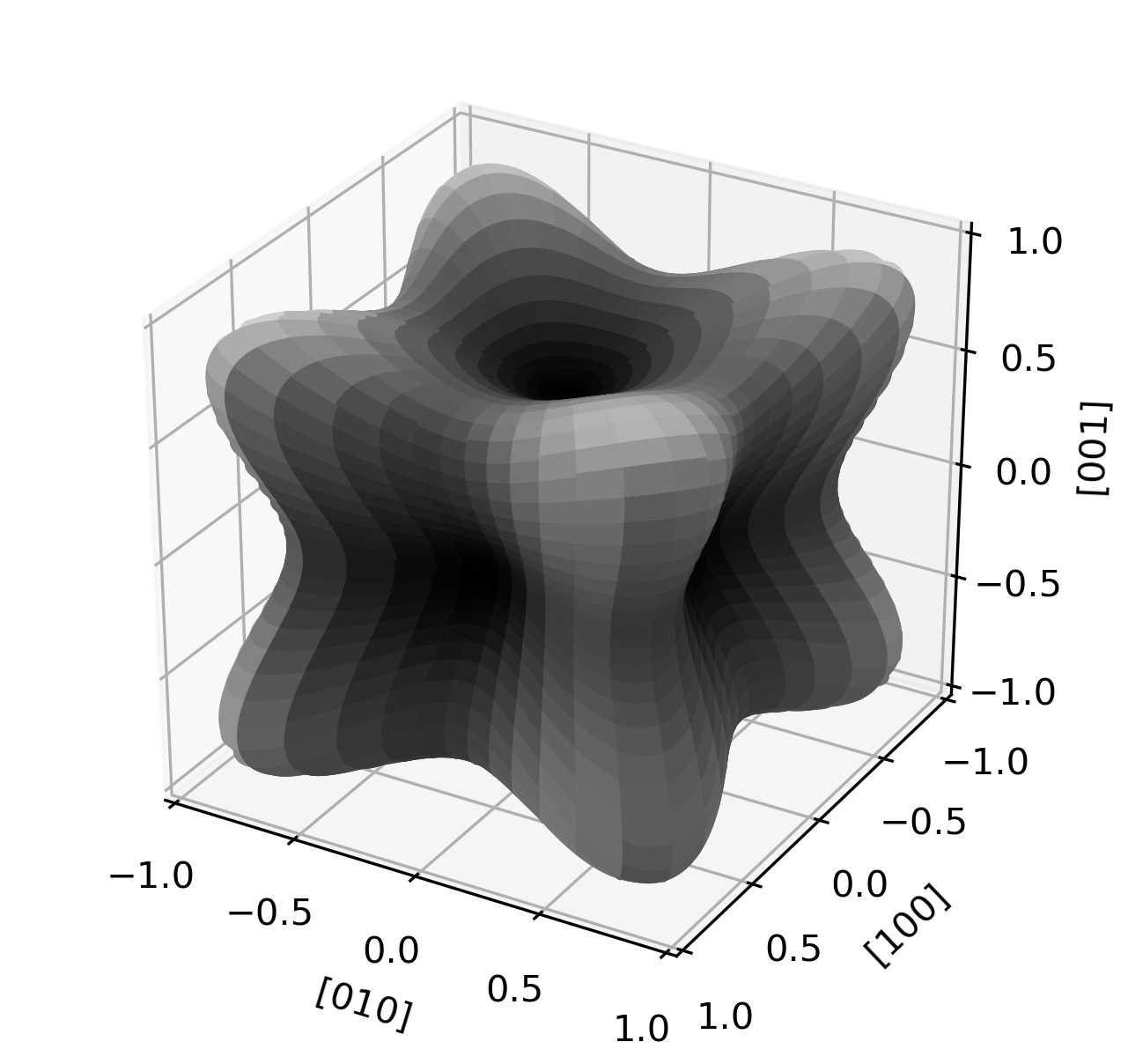}
			\subcaption{Thorium}
		\end{subfigure}
		\begin{subfigure}{0.24\textwidth}
			\centering 
			\includegraphics[trim=30 0 0 0,clip, width=\textwidth]{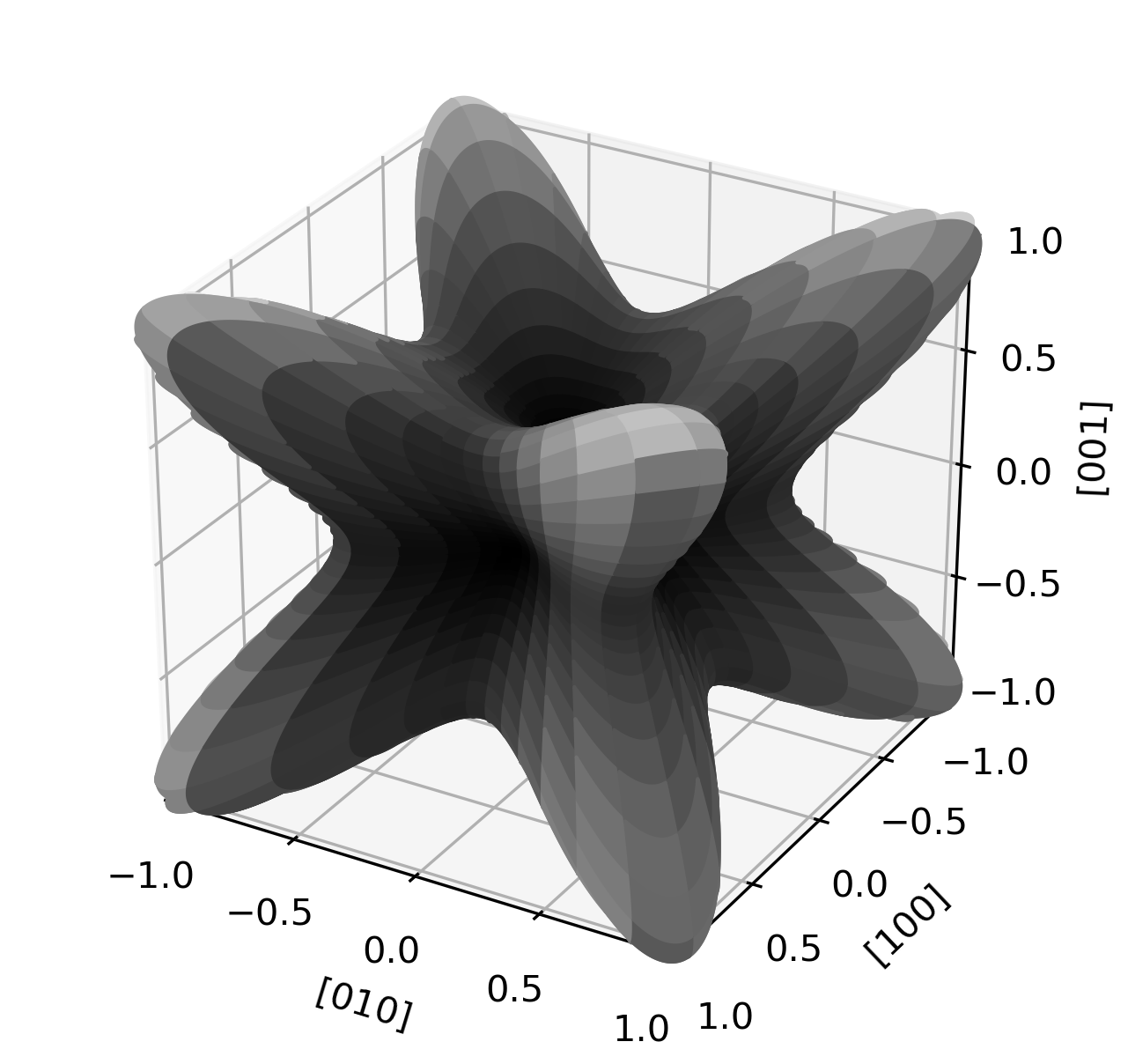}
			\subcaption{$\delta$-Plutonium}
		\end{subfigure}

			\hspace{4em}\includegraphics[width=0.5\textwidth]{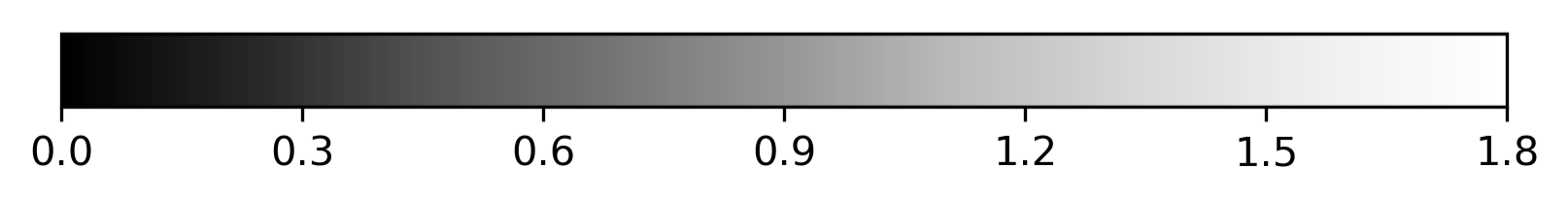}
			\newline
			$E(\fd)/E^{\mathrm{eff}}$
	\end{center}
\caption{Young's modulus of fcc materials}
\label{fig:young}
\end{figure}

In the following, we investigate the stress distribution of fcc polycrystals in the linear elastic range, focusing on the effect of anisotropy. To this end, we consider a set of 4 materials with increasingly anisotropic stiffness, see Tab.~\ref{tab:parameters}, with aluminum being nearly isotropic ($A=1.2$) and $\delta$-plutonium as the most anisotropic example.
 \begin{table}[h]
 	\centering
 	\begin{tabular}{llllll}
 		\toprule
 		Material & $C_{11}$ in GPa & $C_{12}$ in GPa & $C_{44}$ in GPa & A & Reference\\
 		\midrule
 		Al& $107$&$61$&$28$&$1.2$&\cite{paufler1978physikalische}\\
 		Cu& $170.2$&$114.9$&$61.0$&$2.2$&\cite{Simmons1971} \\	
 		Th& $77$&$51$&$46$&$3.5$ &\cite{every1992second}\\
 		$\delta$-Pu&$36$&$27$&$34$&$7.6$&\cite{every1992second}\\	
 		\bottomrule
 	\end{tabular}
 	\caption{Elastic moduli of fcc materials with respective sources}
 	\label{tab:parameters}
 \end{table}
For illustrating the different degrees of anisotropy, the Young's modulus as a function of load direction is visualized in Fig.~\ref{fig:young} in the crystal coordinate system. For better comparability, all values have been normalized with the effective Young's modulus of the respective polycrystal, obtained by averaging the computed Young's modulus over all 200 realizations of the ensemble.

\begin{figure}[htb]
	\begin{subfigure}[c]{\textwidth}
		\begin{subfigure}[c]{0.33\textwidth}
			\centering 
			\large $\la 100\ra$
			\includegraphics[width=\textwidth]{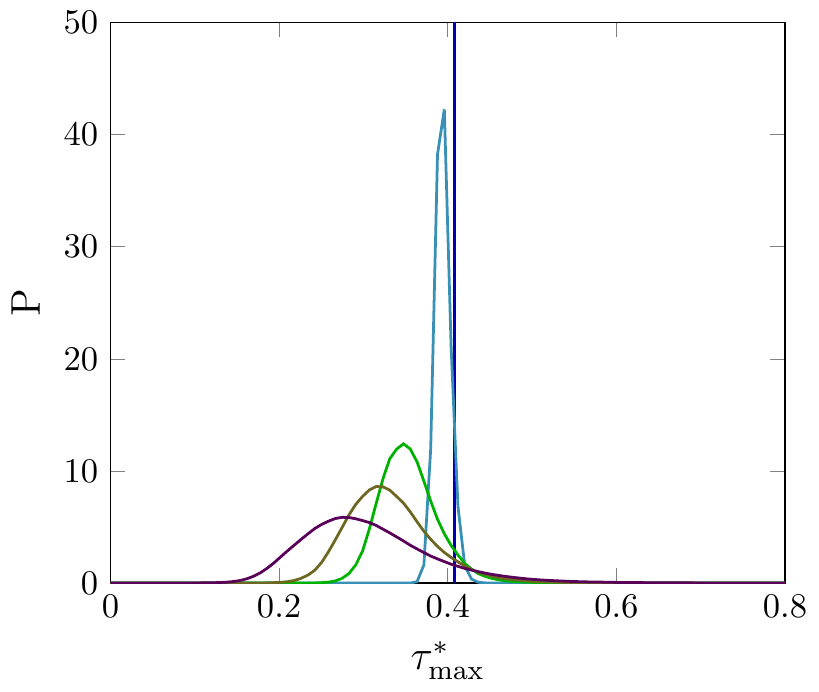}
		\end{subfigure}
		\begin{subfigure}[c]{0.33\textwidth}
			\centering 
			\large $\la 110\ra$
			\includegraphics[width=\textwidth]{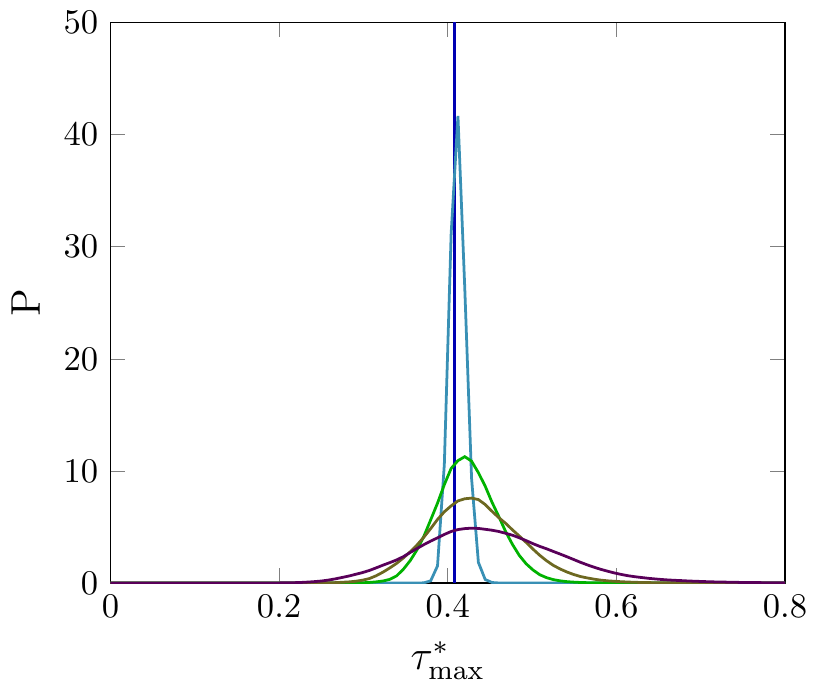}
		\end{subfigure}
		\begin{subfigure}[c]{0.33\textwidth}
			\centering 
			\large $\la 111\ra$
			\includegraphics[width=\textwidth]{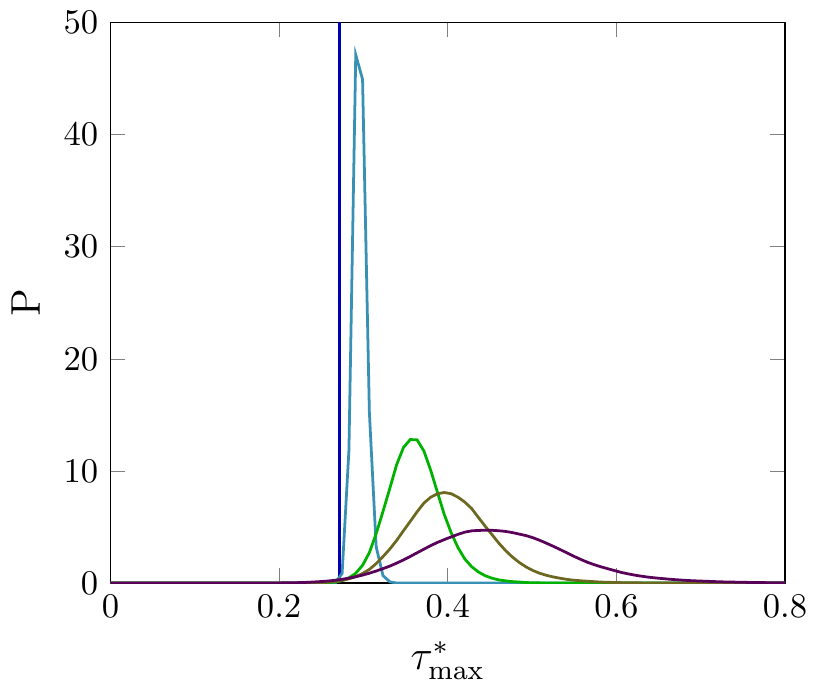}
		\end{subfigure}
	\end{subfigure}
	\centering
	\includegraphics[width=0.7\textwidth]{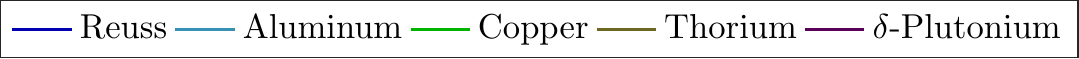}
	\caption{Material specific stress distribution in the linear elastic range}
	\label{fig:KDE_elastic}
\end{figure}

\begin{figure}[htb]
	\begin{subfigure}[c]{0.4\textwidth}
		\centering 
		\includegraphics[width=\textwidth]{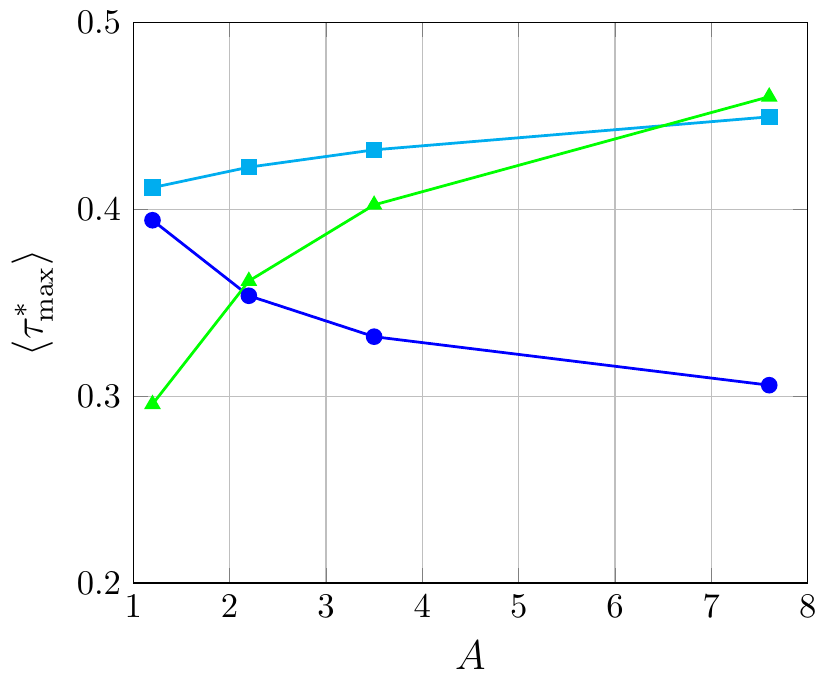}
		\subcaption{Mean value}
		\label{fig:MV_elastic}
		\vspace{5pt}
	\end{subfigure}
	\begin{subfigure}[c]{0.4\textwidth}
		\centering 
		\includegraphics[width=\textwidth]{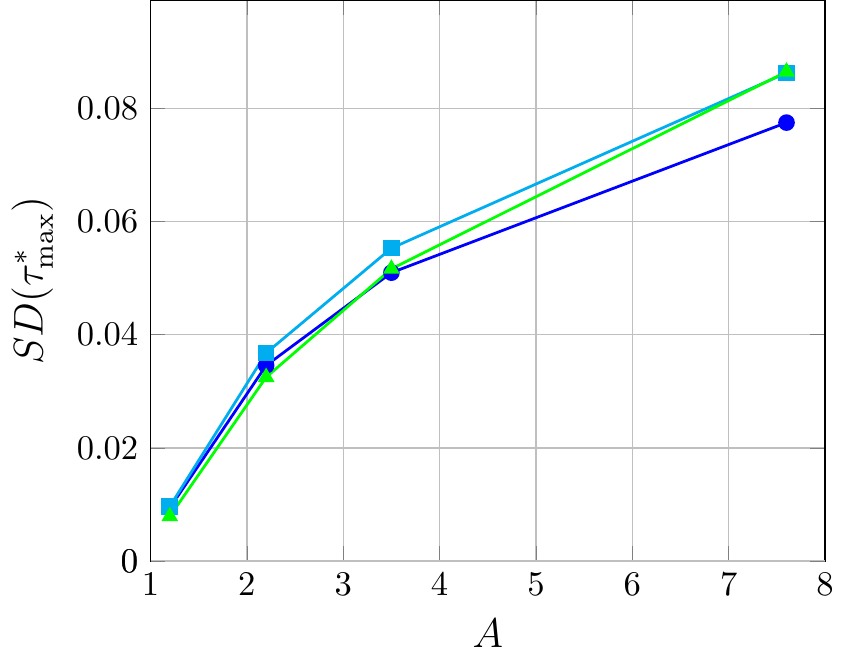}
		\subcaption{Standard deviation}
		\label{fig:SD_elastic}
		\vspace{5pt}
	\end{subfigure}
	\centering
	\begin{subfigure}[c]{0.5\textwidth}
		\vspace{5pt}
		\includegraphics[width=\textwidth]{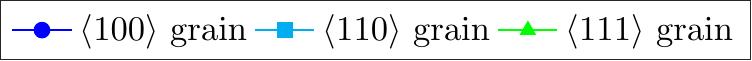}
	\end{subfigure}
	\caption{Characteristics of the stress distribution in the linear elastic range}
	\label{fig:MV,SD_elastic}
\end{figure}

With the material parameters at hand, we evaluate the distribution of the maximum shear stress $\tau^*_\textrm{max}$ in the designated $\la 100\ra$, $\la 110\ra$ and $\la 111\ra$ grains, see Fig.~\ref{fig:KDE_elastic}. As additional reference, the maximum shear stress $\tau^*_\textrm{max}$ for an assumed homogeneous stress field, i.e., the Reuss hypothesis, has been included, coinciding with the Schmid factor of the respective orientation.
In general, we notice broader stress distributions for higher anisotropy, consistent with the trends observed by Brenner et al. \cite{BRENNER20093018} in the elastic regime and by Castelnau et al. \cite{castelnau2008micromechanical} and Bhattacharyya et al. \cite{bhattacharyya2021elastoplastic} in the plastic regime. Indeed, the standard deviation of the distribution appears to be primarily a function of $A$, with the orientation only playing a minor role, see Fig.~\ref{fig:SD_elastic}.
In contrast, the trends observed for the mean value of $\tau_\mathrm{max}^*$ are closely tied to the specific grain orientation. 
For low anisotropy, $\la\tau_\mathrm{max}^*\ra$ starts close to the Reuss prediction $\tau_\mathrm{max}^* = 0.408$ for the $\la 100 \ra$ and $\la110\ra$ grains and $\tau_\mathrm{max}^* = 0.272$ for the $\la111\ra$ grain, see Fig.~\ref{fig:MV_elastic}. With increasing anisotropy, the stress distribution shifts to the left in case of the  $\la 100\ra$ grains and to the right for the $\la 111\ra$ orientation. 
For the $\la 110\ra$ grains, there is little change in $\la\tau_\mathrm{max}^*\ra$ with respect to anisotropy and all distributions stay fairly centered around the corresponding Schmid factor. The observed trends can be related to the stiffness of the crystals in the respective directions, see Fig.~\ref{fig:young} and Tab.~\ref{tab:Emodul}. For all investigated materials, the Young's modulus is lowest in $\la 100\ra$ direction and highest in $\la 111\ra$ direction. More precisely, the ratio $E_{\la 100\ra}/E_\mathrm{eff}$ is smaller than one throughout, i.e., the $\la 100 \ra$ grain is more compliant than its surroundings and therefore carries lower stresses. The effect is amplified for higher anisotropy, hence the gradual shift of $\la\tau_\mathrm{max}^*\ra$. The opposite is true for the $\la 111\ra$ grains, which are increasingly stiffer than their surroundings. For the picked materials, $E_{\la 110\ra}\approx E_\mathrm{eff}$, hence, no strong trend is noted. 

\begin{table}[htb]
	\centering
	\begin{tabular}{lcccc}
		\toprule
		Material    & $E_{\la 100\ra}$ in GPa  &$E_{\la 110\ra}$ in GPa  & $E_{\la 111\ra}$ in GPa& $E_\mathrm{eff}$ in GPa\\
		\midrule
		Al & $ 62.7$&$71.4$&$74.8$&$69.8$\\
		Cu&$77.6$&$126.9$&$161.0$&$122.0$\\
		Th&$36.4$&$73.0$&$109.8$&$73.2$\\
		$\delta$-Pu &$12.9$&$33.8$&$74.0$&$40.7$\\
		\bottomrule
	\end{tabular}
	\caption{Young's modulus for specific orientations}
	\label{tab:Emodul}
\end{table}

\begin{figure}[htbp]
	\begin{subfigure}[c]{0.8\textwidth}
		\centering 
		\large $\la 100\ra$
		\includegraphics[width=\textwidth]{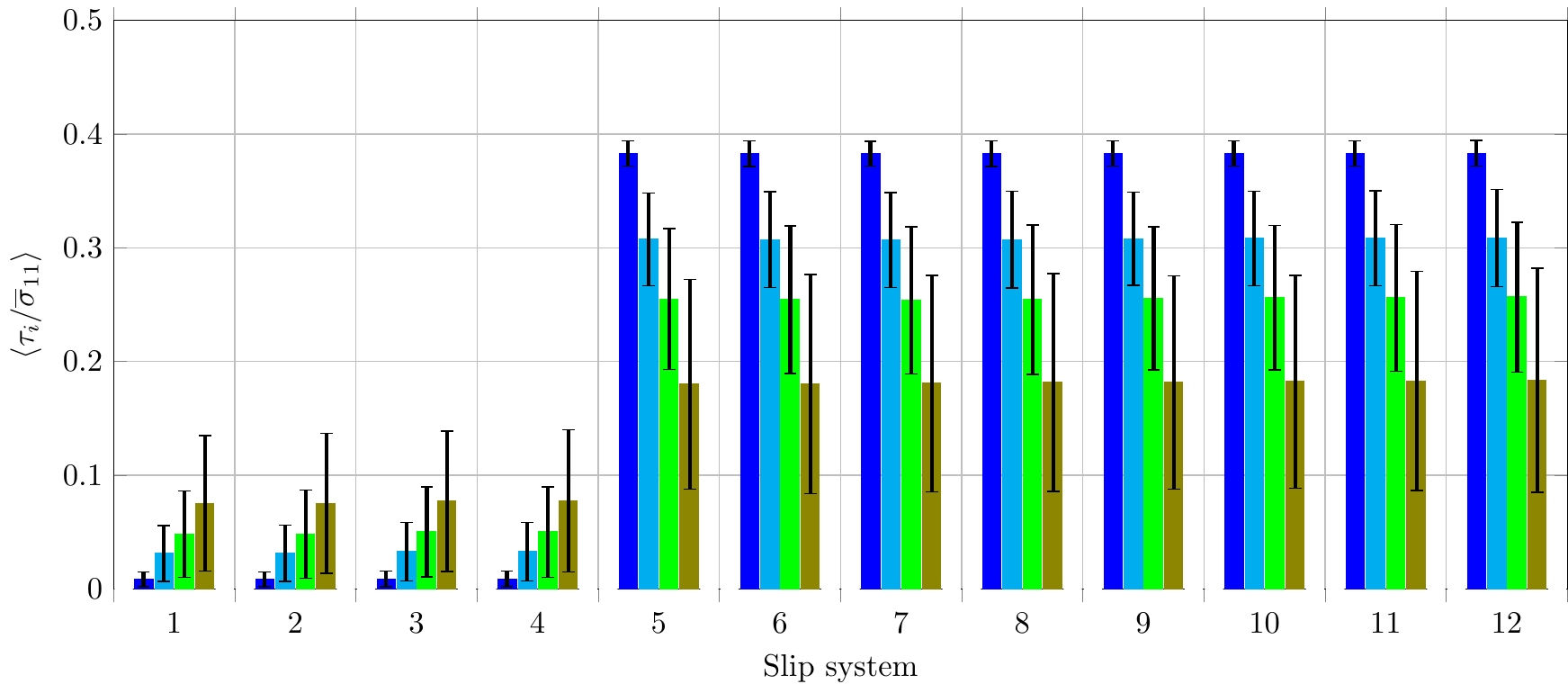}
	\end{subfigure}
	\begin{subfigure}[c]{0.8\textwidth}
		\centering 
		\large $\la 110\ra$
		\includegraphics[width=\textwidth]{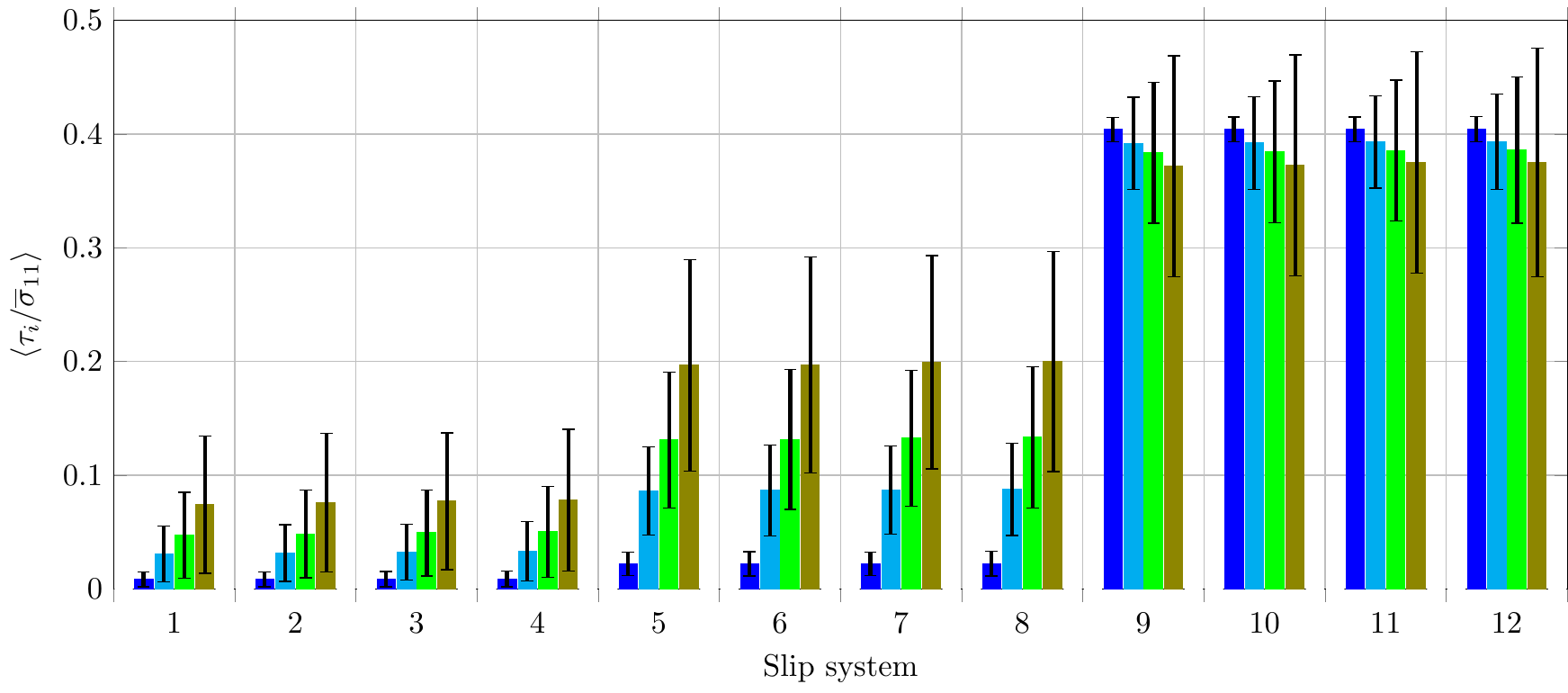}
	\end{subfigure}
	\begin{subfigure}[c]{0.8\textwidth}
		\centering 
		\large $\la 111\ra$
		\includegraphics[width=\textwidth]{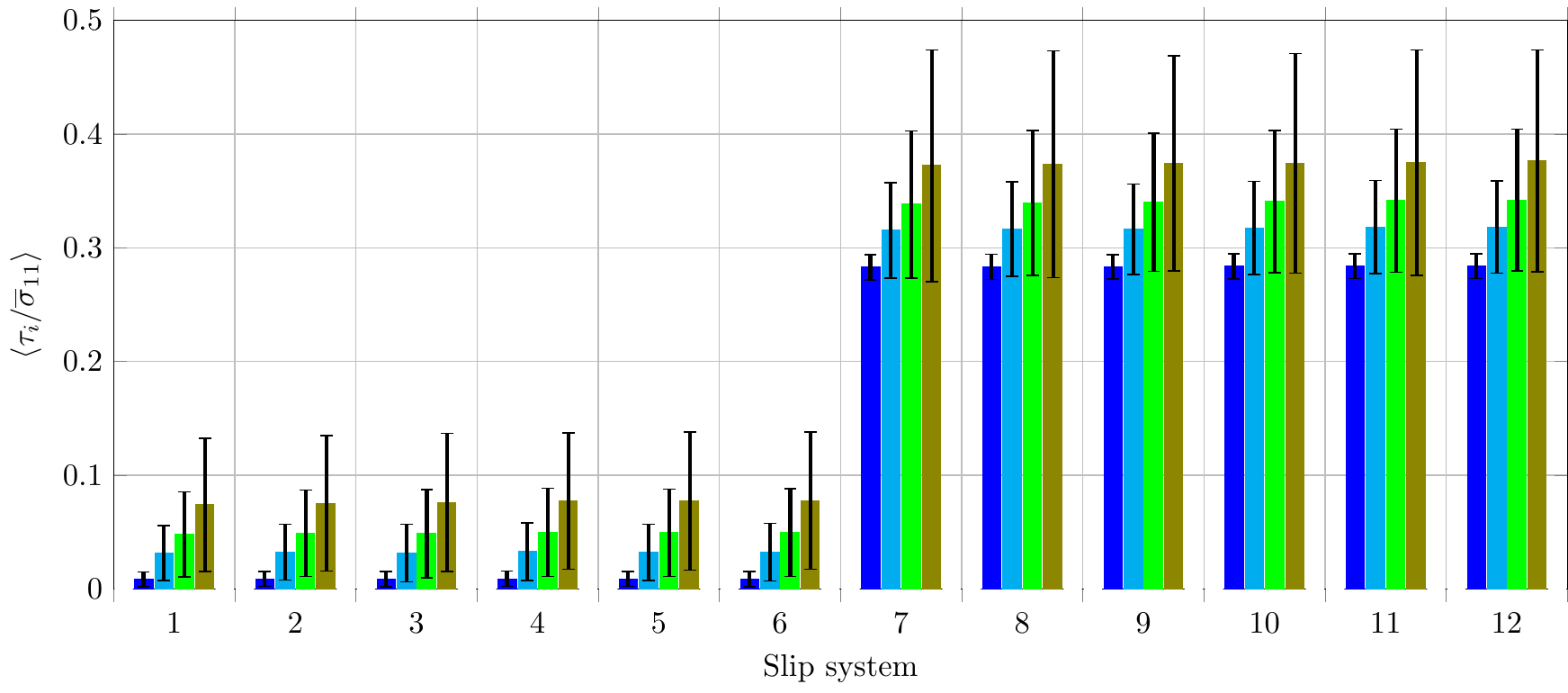}
	\end{subfigure}
	\centering
	\includegraphics[height=0.04\textwidth]{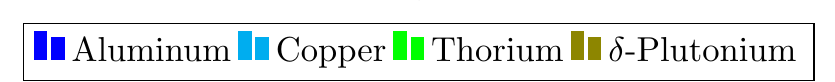}
	\caption{Average standardized shear stress in the slip systems}
	\label{fig:Slip_systems_binned_sorted}
\end{figure}

Up to this point, we have focused on the maximum shear stress as indicator/initiator of plastic activity. However, Henning and Vehoff \cite{henning2005local} claim that the initially active slip systems influence strongly the activation of other slip systems and thus the development of inelastic deformation. To get a more detailed picture of the slip system activity for each grain orientation,
the mean values of the shear stresses in all twelve fcc slip systems is plotted in Fig.~\ref{fig:Slip_systems_binned_sorted}, where the black bar represents a single standard deviation. For a better overview, the slip systems have been sorted by magnitude of mean shear stress, see Tab.~\ref{tab:activeSlip} for a detailed breakdown. As a general trend, carrying over from our previous investigations, the observed values for aluminum are close to the Reuss prediction, with more pronounced deviations and higher scatter for increasing anisotropy. For an idealized homogeneous stress field, eight slip systems are predicted to be active in the $\la100\ra$ grain, four in the $\la110\ra$ grain and six in the $\la111\ra$ grain. This is reflected in the computational results, with a corresponding number of slip systems displaying significantly higher activity in the respective grains. However, the difference between active and inactive systems becomes less distinct for higher anisotropy. For instance, the shear stress values of active and inactive systems overlap within one standard deviation for $\delta$-plutonium in the $\la100\ra$ grain. In the $\la110\ra$ grain, four distinct slip systems with intermediary stresses appear with increasing anisotropy. The effect may be traced back to the muliti-axiality of the stress state in the individual grains, caused by the fluctuation of mechanical properties in the polycrystal. Take a look at the average stress tensor in each grain
\begin{alignat*}{4}
\la\fsigma\ra_{\la 110\ra}^{\mathrm{Al}} &=
\begin{pmatrix}
353.14 \\
-10.99 \\
7.62 \\
0.06\\
0.01\\
0.14
\end{pmatrix}\,\unit{MPa},
\qquad
\la\fsigma\ra_{\la 110\ra}^{\mathrm{Cu}} &=
\begin{pmatrix}
633.79 \\
-74.95 \\
52.55 \\
0.28\\
0.07\\
1.07
\end{pmatrix}\,\unit{MPa},
\\
\la\fsigma\ra_{\la 110\ra}^{\mathrm{Th}} &=
\begin{pmatrix}
393.12  \\
-69.00 \\
48.32 \\
0.23\\
0.08\\
1.11
\end{pmatrix}\,\unit{MPa},
\qquad
\la\fsigma\ra_{\la 110\ra}^{\mathrm{Pu}} &=
\begin{pmatrix}
226.33 \\
-57.30 \\
40.42 \\
0.12\\
0.07\\
1.00
\end{pmatrix}\,\unit{MPa},
\label{eq:Sigma_Avg}
\end{alignat*}
given in Voigt notation. For aluminum, the stress state is nearly uniaxial with
${\|(\I - \fe_1\otimes^4\fe_1):\fsigma \| / \|\fsigma\| =3.79\%},$ whereas plutonium exhibits a pronounced deviation from uniaxiality with $29.60\%$. Thus, the projection of the off-axis components onto the respective Schmid tensors leads to the intermediary activity in four slip systems in the $\la110\ra$ grain.

\begin{table}[h]
	\centering
	\begin{tabular}{llll}
		\toprule
		Slip system    & $\la 100\ra$&$\la 110\ra$  &$\la 111\ra$\\
		\midrule
		1&$(\ms1,\ms1,\ms1) \lbrack \ms0,-1,\ms1\rbrack  $&$(\ms1,-1,\ms1) \lbrack \ms1,\ms1,\ms0\rbrack $ &$(\ms1,\ms1,\ms1) \lbrack \ms1,\ms0,-1\rbrack $ \\
	2& $(\ms1,-1,\ms1)\lbrack \ms0,-1,-1\rbrack $&$(\ms1,\ms1,\ms1) \lbrack -1,\ms1,\ms0\rbrack $ &$(\ms1,\ms1,\ms1) \lbrack \ms0,-1,\ms1\rbrack $ \\
	3& $(-1,\ms1,\ms1) \lbrack \ms0,\ms1,-1\rbrack $&$(-1,-1,\ms1) \lbrack \ms1,-1,\ms0\rbrack $ &$(\ms1,\ms1,\ms1) \lbrack -1,\ms1,\ms0\rbrack $  \\
	4& $(-1,-1,\ms1) \lbrack \ms0,\ms1,\ms1\rbrack $&$(-1,\ms1,\ms1) \lbrack -1,-1,\ms0\rbrack $ &$(\ms1,-1,\ms1) \lbrack -1,\ms0,\ms1\rbrack $\\ 
	5& $(\ms1,\ms1,\ms1) \lbrack \ms1,\ms0,-1\rbrack $&$(\ms1,-1,\ms1) \lbrack \ms0,-1,-1\rbrack  $ & $(-1,\ms1,\ms1) \lbrack \ms0,\ms1,-1\rbrack $\\
	6& $(\ms1,\ms1,\ms1) \lbrack -1,\ms1,\ms0\rbrack $&$(\ms1,-1,\ms1) \lbrack -1,\ms0,\ms1\rbrack $ &$(-1,-1,\ms1) \lbrack \ms1,-1,\ms0\rbrack $\\
	7&$(\ms1,-1,\ms1) \lbrack \ms1,\ms1,\ms0\rbrack $&$(-1,\ms1,\ms1) \lbrack \ms1,\ms0,\ms1\rbrack$ & $(\ms1,-1,\ms1) \lbrack \ms1,\ms1,\ms0\rbrack $ \\
	8& $(\ms1,-1,\ms1) \lbrack -1,\ms0,\ms1\rbrack $&$(-1,\ms1,\ms1) \lbrack \ms0,\ms1,-1\rbrack $ &$(\ms1,-1,\ms1) \lbrack \ms0,-1,-1\rbrack $\\
	9& $(-1,\ms1,\ms1) \lbrack \ms1,\ms0,\ms1\rbrack $&$(\ms1,\ms1,\ms1) \lbrack \ms1,\ms0,-1\rbrack $ &$(-1,\ms1,\ms1) \lbrack \ms1,\ms0,\ms1\rbrack $ \\
	10&$(-1,\ms1,\ms1) \lbrack -1,-1,\ms0\rbrack $ &$(\ms1,\ms1,\ms1) \lbrack \ms0,-1,\ms1\rbrack $ &$(-1,\ms1,\ms1) \lbrack -1,-1,\ms0\rbrack $ \\
	11& $(-1,-1,\ms1) \lbrack \ms1,-1,\ms0\rbrack $&$(-1,-1,\ms1) \lbrack \ms0,\ms1,\ms1\rbrack $ &$(-1,-1,\ms1) \lbrack \ms0,\ms1,\ms1\rbrack $ \\
	12& $(-1,-1,\ms1) \lbrack -1,\ms0,-1\rbrack $ &$(-1,-1,\ms1) \lbrack -1,\ms0,-1\rbrack $ & $(-1,-1,\ms1) \lbrack -1,\ms0,-1\rbrack $\\
		\bottomrule
	\end{tabular}
	\caption{Slip systems in a fcc single crystal}
	\label{tab:activeSlip}
\end{table}

\subsection{On the shape of the shear stress distribution}
\label{sec:normality}

\begin{figure}[htb]
	\begin{subfigure}[c]{0.4\textwidth}
		\centering 
		\includegraphics[width=\textwidth]{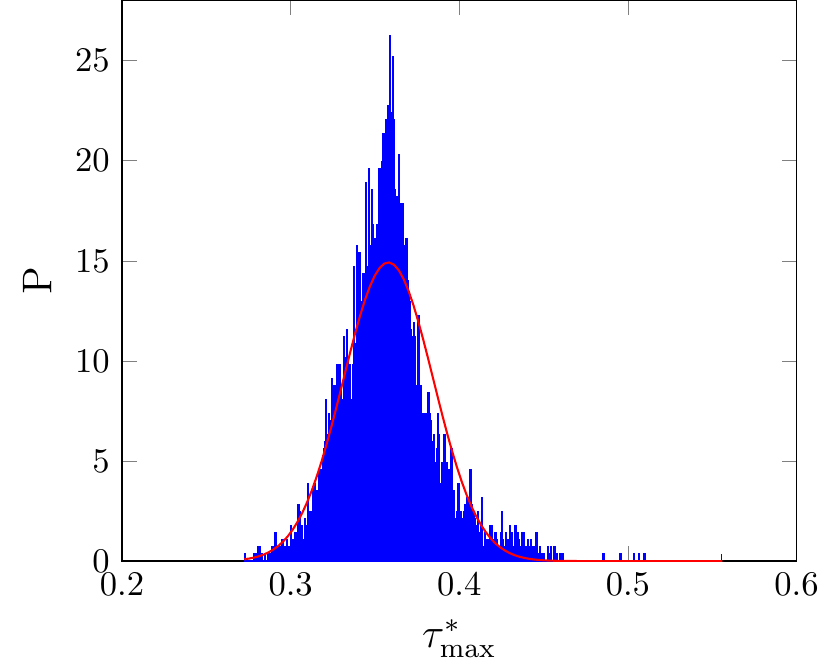}
		\subcaption{1 realization}
		\vspace{5pt}
		\label{fig:histo_1}
	\end{subfigure}
	\begin{subfigure}[c]{0.4\textwidth}
		\centering 
		\includegraphics[width=\textwidth]{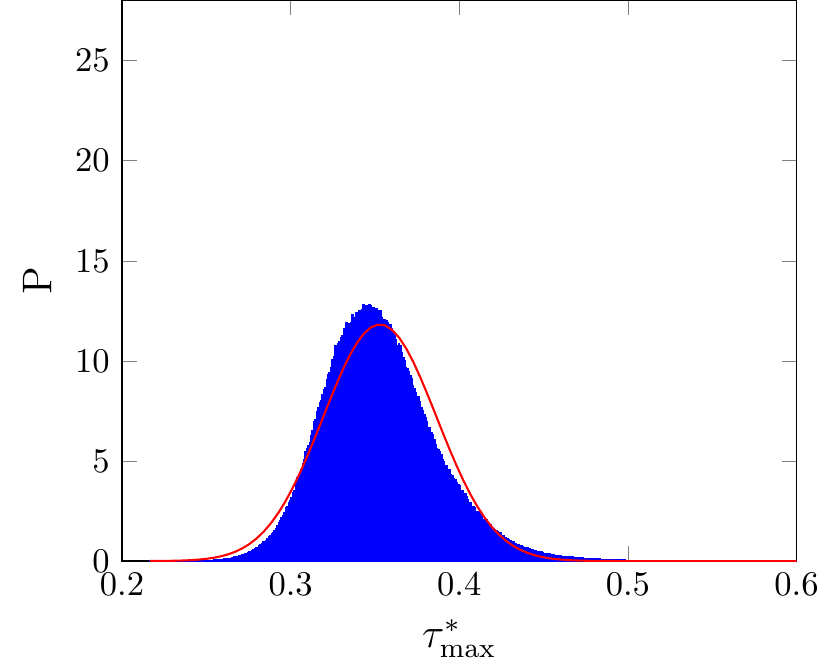}
		\subcaption{200 realizations}
		\vspace{5pt}
		\label{fig:histo_200}
	\end{subfigure}
	\centering
	\includegraphics[width=0.4\textwidth]{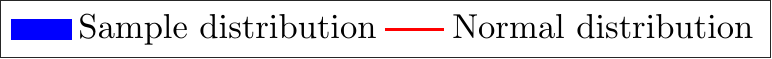}
	\caption{Comparison of the shape of the stress distribution in the $\la 100\ra$ grains}
	\label{fig:realization_normality}
\end{figure}
\begin{figure}[htb]
	\centering
	\begin{subfigure}[c]{0.4\textwidth}
		\begin{subfigure}[c]{\textwidth}
			\centering 
			\includegraphics[width=\textwidth]{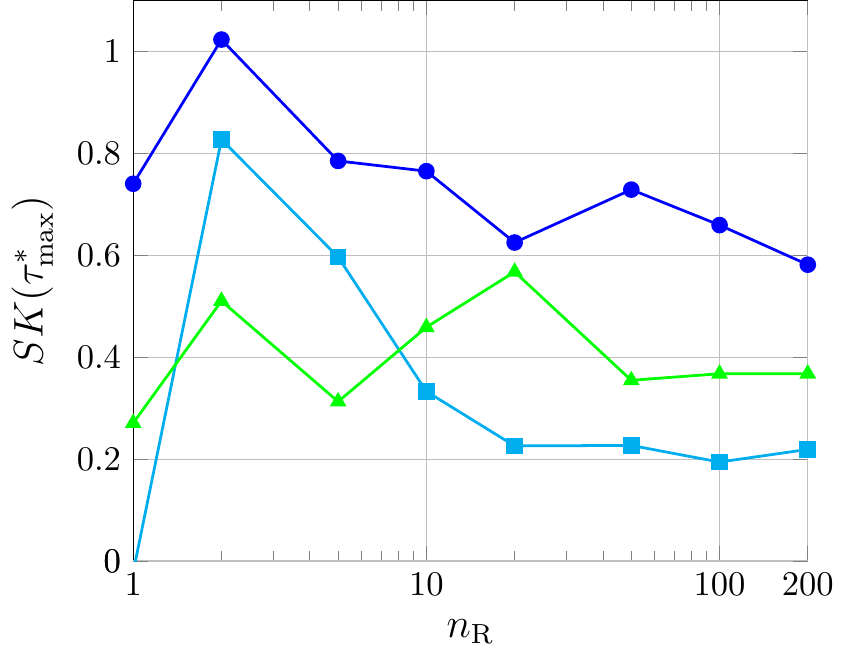}
		\end{subfigure}
		\subcaption{Skewness}
		\vspace{5pt}
	\end{subfigure}
	\begin{subfigure}[c]{0.4\textwidth}
		\begin{subfigure}[c]{\textwidth}
			\centering 
			\includegraphics[width=\textwidth]{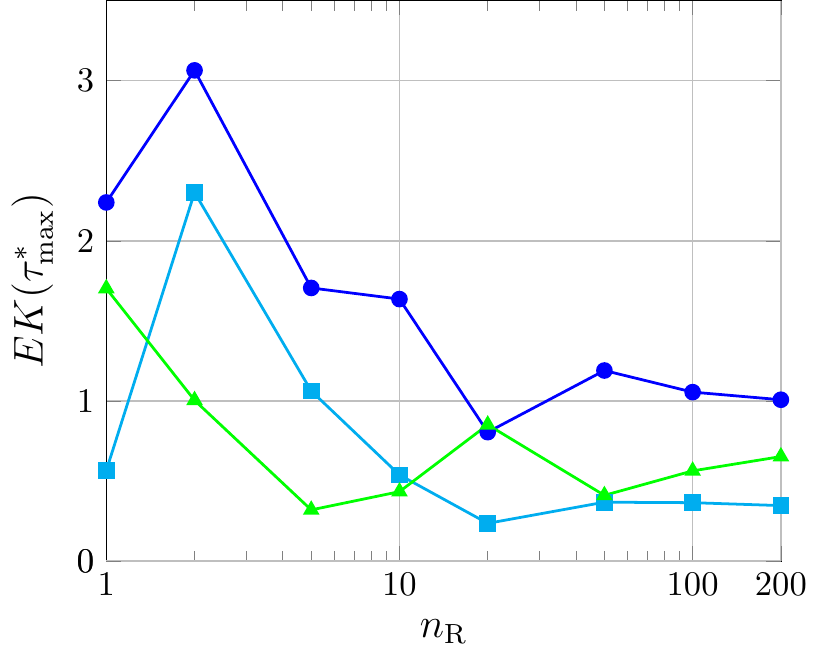}
		\end{subfigure}
		\subcaption{Excess kurtosis}
		\vspace{5pt}
	\end{subfigure}
	\centering
	\includegraphics[width=0.5\textwidth]{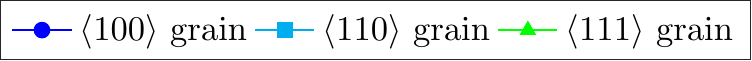}
	\caption{Higher statistical moments of the stress distribution of copper in the realization study}
	\label{fig:realization_study_elastic_skew,kurt}
\end{figure}
In studies on the stress localization in polycrystals, a normal distribution of stresses is often assumed or, at least, suspected \cite{rollett2010stress,BRENNER20093018,el2018intergranular,hure2016intergranular,gallardo2021lath}. For instance, Rollet et al. \cite{rollett2010stress} note that the von Mises stress distribution is approximate normal in shape, while Gallardo-Basile et al. \cite{gallardo2021lath} found the same resemblance for the axial stress distribution. In a more detailed study, Brenner et al. \cite{BRENNER20093018} compared the stress distribution in a grain for a single microstructure and an ensemble of 100 polycrystals and noted that "[the stress distribution] gets closer to Gaussian when many random realizations are considered although there is no proof that it becomes really Gaussian". To illustrate this point, we follow Brenner et al. \cite{BRENNER20093018} and take a look at the shear stress distribution resulting from a single realization and the full ensemble in Fig.~\ref{fig:realization_normality}. The associated normal distributions with the same mean and standard deviation as the sample data are included as red curves. As in the aforementioned study \cite{BRENNER20093018}, the distribution for a larger ensemble is notably smoother and closer to the normal distribution than for the single realization. However, the sample data displays a small but noticeable asymmetry, with a slightly longer tail on the right. Overall, it is difficult to judge from visual inspection, whether the sample data meaningfully deviates from a normal distribution. 
Having evaluated a large data set of 200 microstructures, this motivates us to inspect the distribution of $\tau^*_\textrm{max}$ more closely.
\begin{figure}[htb]
	\begin{subfigure}[c]{0.4\textwidth}
		\centering 
		\includegraphics[width=\textwidth]{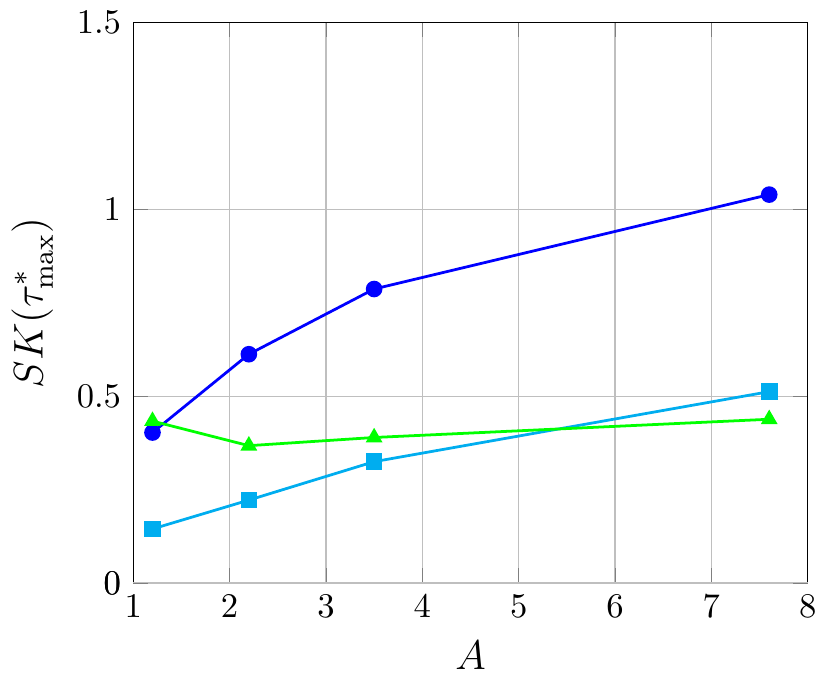}
		\subcaption{Skewness}
		\vspace{5pt}
	\end{subfigure}
	\begin{subfigure}[c]{0.4\textwidth}
		\centering 
		\includegraphics[width=\textwidth]{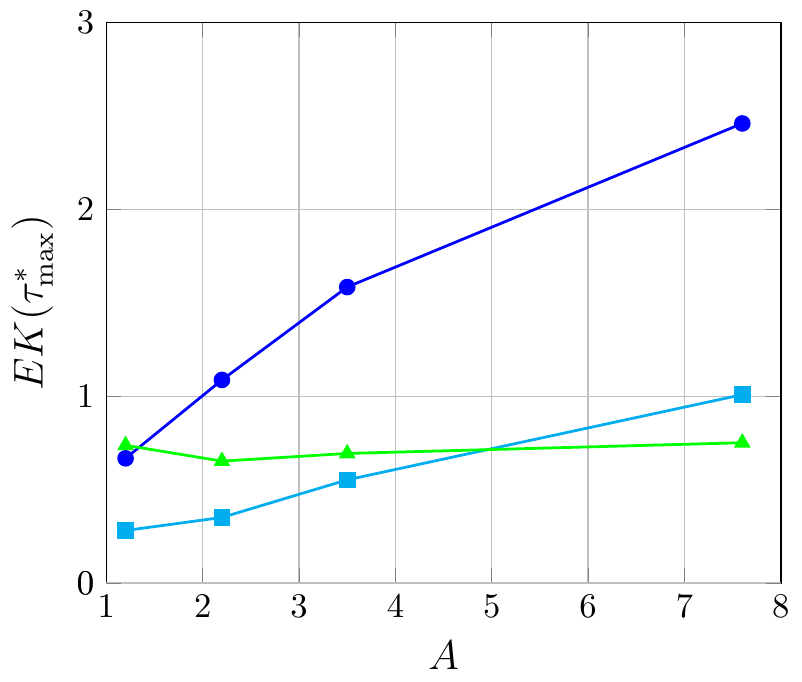}
		\subcaption{Excess kurtosis}
		\vspace{5pt}
	\end{subfigure}
	\centering
	\includegraphics[width=0.5\textwidth]{figures/Tau_ID0-199_grains512_grainres5000material_orient_a_tau_normed_MV_legend.pdf}
	\caption{Higher statistical moments of the stress distribution in the linear elastic range}
	\label{fig:KU,SK_elastic}
\end{figure}
Firstly, we revisit the topic of ensemble size, as a larger ensemble is expected to yield a more normal distribution. To this end, we investigate the higher statistical moments in the form of skewness and excess kurtosis with respect to the number of microstructures in the ensemble, see Fig.~\ref{fig:realization_study_elastic_skew,kurt}. Recall that both quantities are zero for a normal distribution. At first glance, Fig.~\ref{fig:realization_study_elastic_skew,kurt} confirms the initial hypothesis, as a general decrease in skewness and kurtosis with ensemble size is observed. However, the values mostly stabilize for more than $50$ realizations, indicating that the converged distribution is indeed slightly right-skewed and more tail-heavy than a normal distribution.

\begin{figure}[htb]
	\begin{subfigure}[c]{0.4\textwidth}
		\centering 
		\includegraphics[width=\textwidth]{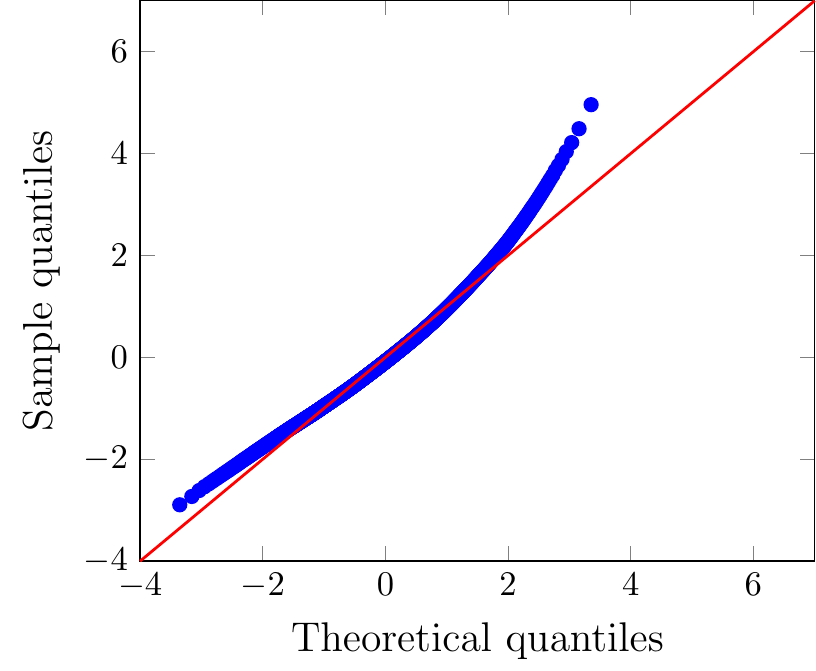}
		\subcaption{Copper}
		\vspace{5pt}
		\label{fig:Q-Q_plot_Cu_norm}
	\end{subfigure}
	\begin{subfigure}[c]{0.4\textwidth}
		\centering 
		\includegraphics[width=\textwidth]{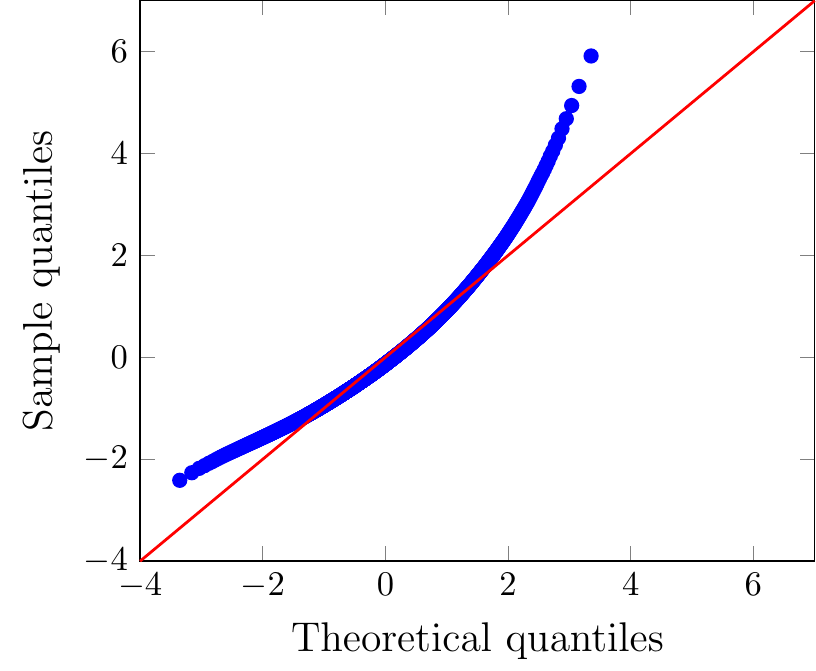}
		\subcaption{$\delta$-Plutonium}
		\vspace{5pt}
		\label{fig:Q-Q_plot_Pu_norm}
	\end{subfigure}
	\centering
	\includegraphics[width=0.4\textwidth]{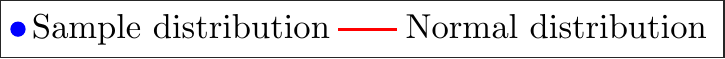}
	\caption{Material specific normal Q-Q plots for $\la 100\ra$ grains}
	\label{fig:Q-Q_plot_Pu}
\end{figure}
\begin{figure}[htb]
	\begin{subfigure}[c]{0.4\textwidth}
		\centering 
		\includegraphics[width=\textwidth]{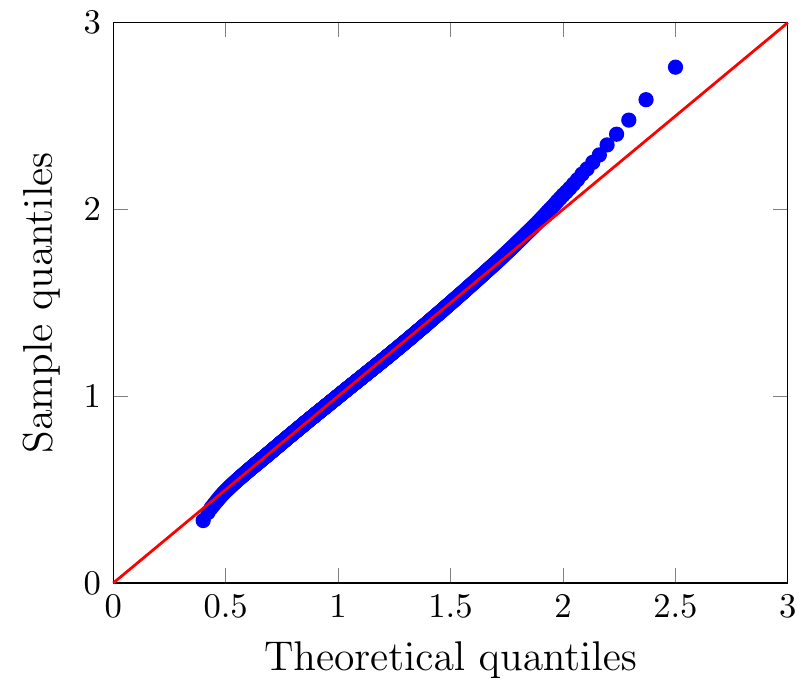}
		\subcaption{Q-Q plot}
		\label{fig:Q-Q_plot_Pu_lognorm}
		\vspace{5pt}
	\end{subfigure}
	\begin{subfigure}[c]{0.4\textwidth}
		\centering 
		\includegraphics[width=\textwidth]{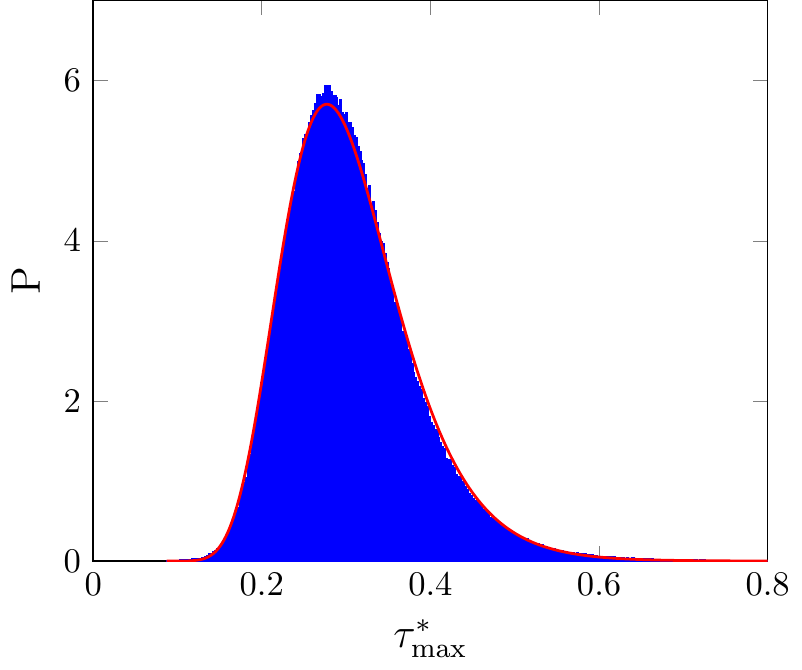}
		\subcaption{Histogram}
		\label{fig:Histo_Pu_lognorm}
		\vspace{5pt}
	\end{subfigure}
	\centering
	\includegraphics[width=0.4\textwidth]{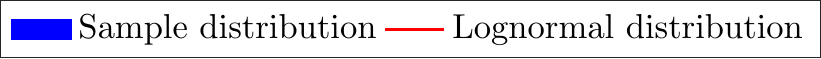}
	\caption{Comparison of sample data to log-normal distribution for $\la 100\ra$ grains of copper}
	\label{fig:Pu_lognorm}
\end{figure}

For the full ensemble, both skewness and kurtosis have a magnitude smaller than one, which may lead to the conclusion that the remaining deviation from a normal distribution is insignificant either way. However, up to this point, the investigations in this sections are based on the material parameters of copper. Taking materials with higher anisotropy into account, we observe a general increase in skewness and kurtosis, see Fig.~\ref{fig:KU,SK_elastic}, with the $\la111\ra$ grain as a notable exception. The resulting deviation from a normal distribution can be further illustrated using a so-called quantile-quantile (Q-Q) plot, a common tool for inspecting if a sample agrees with some assumed distribution \cite{thode2002testing,d1990suggestion,das2016brief}.
Plotting the quantiles of the sample over the quantiles of the tested distribution, a good agreement is characterized if most points lie close to the $45^\circ$ line. Taking a look at Fig.~\ref{fig:Q-Q_plot_Pu}, where the data sets of copper and plutonium are compared to a normal distribution, the deviation can be readily seen for both cases.

Overall, we can conclude that the stress distribution does not converge to a normal distribution, even though the deviation may be small for materials with low anisotropy. If a more accurate statistical description of the data is necessary, e.g., due to high anisotropy, a fit by a log-normal distribution may be more appropriate, see Fig.~\ref{fig:Pu_lognorm} for the case of plutonium.

 \subsection{Comparison with MEM results}
 \label{sec:compare_to_mem}
  In a previous comparison of the MEM with FFT-computations for a single copper polycrystal \cite{krause:20}, the MEM results closely matched the stress distribution of the full-field simulation over the whole polycrystal. However, the MEM was found to significantly overestimate the stress fluctuations for specific orientations. It was hypothesized that this discrepancy is due partly to the effect of the crystal being embedded in 
  \begin{figure}[H]
 	\begin{subfigure}[c]{\textwidth}
 		\begin{subfigure}[c]{0.33\textwidth}
 			\centering 
 			\large $\la 100\ra$
 			\includegraphics[width=\textwidth]{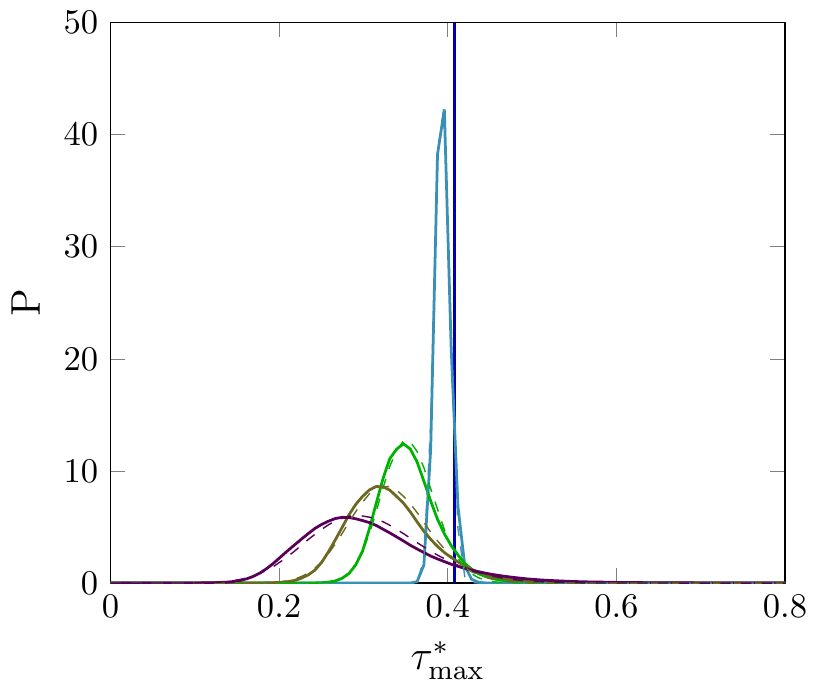}
 		\end{subfigure}
 		\begin{subfigure}[c]{0.33\textwidth}
 			\centering 
 			\large $\la 110\ra$
 			\includegraphics[width=\textwidth]{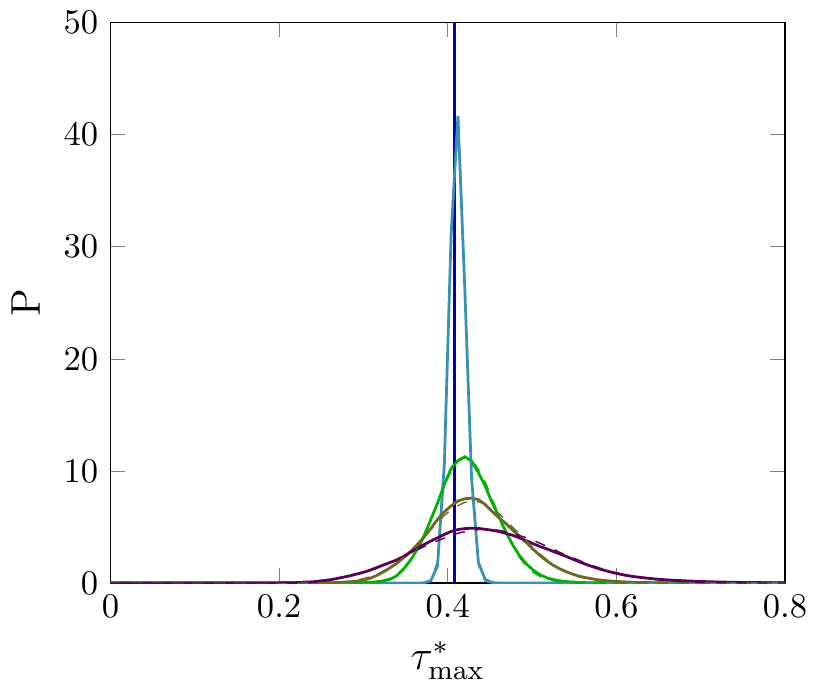}
 		\end{subfigure}
 		\begin{subfigure}[c]{0.33\textwidth}
 			\centering 
 			\large $\la 111\ra$
 			\includegraphics[width=\textwidth]{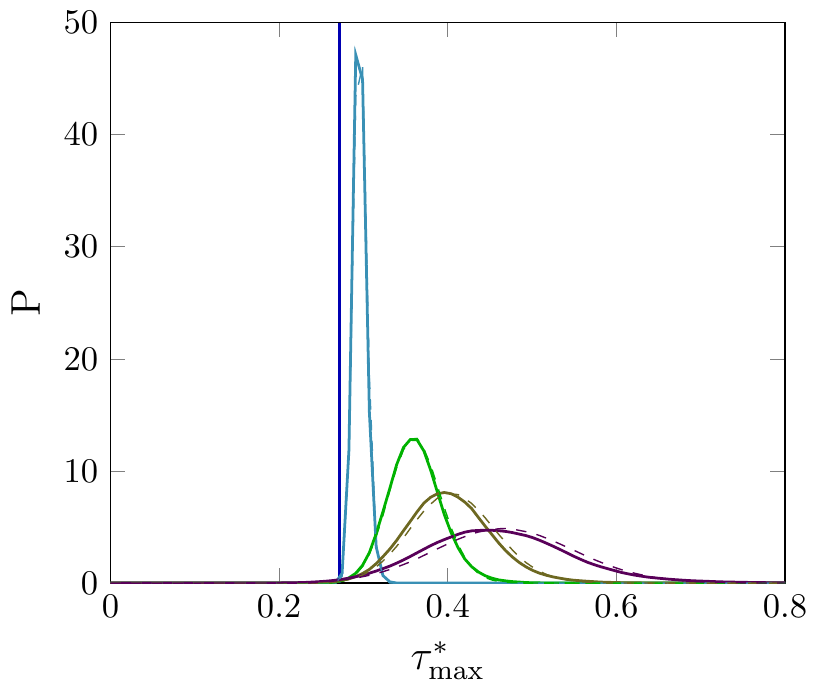}
 		\end{subfigure}
 	\end{subfigure}
 	\centering
 	\includegraphics[width=0.7\textwidth]{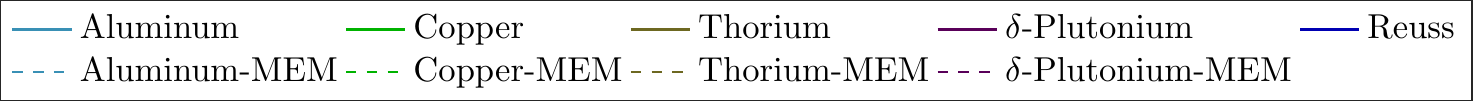}
 	\caption{Comparison of MEM and FFT-material specific stress distribution}
 	\label{fig:KDE_elastic_MEM}
 \end{figure}
  \begin{figure}[H]
	\begin{subfigure}[c]{0.40\textwidth}
		\centering 
		\includegraphics[width=\textwidth]{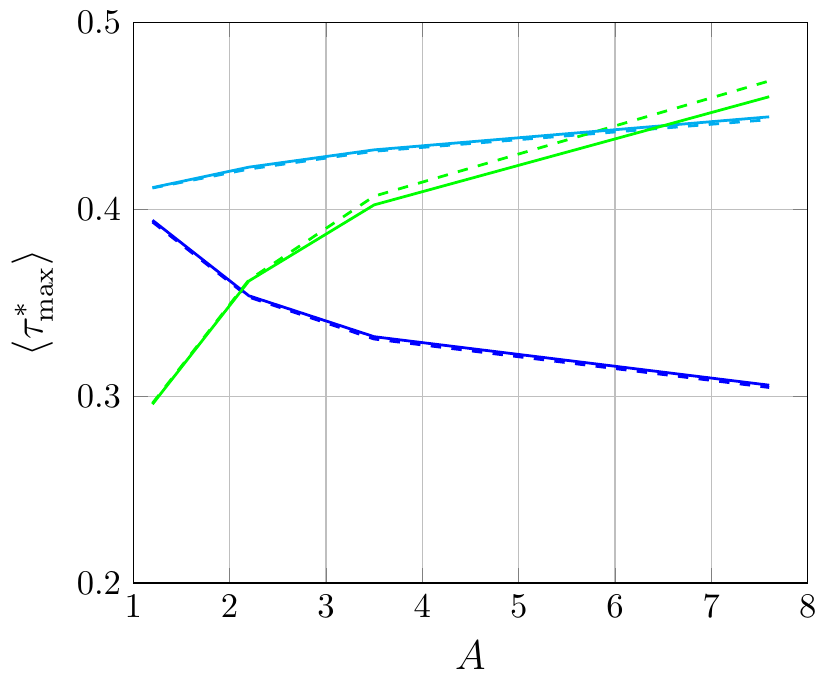}
		\subcaption{Mean value}
		\vspace{5pt}
	\end{subfigure}
	\begin{subfigure}[c]{0.40\textwidth}
		\centering 
		\includegraphics[width=\textwidth]{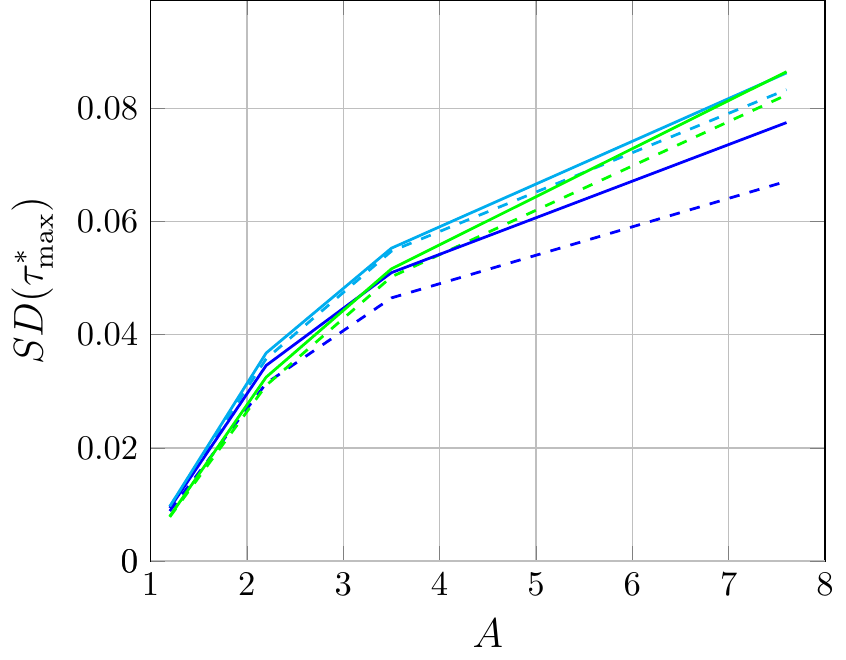}
		\subcaption{Standard deviation}
		\vspace{5pt}
	\end{subfigure}
	\begin{subfigure}[c]{0.40\textwidth}
		\centering 
		\includegraphics[width=\textwidth]{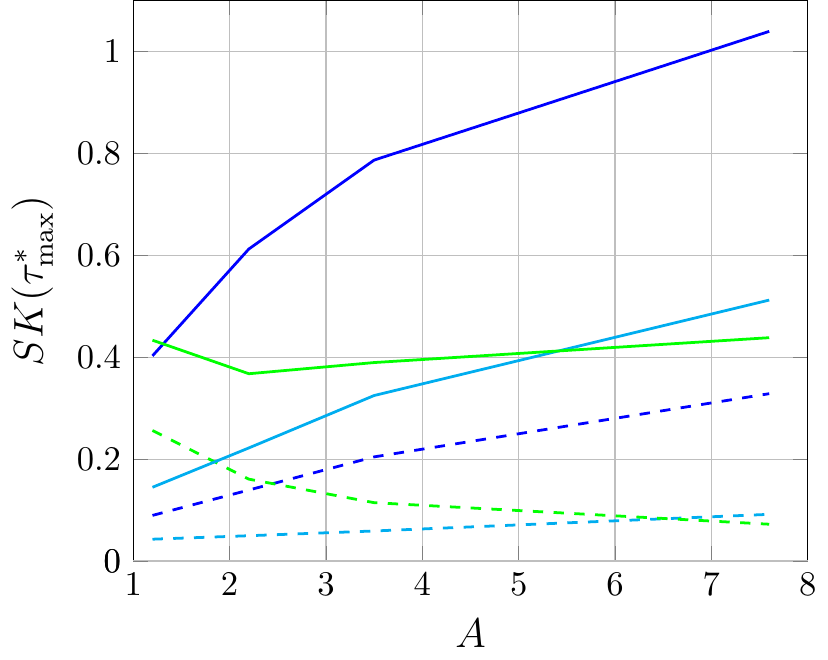}
		\subcaption{Skewness}
		\vspace{5pt}
	\end{subfigure}
	\begin{subfigure}[c]{0.40\textwidth}
		\centering 
		\includegraphics[width=\textwidth]{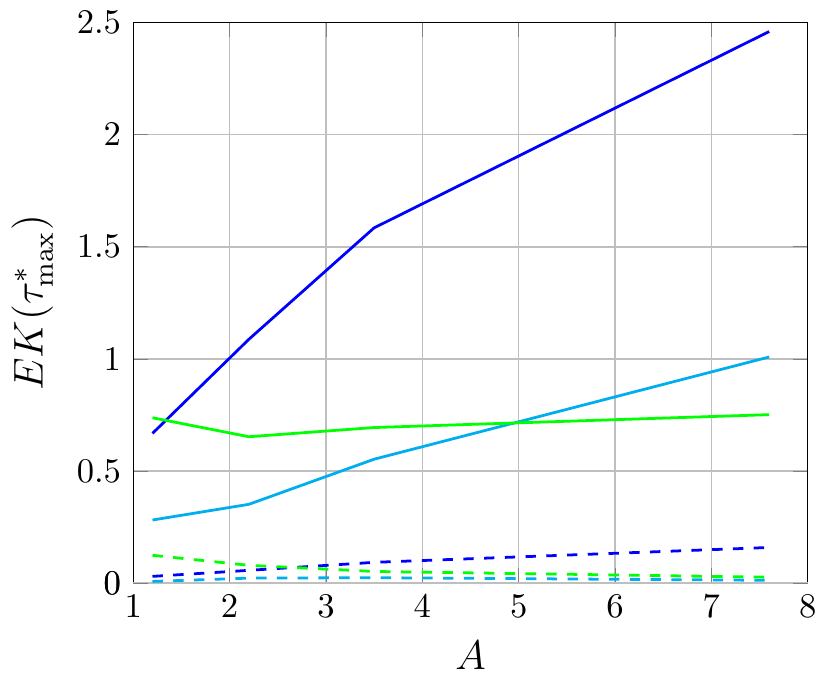}
		\subcaption{Excess kurtosis}
		\vspace{5pt}
	\end{subfigure}
	\centering
	\includegraphics[width=0.5\textwidth]{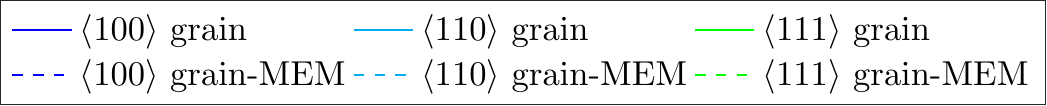}
	\caption{Comparison of MEM and FFT-characteristics of the stress distribution}
	\label{fig:MV,SD_elastic_MEM}
\end{figure}
a single determined neighborhood in a single-microstructure simulation. Thus, we are interested if the MEM proves to be more accurate when considering a more representative ensemble.
 Applying the MEM formulas described in Sec.~\ref{sec:MEM}, mean and covariance of the local stresses are calculated using the effective stiffness of the entire ensemble of polycrystals. The resulting multivariate normal distribution of the single grain with the considered orientation is sampled for computing the distribution of $\tau^*_{\rm max}$ via \eqref{eq:Tau_Stern} using a Monte Carlo approach. 
 In Fig.~\ref{fig:KDE_elastic_MEM}, the material specific stress distributions are shown. The numerically resolved stress distributions, identical to Fig.~\ref{fig:KDE_elastic}, are plotted as lines. The MEM estimate is drawn as a dashed line.
 Overall, the MEM results and the FFT-based computations are in excellent agreement, with both distributions being nearly visually indistinguishable. A closer look at the statistical moments in Fig.~\ref{fig:MV,SD_elastic_MEM} shows that the means of the MEM estimates are very close to the numerical results. The estimates of the standard deviation exhibit a larger discrepancy, especially for high anisotropy.
 By comparing the skewness and excess kurtosis, some limitations of the MEM come to light. The MEM predicts a multivariate normal distribution for the stress tensor $\fsigma$. Even though $\tau^*_\textrm{max}$ is not a linear function of $\fsigma$, the resulting MEM predictions are still very close to normal, as signified by the low values of skewness and kurtosis. Thus, for higher anisotropy, the shape of the shear stress distribution may not be precisely matched by the MEM. Overall, the MEM still proves to be a powerful analytical tool for predicting the stress distribution in polycrystalline materials. Indeed, for most investigated cases, the relative error in the mean is below $1\%$ and in the standard deviation below $10\%$.

 \subsection{On the spatial stress distribution}

 \label{sec:lin_elast_space}
  
  \begin{wrapfigure}{L}{0.4\textwidth}
  	\includegraphics[width=0.4\textwidth]{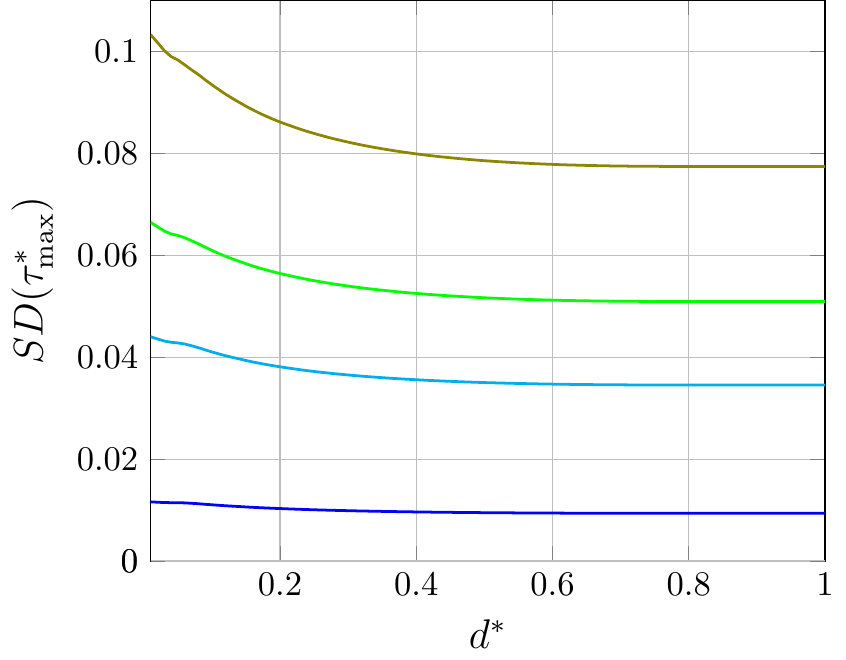}
  	\includegraphics[width=0.4\textwidth]{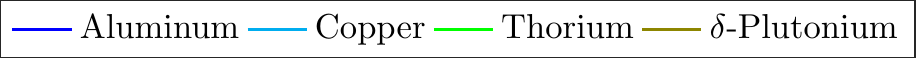}
  	\caption{Standard deviation over grain boundary distance for $\la 100\ra$ grains}
  	\label{fig:sd_elastic_distance}
  \end{wrapfigure}

\begin{figure}[p]
 	\begin{subfigure}[c]{\textwidth}
 		\begin{subfigure}[c]{0.3\textwidth}
 			\centering 
 			\large $\la 100\ra$
 			\includegraphics[width=\textwidth]{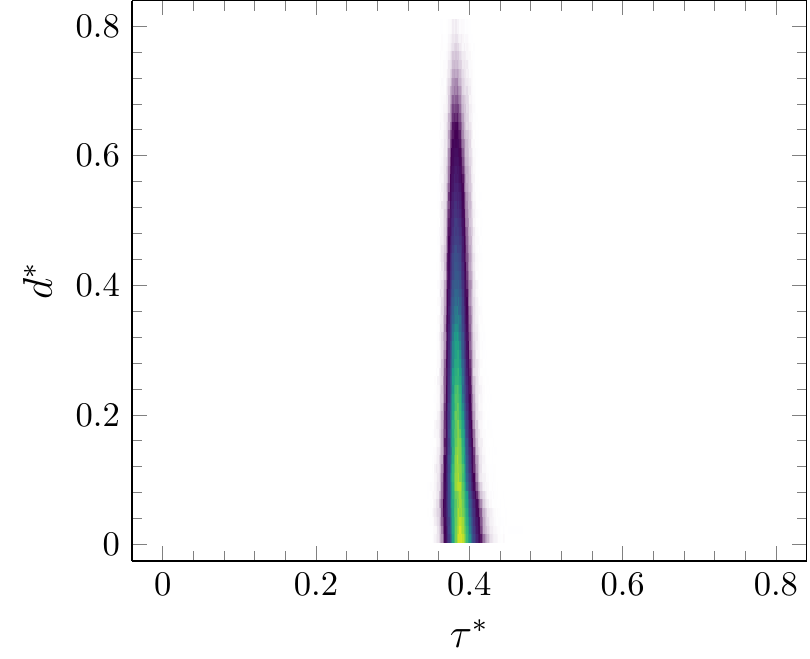}
 		\end{subfigure}
 		\begin{subfigure}[c]{0.3\textwidth}
 			\centering 
 			\large $\la 110\ra$
 			\includegraphics[width=\textwidth]{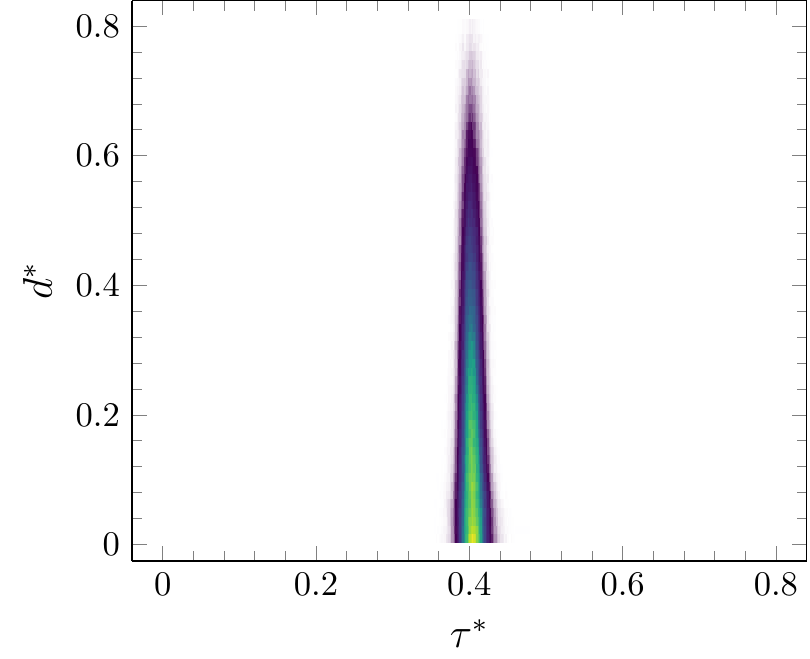}
 		\end{subfigure}
 		\begin{subfigure}[c]{0.3\textwidth}
 			\centering 
 			\large $\la 111\ra$
 			\includegraphics[width=\textwidth]{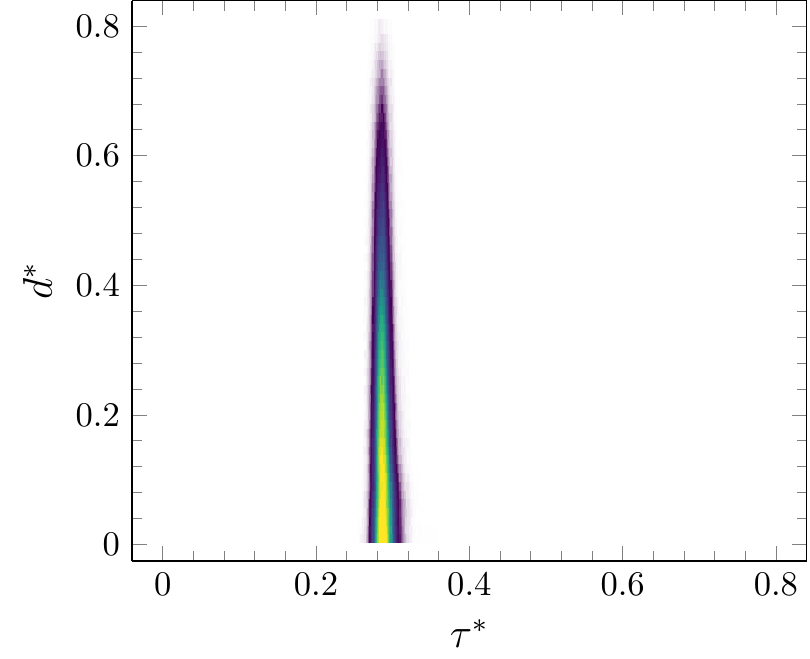}
 		\end{subfigure}
 		\begin{subfigure}[c]{0.075\textwidth}
 			\centering 
 			\includegraphics[width=\textwidth]{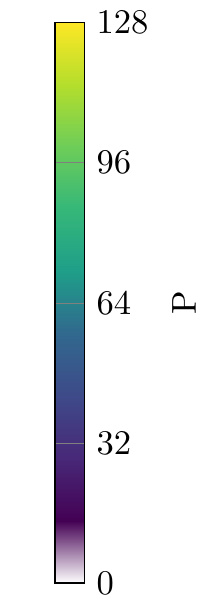}
 		\end{subfigure}
 		\subcaption{Aluminum}
 		\vspace{5pt}
 	\end{subfigure}
 	\begin{subfigure}[c]{\textwidth}
 		\begin{subfigure}[c]{0.3\textwidth}
 			\centering 
 			\includegraphics[width=\textwidth]{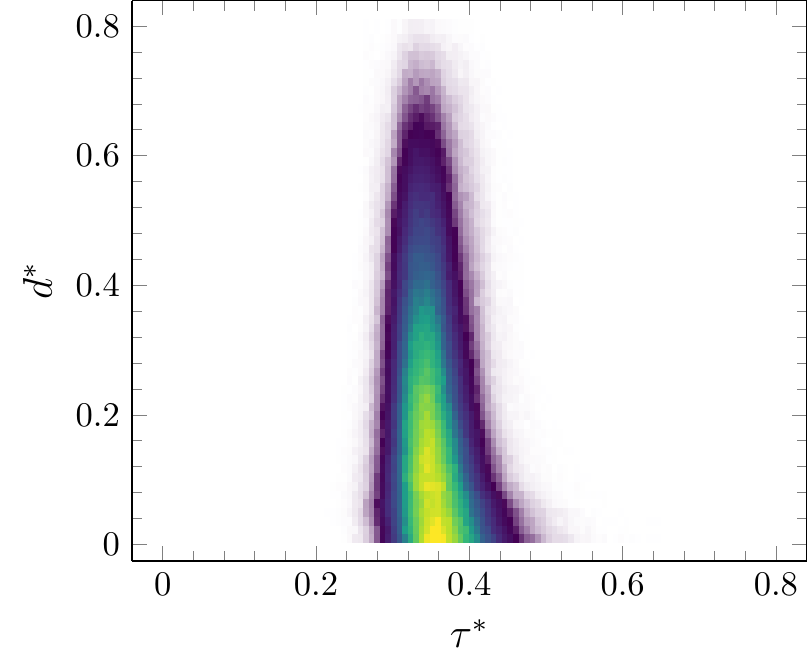}
 		\end{subfigure}
 		\begin{subfigure}[c]{0.3\textwidth}
 			\centering 
 			\includegraphics[width=\textwidth]{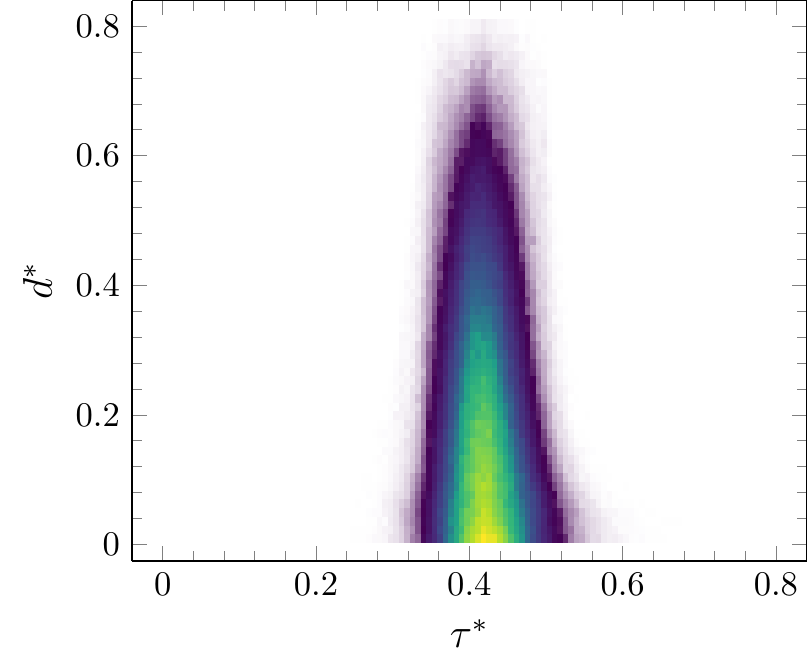}
 		\end{subfigure}
 		\begin{subfigure}[c]{0.3\textwidth}
 			\centering 
 			\includegraphics[width=\textwidth]{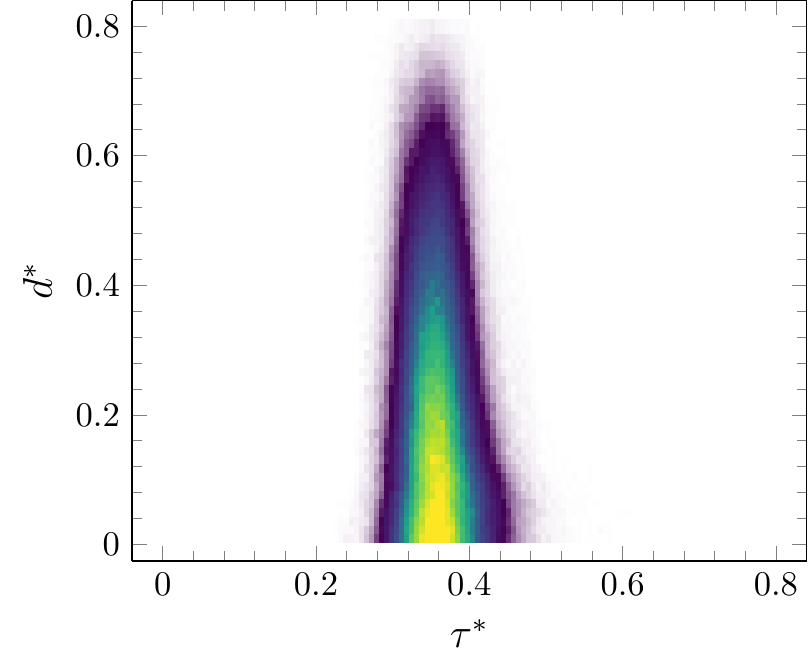}
 		\end{subfigure}
 		\begin{subfigure}[c]{0.07\textwidth}
 			\centering 
 			\includegraphics[width=\textwidth]{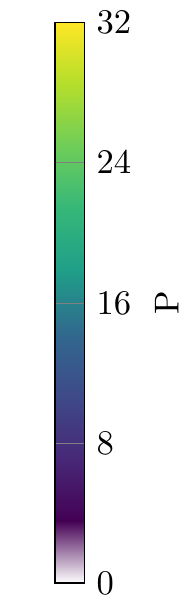}
 		\end{subfigure}
 		\subcaption{Copper}
 		\vspace{5pt}
 	\end{subfigure}
 	\begin{subfigure}[c]{\textwidth}
 		\begin{subfigure}[c]{0.3\textwidth}
 			\centering 
 			\includegraphics[width=\textwidth]{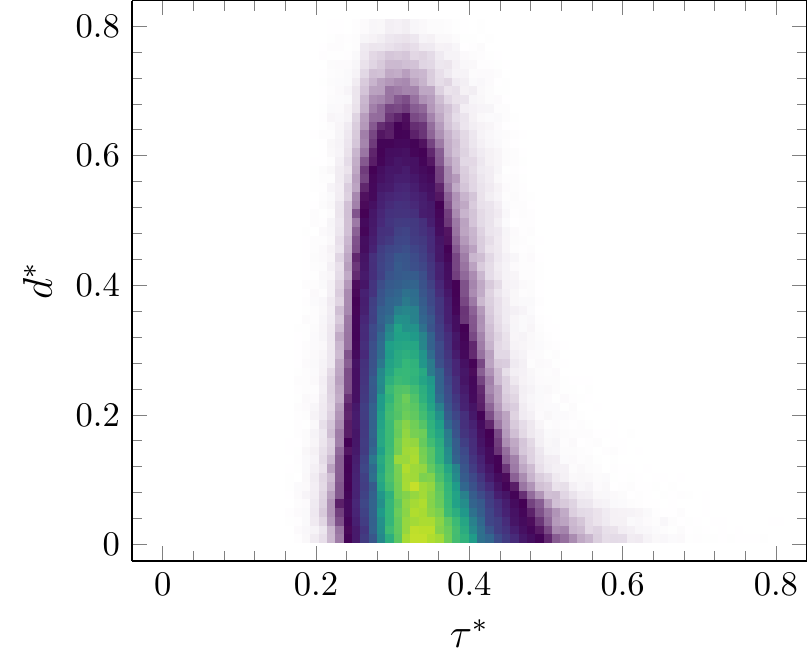}
 		\end{subfigure}
 		\begin{subfigure}[c]{0.3\textwidth}
 			\centering 
 			\includegraphics[width=\textwidth]{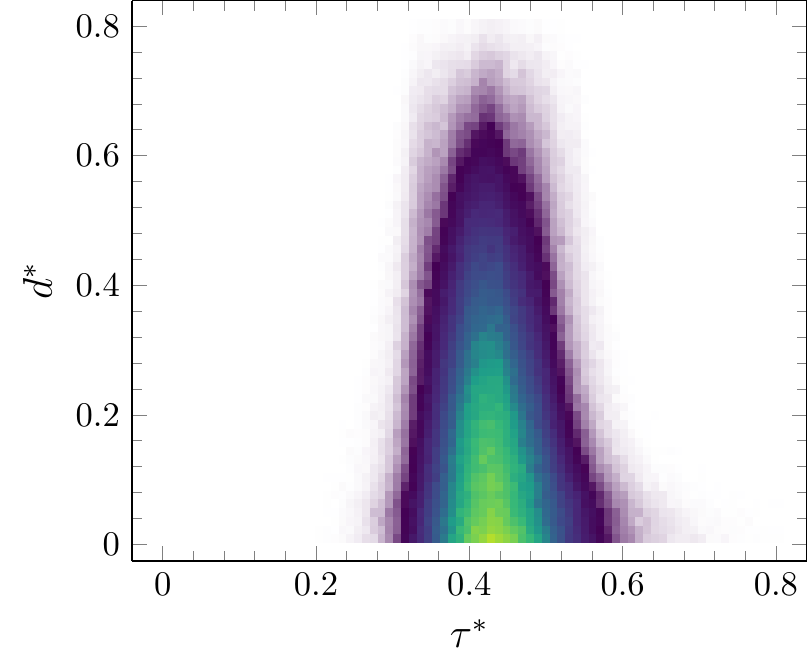}
 		\end{subfigure}
 		\begin{subfigure}[c]{0.3\textwidth}
 			\centering 
 			\includegraphics[width=\textwidth]{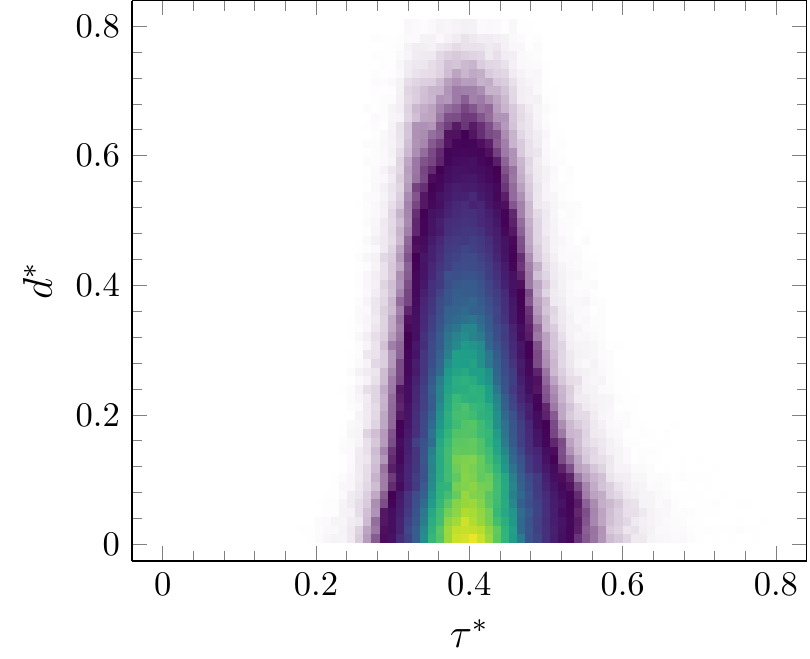}
 		\end{subfigure}
 		\begin{subfigure}[c]{0.07\textwidth}
 			\centering 
 			\includegraphics[width=\textwidth]{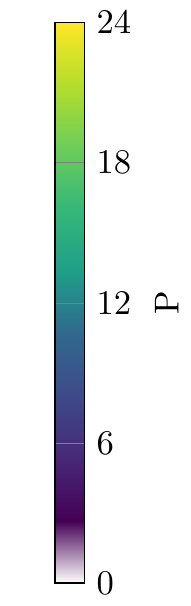}
 		\end{subfigure}
 		\subcaption{Thorium}
 		\vspace{5pt}
 	\end{subfigure}
 	\begin{subfigure}[c]{\textwidth}
 		\begin{subfigure}[c]{0.3\textwidth}
 			\centering
 			\includegraphics[width=\textwidth]{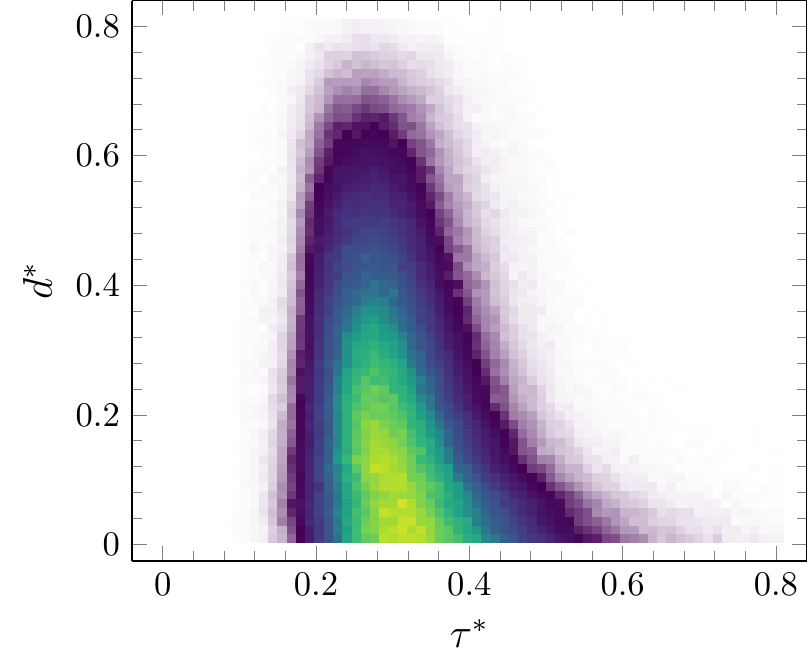}
 		\end{subfigure}
 		\begin{subfigure}[c]{0.3\textwidth}
 			\centering 
 			\includegraphics[width=\textwidth]{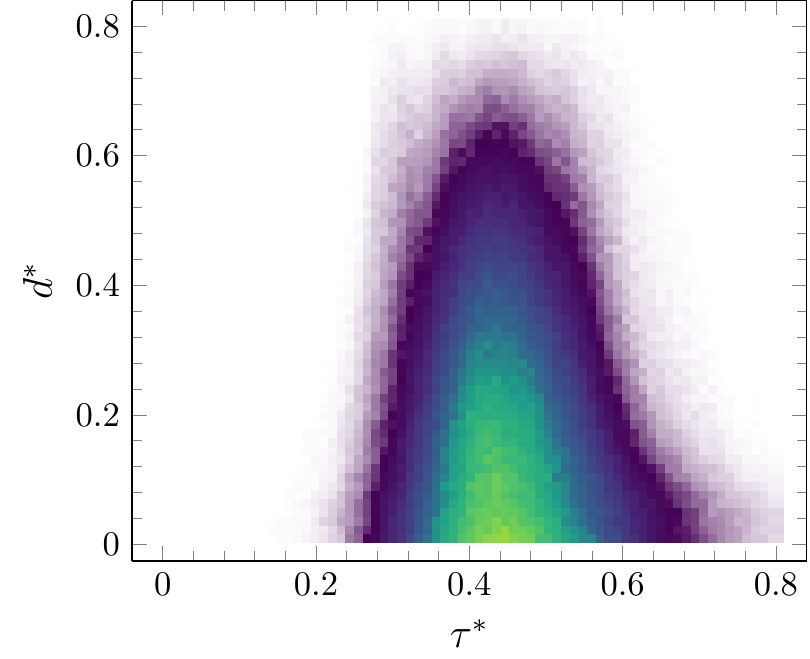}
 		\end{subfigure}
 		\begin{subfigure}[c]{0.3\textwidth}
 			\centering 
 			\includegraphics[width=\textwidth]{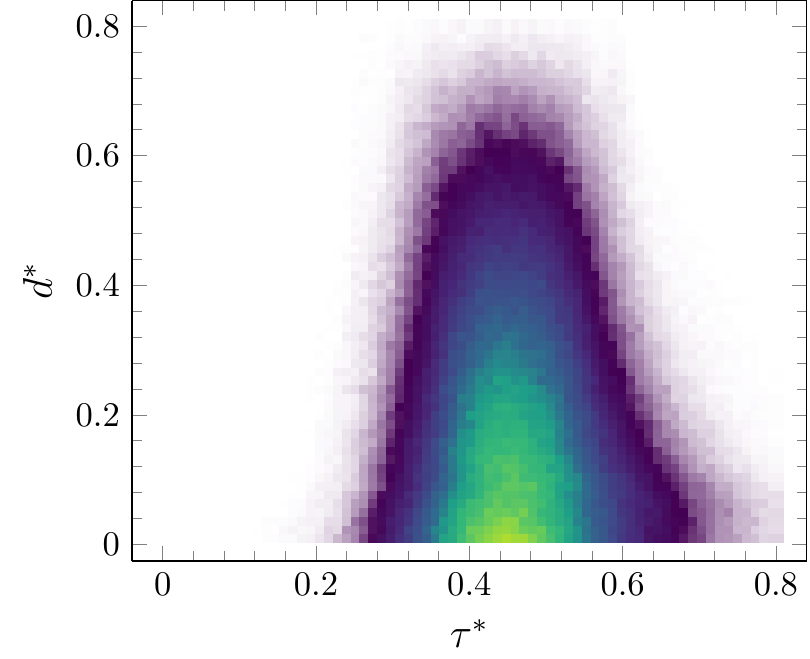}
 		\end{subfigure}
 		\begin{subfigure}[c]{0.07\textwidth}
 			\centering 
 			\includegraphics[width=\textwidth]{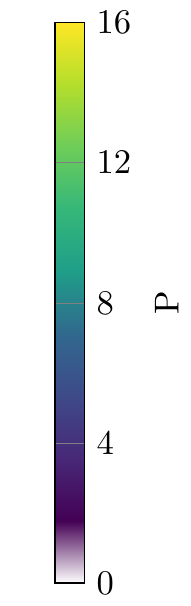}
 		\end{subfigure}
 		\subcaption{$\delta$-Plutonium}
 		\vspace{5pt}
 	\end{subfigure}
 	\caption{Spatial probability density in the linear elastic range}
 	\label{fig:scatter_elastic_pgf}
\end{figure}

Due to the local mismatch in material properties, pronounced stress and strain hot spots are commonly observed close to grain boundaries \cite{castelnau2020multiscale,rollett2010stress,henning2005local,berger2022experimental}, serving as sites for incipient microplasticity and/or damage. For example, Berger et al. \cite{berger2022experimental} recognized that the resulting pattern of plastic deformation in $\alpha$-FE polycrystals strongly depends on the initial strain localization, which is located at the grain boundary. Thus, the spatial distribution of strains and stresses in the grains has attracted increased research interest. For instance, Rollett et al. \cite{rollett2010stress} used Euclidean distance maps to relate stress to grain boundary distance and investigated stress hot spots in two polycrystalline microstructures. Similar strategies were used by Castelnau et al. \cite{castelnau2020multiscale} for studying Wadsleyite (10 realizations) and Gallardo-Basile et al. \cite{gallardo2021lath} for lath martensite (10 realizations).
 Motivated by these studies, in the present section, we focus on the spatial stress distribution in relation to the degree of elastic anisotropy. 
 Building upon these results, the spatial stress distribution during the initial stages of plastification is further discussed in Sec.~\ref{sec:onset}.
 
Following Rollett et al. \cite{rollett2010stress}, the distance $d$ to the nearest grain boundary in each voxel is stored as a distance map. For a dimensionless representation of the results, we use $d^*=d/r_0$, i.e., the ratio of the boundary distance and the equivalent grain radius $r_0 = \sqrt[3]{0.75\pi V}$ for a grain with volume $V$. Generally, $d^*$ may lie in the range $[0,1]$. However, for the generated microstructures, we find that the maximum is around $0.8$ as the actual grains are not perfectly spherical . Note that, in the discretized structure, more data points or voxels are located close to the boundary compared to the center of the grain. Indeed, the region with $d^*\leq0.1$ accounts for roughly $30\%$ of the total volume in our synthetic polycrystals.
For illustrating the spatial distribution of the shear stresses, we rely on heat maps, see Fig.~\ref{fig:scatter_elastic_pgf}, where the frequency of a tuple $(d^*,\tau_\mathrm{max}^*)$ is encoded by a color map. As observed in Sec.~\ref{sec:lin_elast_anisotropy}, a higher elastic anisotropy generally leads to a wider distribution. This is true for both core and boundary regions, which follow similar trends. 
 However, for $d^*\leq0.1$, there is a marked increase in scatter, leading to a standard deviation up to $30\%$ higher than in the remainder of the grain, see Fig.~\ref{fig:sd_elastic_distance}. Consequently, conforming to the observations by Rollett et al., the stress maxima and minima are found close to the grain boundary. 
This matches the experimental findings of Berger et al. \cite{berger2022experimental} according to which some grain boundaries show intense deformation, while others only deform quasi-elastically.
Furthermore, it can be seen in Fig.~\ref{fig:scatter_elastic_pgf} that the spatial distribution of the shear stresses is not necessarily symmetrical, as exemplified by the $\la100\ra$ grains. Close to the boundary, the stress distribution is notably right-skewed, with a pronounced tail at the end of high stress values. Thus, stress hot-spots are more likely to occur than low-stress regions and the stress minima are closer to the mean than the stress maxima. These observations still hold for the $\la110\ra$ and $\la111\ra$ grains but are much less pronounced.

 \section{Onset of plastification}
 \subsection{Statistical moments}
\label{sec:onset}
\begin{wrapfigure}{htb}{0.4\textwidth}
	\centering 
	\includegraphics[width=0.4\textwidth]{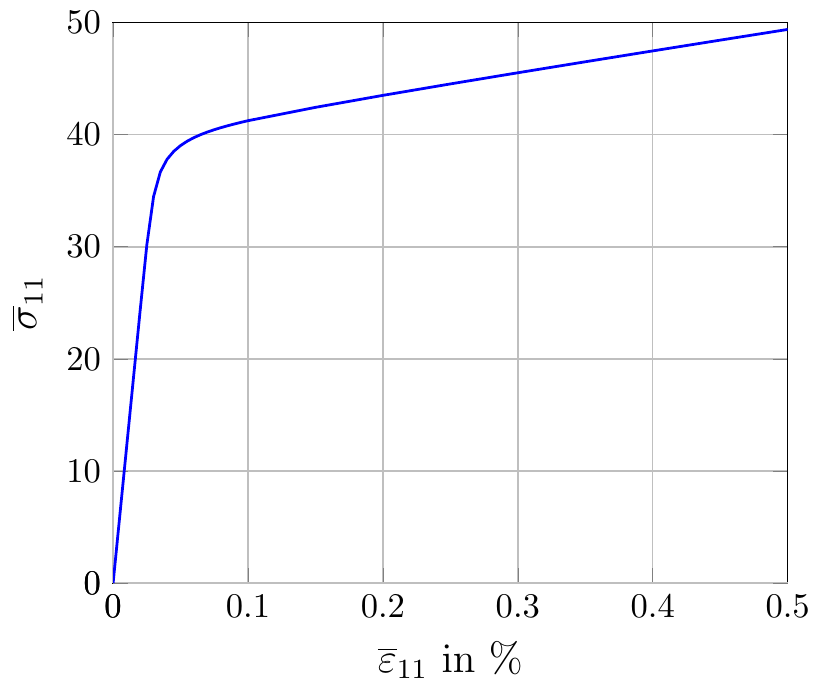}
	\caption{Macroscopic stress-strain curve}
	\label{fig:macro_stress_strain1}
\end{wrapfigure}
 In the following, we expand our investigation of the stress distribution into the elastic-plastic range. To keep computational costs manageable, we restrict to a single set of material parameters corresponding to copper, see Tab.~\ref{tab:copper_parameters}, taken from Eghtesad et al. \cite{eghtesad2018openmp} and adapted to the material model in Sec.~\ref{sec:crystal_plasticity} by Wicht et al. \cite{wicht2020efficient}. For the investigation at hand, we focus on the initial stages of plastification, i.e. the transition from the elastic to the plastic regime, building upon the results of Sec.~\ref{sec:lin_elast}. 
  \begin{table}[b]
 	\centering
 	\begin{tabular}{lll}
 		\toprule
 		Elastic moduli & Flow rule (Chaboche) & Linear exponential\\ &&Hardening\\ 
 		\midrule
 		$C_{11} = 170.2$ GPa&$\dot{\gamma}_{\mathrm{0}}=0.001$ s$^{-1}$&$\Theta_0=250$ MPa\\
 		$C_{12} = 114.9$ GPa&$m=20$&$\Theta_{\infty}=14$ MPa\\
 		$C_{44} = 61.0$ GPa&$\tau_0=6.5$ MPa&$\tau_{\infty}=113.5$ MPa\\
 		&$\tau_{\mathrm{D}}=8$ MPa& \\
 		\bottomrule
 	\end{tabular}
 	\caption{Material parameters of copper in an elasto-viscoplastic material model}
 	\label{tab:copper_parameters}
 \end{table}
 More precisely, we consider uniaxial extension up to $0.5\%$, subdivided into $20$ steps of $0.005\%$ up to $0.1\%$ followed by $8$ steps of $0.05\%$, see Fig.~\ref{fig:macro_stress_strain1} for the effective stress-strain curve. Note that, for larger plastic strains, the particular formulation for flow rule and hardening law are expected to have a significant impact on the results, as the shear stress distribution is governed by the evolution of the critical shear stresses on the slip systems. However, during the initial stages of loading, all suitable crystal plasticity models should cover the basic phenomena such as plastification upon reaching the initial yield stress. Thus, even for the simple model in Sec.~\ref{sec:crystal_plasticity}, we expect qualitatively meaningful results.
\begin{figure}[htb]
	\begin{subfigure}[c]{0.40\textwidth}
		\centering 
		\includegraphics[width=\textwidth]{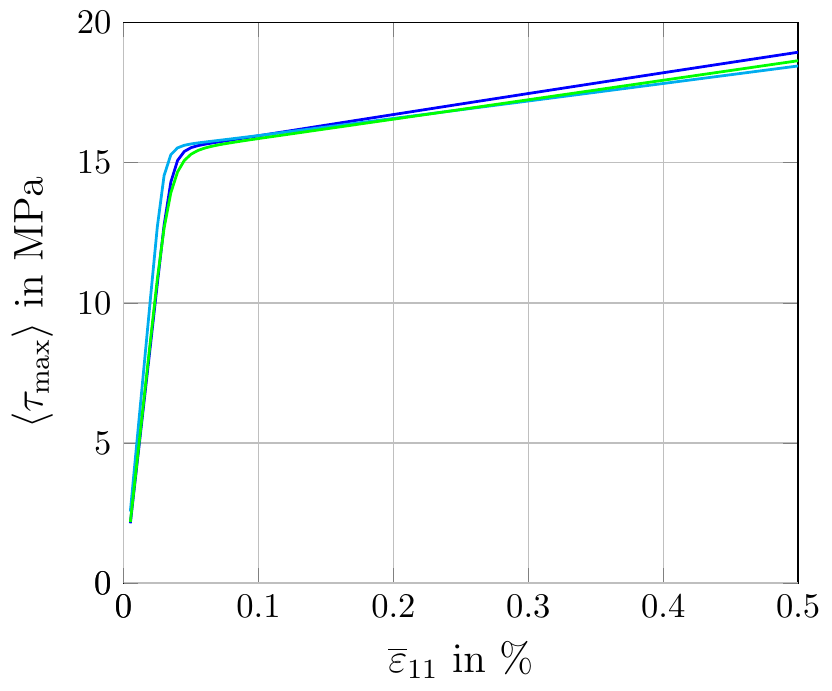}
		\subcaption{Mean value}
		\vspace{5pt}
	\end{subfigure}
	\begin{subfigure}[c]{0.40\textwidth}
		\centering 
		\includegraphics[width=\textwidth]{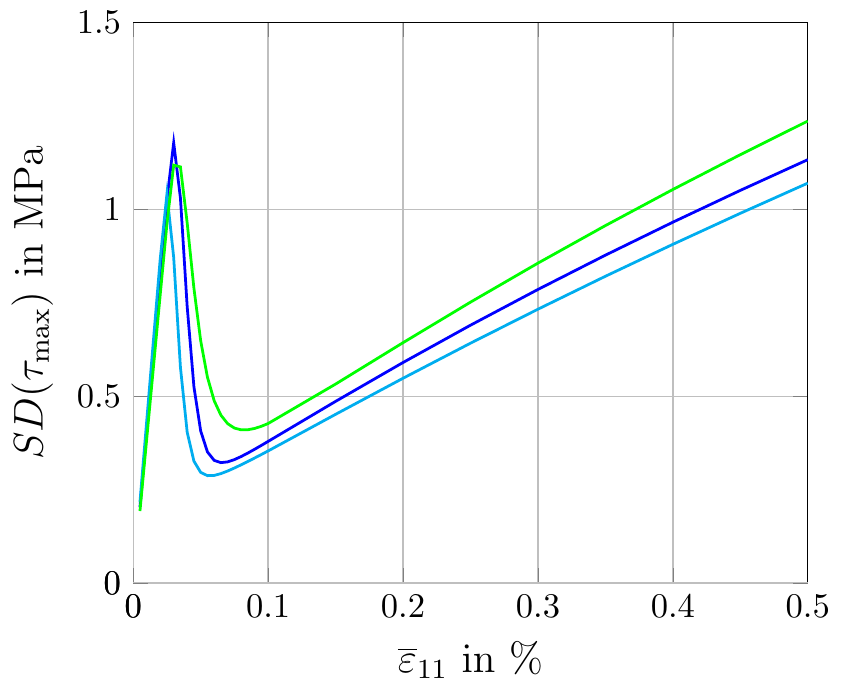}
		\subcaption{Standard deviation}
		\vspace{5pt}
	\end{subfigure}
	\begin{subfigure}[c]{0.40\textwidth}
		\centering 
		\includegraphics[width=\textwidth]{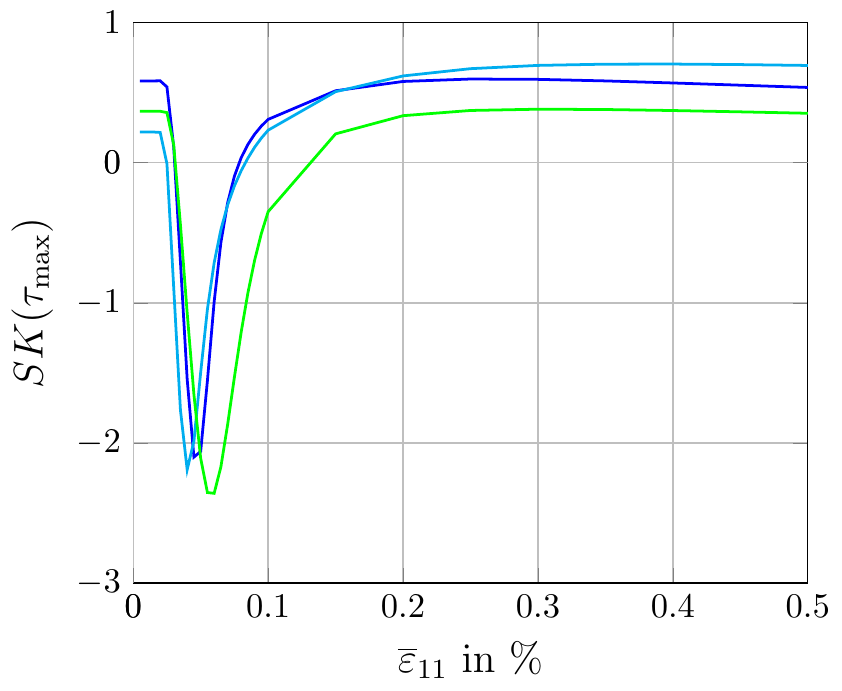}
		\subcaption{Skewness}
		\vspace{5pt}
	\end{subfigure}
	\begin{subfigure}[c]{0.40\textwidth}
		\centering 
		\includegraphics[width=\textwidth]{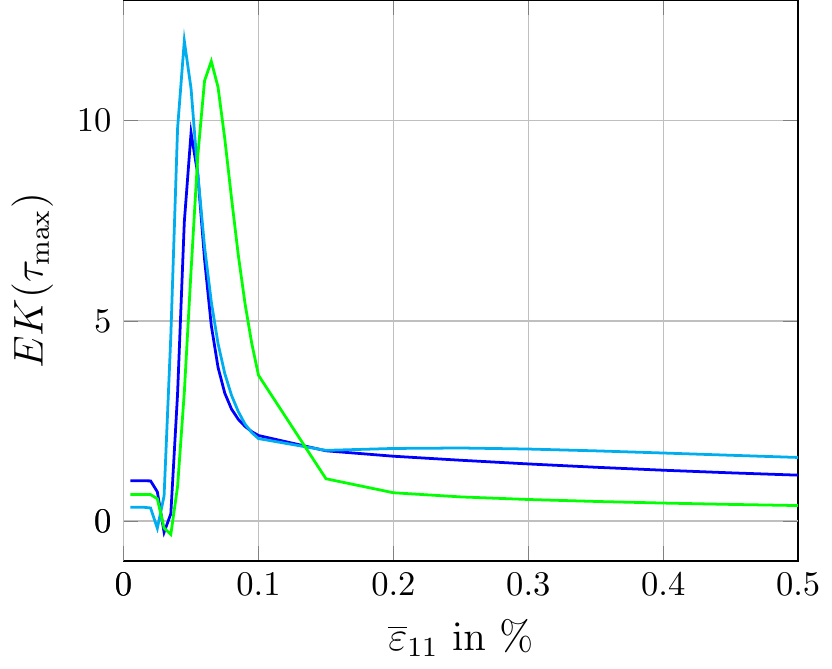}
		\subcaption{Excess kurtosis}´
		\vspace{5pt}
	\end{subfigure}
	\centering
	\includegraphics[height = 0.022\textheight]{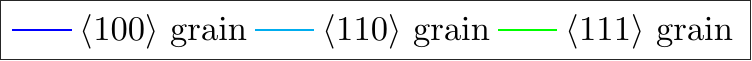}
	\caption{Characteristics of the stress distribution in the elastic-plastic range}
	\label{fig:MV,SD,SK,KU_plastic}
\end{figure}

As a first step, let us take a look at the evolution of the statistical moments in Fig.~\ref{fig:MV,SD,SK,KU_plastic}. To give a clearer impression of the stress evolution during extension, we plot the  values corresponding to the distribution of the maximum shear stress $\tau_\textrm{max}$, and \emph{not} its dimensionless counterpart $\tau^*_\textrm{max}$.
\begin{wrapfigure}{htb}{0.4\textwidth}
	\includegraphics[width=0.4\textwidth]{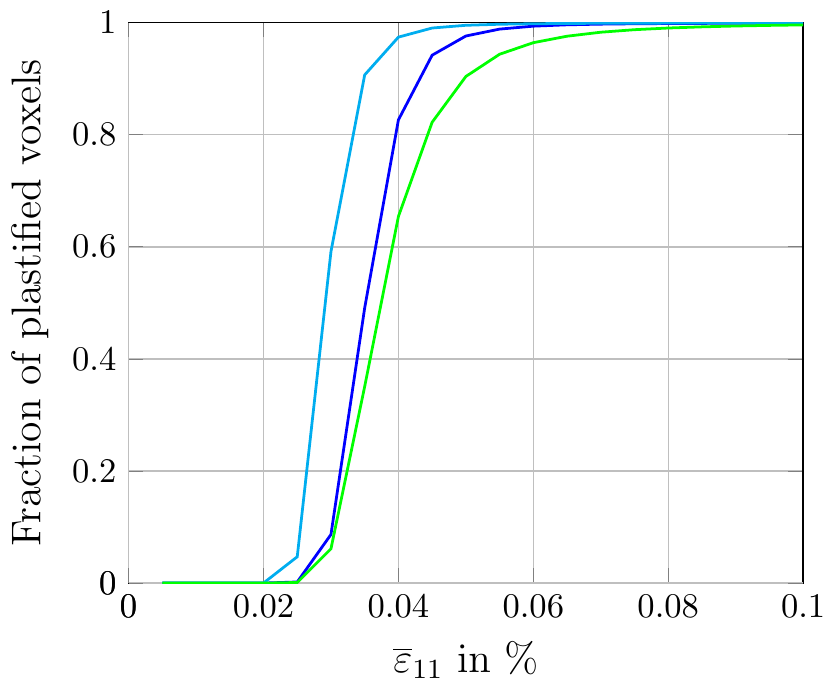}
	\includegraphics[width=0.4\textwidth]{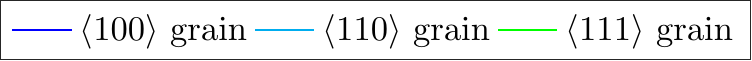}
	\caption{Fraction of plastified voxels during uniaxial extension}
	\label{fig:micro_macro_yield}
\end{wrapfigure}
In the elastic regime, mean and standard deviation scale according to the results of Sec.~\ref{sec:lin_elast_anisotropy}. Note that, for $\la\tau_\textrm{max}\ra$, the vertical difference between the different orientations (recall Fig.~\ref{fig:MV,SD_elastic}) does not translate to a pronounced horizontal offset of the curves, due to the high elastic stiffness. Thus, all grains plastify at similar macroscopic strain. A more detailed view of this process is shown in Fig.~\ref{fig:micro_macro_yield}, where the fraction of plastified voxels is plotted based on the indicator $\tau_\textrm{max} > \tau_\textrm{crit}$ where $\tau_\textrm{crit} = \tau^\textrm{F}_0 + \tau^\textrm{D} = 14.5\,\unit{MPa}$.
Thus, the onset of plasticity can be located at around $0.02\%$, with a transitional range up to $0.08\%$ when virtually all grains are fully plastified.
The onset of plastification is further marked by rapid changes in the higher statistical moments. The standard deviation drops from roughly $10\%$ down to around $2\%$ of the mean stress value, signifying the concentration of stresses around the critical shear stress. 

Looking at the skewness, the distribution shifts from slightly right-skewed to heavily left-skewed, as the critical shear
stress acts as a natural boundary on the higher end of stress values. In addition, the strong deviation from normal is reflected in the excess kurtosis peaking at high values around 10.

For strains beyond $0.1\%$, we observe an increase in the scatter of the shear stress, as the increasing heterogeneity of the plastic slips is reflected in the stress distribution via the hardening law. Indeed, around $0.5\%$, the standard deviation exceeds the values for the linear elastic range. However, we wish to reemphasize that the specific formulation of the hardening law is expected to have an increased influence at this point, which would justify a more detailed study in its own right. Last but not least, we note that skewness and excess kurtosis return to the levels of the elastic domain between $0.1\%$ and $0.2\%$, indicating that the shape of the stress distribution in the plastified material is similar to the one determined in Sec.~\ref{sec:lin_elast}.

\subsection{Spatial stress distribution during plastification}
\label{sec:spatial_plastification}

\begin{figure}[htb]
	\begin{subfigure}[c]{0.3\textwidth}
		\centering 
		\includegraphics[width=\textwidth]{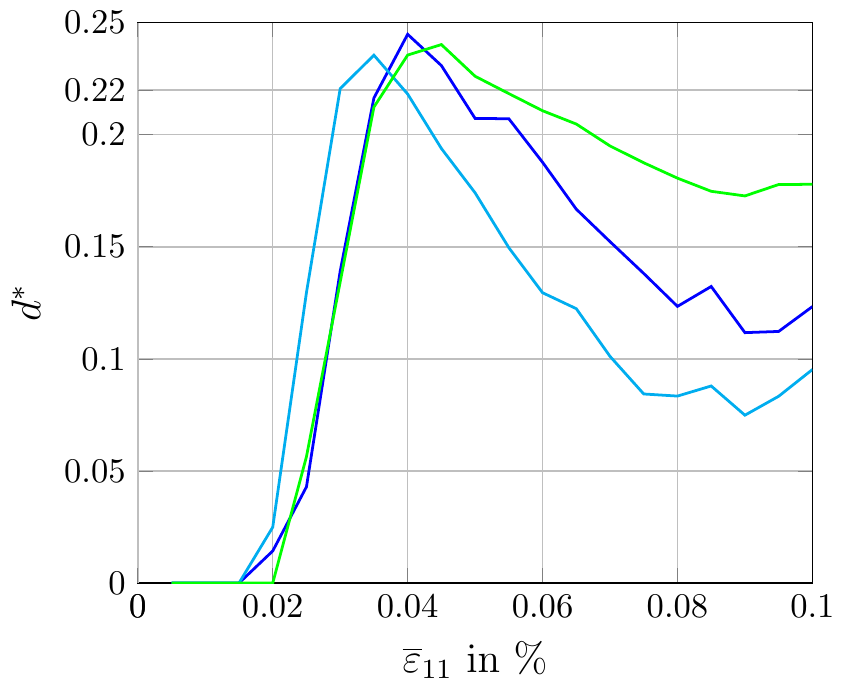}
		\caption{Average distance of newly plastified voxels}
		\label{fig:average_distance_newly_plastified}
	\end{subfigure}
	\begin{subfigure}[c]{0.3\textwidth}
		\centering 
		\includegraphics[width=\textwidth]{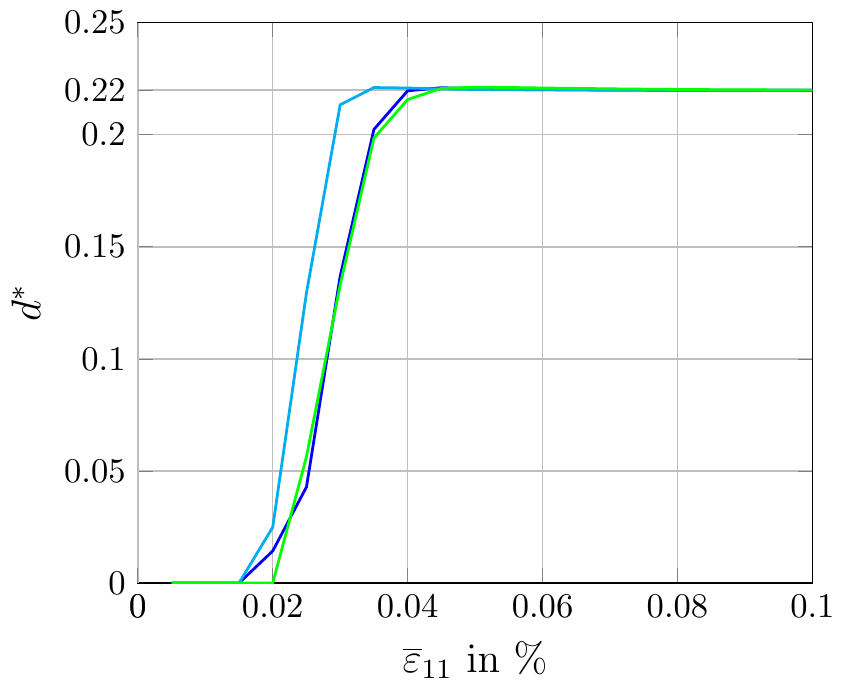}
		\caption{Average distance of plastified voxels}
		\label{fig:average_distance_plastified}
	\end{subfigure}
	\begin{subfigure}[c]{0.3\textwidth}
		\centering 
		\includegraphics[width=\textwidth]{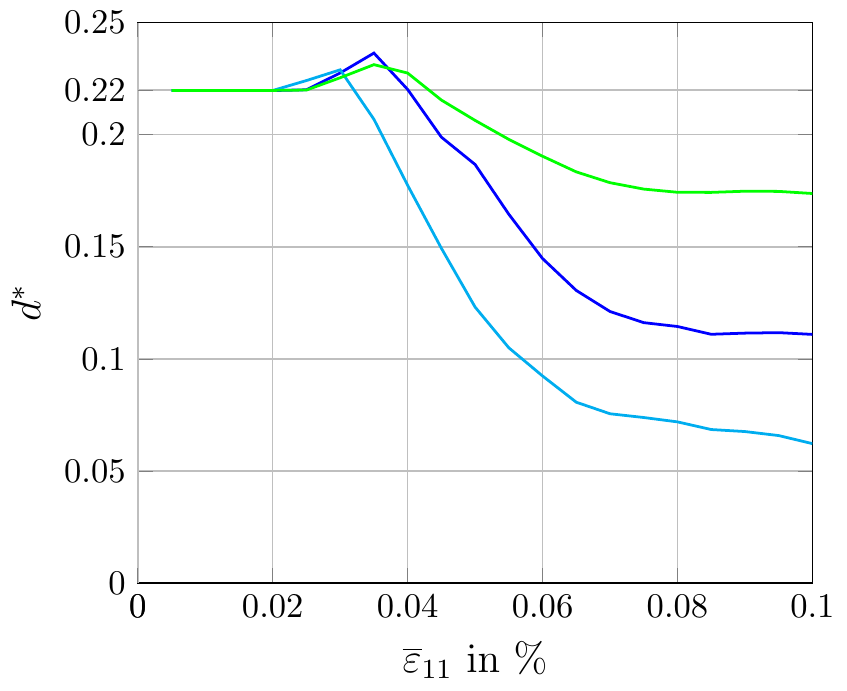}
		\caption{Average distance of unplastified voxels}
		\label{fig:average_distance_not_plastified}
	\end{subfigure}
	\centering
	\includegraphics[width=0.4\textwidth]{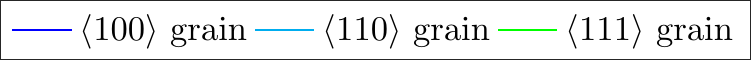}
	\caption{Plastification in relation to grain boundary distance}
	\label{fig:average_distance}
\end{figure}

In Sec.~\ref{sec:lin_elast_space}, we noted that the extremal shear stresses are found close to the grain boundaries. To evaluate, whether this is reflected during plastification, we track the boundary distance of the plastified and unplastified voxels in Fig.~\ref{fig:average_distance}, where the average distance $\la d^*\ra=0.22$ for all voxels is marked as reference. More precisely, in Fig.~\ref{fig:average_distance_newly_plastified}, the average distance of voxels which plastified \emph{in the associated load step} is shown, indicating where plastification occurs at any particular point during loading. The average of \emph{all} plastified voxels is shown in Fig.~\ref{fig:average_distance_plastified}. In agreement to Sec.~\ref{sec:lin_elast_space}, we observe that plastification does indeed start at the grain boundary. Up to $0.04\%$, the plastified region expands to the inside of the grain. At this point, the average distance of plastified voxel is equivalent to the average distance of all voxels, indicating that the grain has homogeneously plastified. Notably, the last voxels to plastify and the last unplastified voxels, see Fig.~\ref{fig:average_distance_not_plastified}, are, on average, located closer to the grain boundary, consistent with the fact that not only stress maxima but also stress minima are found in this region.

\begin{figure}[H]
	\begin{subfigure}[c]{0.33\textwidth}
		\begin{subfigure}[c]{0.8\textwidth}
			\centering 
			\includegraphics[width=\textwidth]{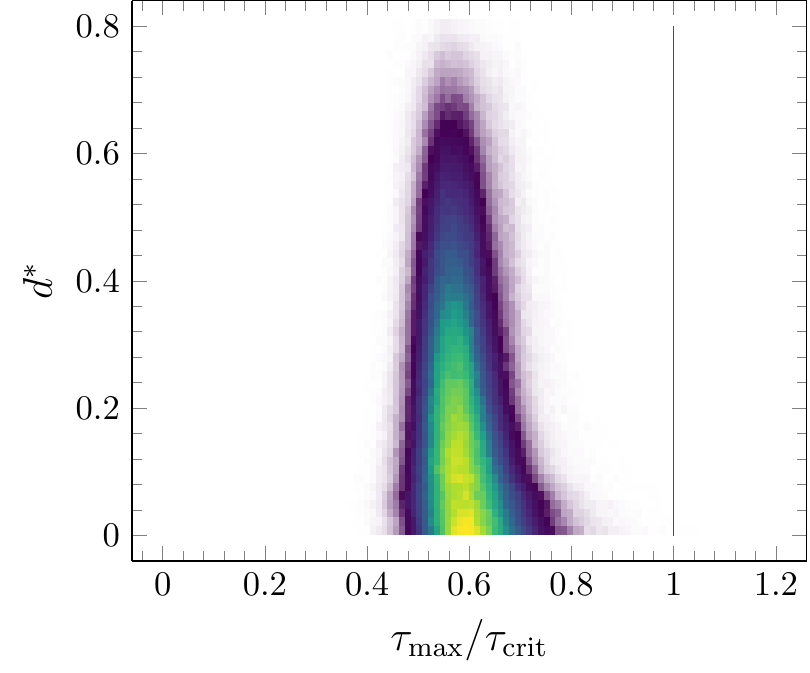}
		\end{subfigure}
		\begin{subfigure}[c]{0.18\textwidth}
			\centering 
			\includegraphics[width=\textwidth]{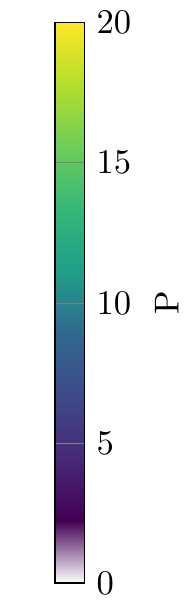}
			\vspace{3pt}
		\end{subfigure}
		\caption{$0.020\%$ strain}
		\vspace{5pt}
		\label{fig:scatter_plastic_pgf_0.02}
	\end{subfigure}
	\begin{subfigure}[c]{0.33\textwidth}
		\begin{subfigure}[c]{0.8\textwidth}
			\centering 
			\includegraphics[width=\textwidth]{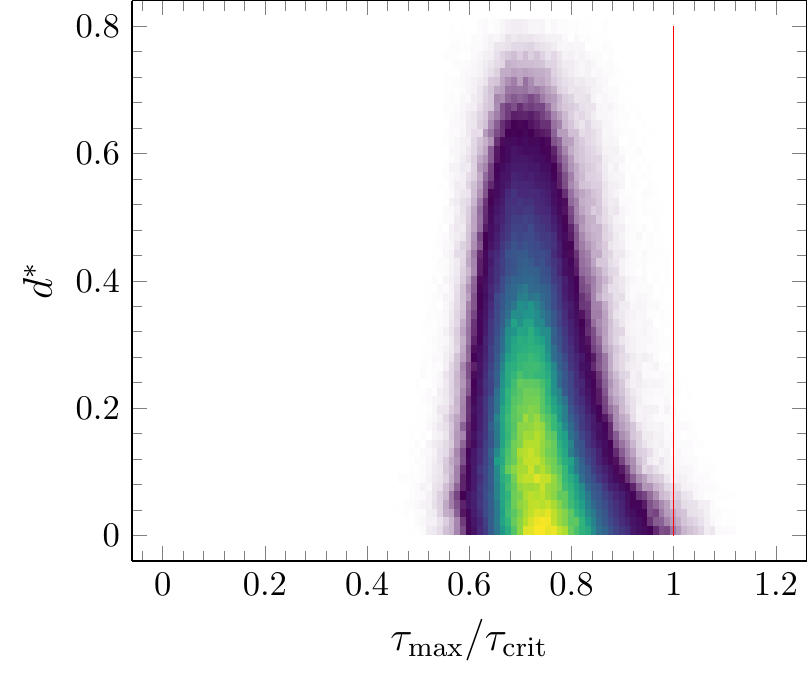}
		\end{subfigure}
		\begin{subfigure}[c]{0.18\textwidth}
			\centering 
			\includegraphics[width=\textwidth]{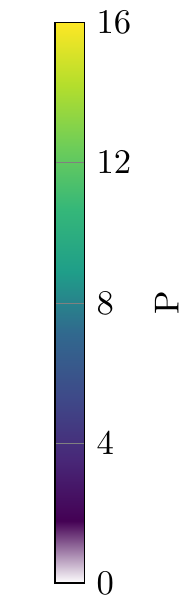}
			\vspace{3pt}
		\end{subfigure}
		\caption{$0.025\%$ strain}
		\vspace{5pt}
		\label{fig:scatter_plastic_pgf_0.025}
	\end{subfigure}
	\begin{subfigure}[c]{0.33\textwidth}
		\begin{subfigure}[c]{0.8\textwidth}
			\centering 
			\includegraphics[width=\textwidth]{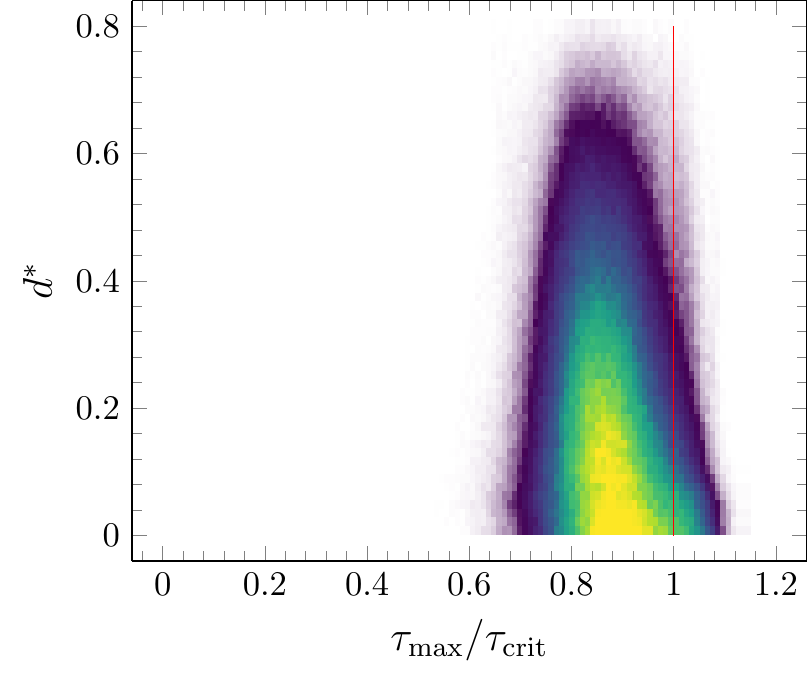}
		\end{subfigure}
		\begin{subfigure}[c]{0.18\textwidth}
			\centering 
			\includegraphics[width=\textwidth]{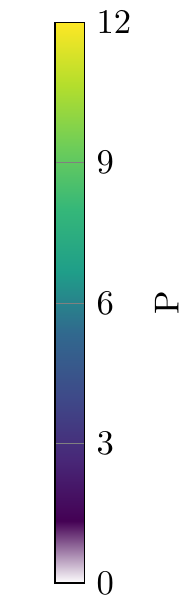}
			\vspace{3pt}
		\end{subfigure}
		\caption{$0.030\%$ strain}
		\vspace{5pt}
		\label{fig:scatter_plastic_pgf_0.03}
	\end{subfigure}
	\begin{subfigure}[c]{0.33\textwidth}
		\begin{subfigure}[c]{0.8\textwidth}
			\centering 
			\includegraphics[width=\textwidth]{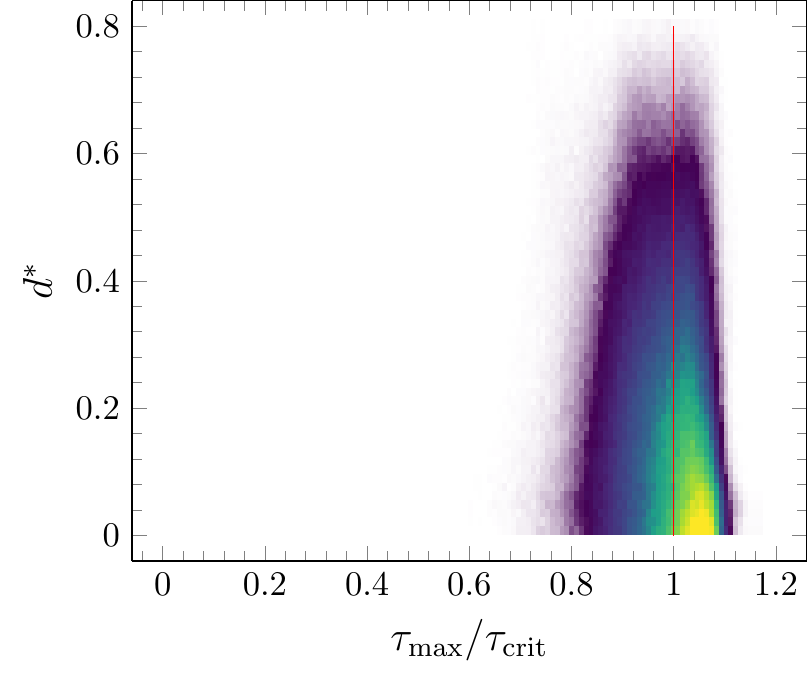}
		\end{subfigure}
		\begin{subfigure}[c]{0.18\textwidth}
			\centering 
			\includegraphics[width=\textwidth]{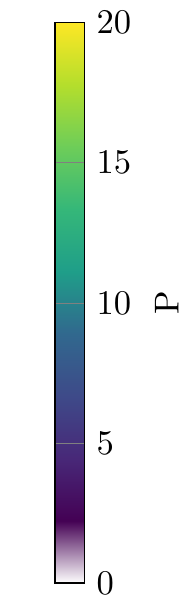}
			\vspace{3pt}
		\end{subfigure}
		\caption{$0.035\%$ strain}
		\vspace{5pt}
		\label{fig:scatter_plastic_pgf_0.035}
	\end{subfigure}
	\begin{subfigure}[c]{0.33\textwidth}
		\begin{subfigure}[c]{0.8\textwidth}
			\centering 
			\includegraphics[width=\textwidth]{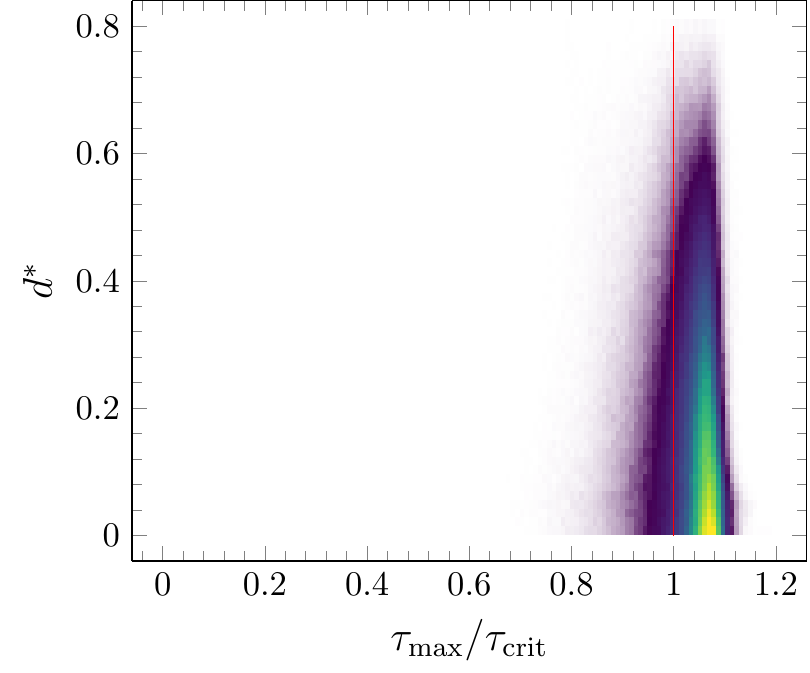}
		\end{subfigure}
		\begin{subfigure}[c]{0.18\textwidth}
			\centering 
			\includegraphics[width=\textwidth]{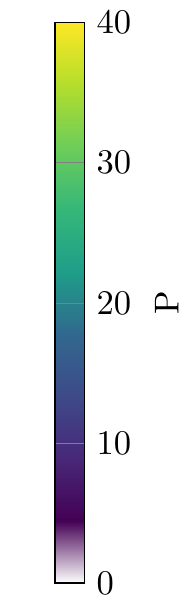}
			\vspace{3pt}
		\end{subfigure}
		\caption{$0.040\%$ strain}
		\vspace{5pt}
		\label{fig:scatter_plastic_pgf_0.04}
	\end{subfigure}
	\begin{subfigure}[c]{0.33\textwidth}
		\begin{subfigure}[c]{0.8\textwidth}
			\centering 
			\includegraphics[width=\textwidth]{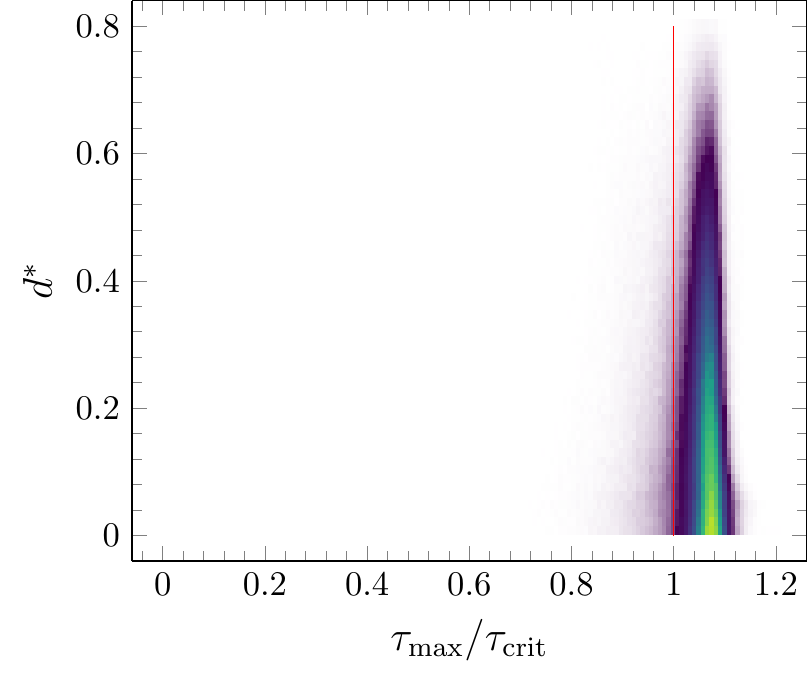}
		\end{subfigure}
		\begin{subfigure}[c]{0.18\textwidth}
			\centering 
			\includegraphics[width=\textwidth]{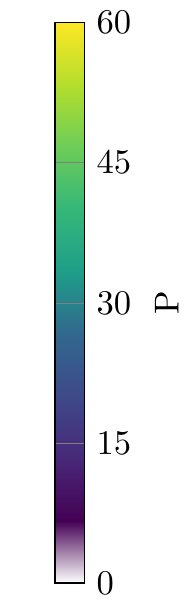}
			\vspace{3pt}
		\end{subfigure}
		\caption{$0.045\%$ strain}
		\vspace{5pt}
		\label{fig:scatter_plastic_pgf_0.045}
	\end{subfigure}
	\caption{Spatial probability density during the elastic-plastic transition}
	\label{fig:scatter_plastic_pgf}
\end{figure}

For a more vivid illustration of the plastification progress, Fig.~\ref{fig:scatter_plastic_pgf} shows the heat maps of the spatial stress distributions for the $\la100\ra$ grain at different strain levels. More precisely, the x-axis is scaled with $\tau_\textrm{crit}$ so that values over one indicate plastification. Indeed, around $0.025\%$ the right tail of the distribution starts to exceed the critical stress close to the grain boundary. As the whole distribution shifts further to the right, larger regions plastify. Whereas the part of the distribution left of $\tau_\textrm{max}/\tau_\textrm{crit} = 1$ roughly retains its shape, the stresses in the plastified regions concentrate around the critical shear stress, shedding light onto the evolution of the statistical moments in Fig.~\ref{fig:MV,SD,SK,KU_plastic}. At $0.045\%$ the grain is virtually fully plastified, with only some faint regions close to the grain boundary with stresses below $\tau_\textrm{crit}$.

 \section{Conclusions}

In the present work, the local shear stress distribution in polycrystalline metals with fcc crystal structure and anisotropic stiffness is investigated numerically using FFT-based solvers. In particular, we focus on characteristically oriented grains in the linear elastic range and at incipient elasto-viscoplastic deformation. All studies are carried out on a data set of 200 microstructures.

Regarding the influence of anisotropy on the stress distribution in the linear elastic range, an increase in scatter is observed with increasing anisotropy, irrespective of grain orientation. However, the grain orientation has a marked influence on the average stress level as a higher Young's modulus with respect to loading direction leads to higher stresses ($\la 100\ra$ grains) and vice versa ($\la 111\ra$ grains).
The aforementioned behavior is confirmed by a more granular analysis of the standardized shear stresses in the 12 slip systems. In this context, the number of active slip systems is found to correspond well to the commonly used Reuss hypothesis, i.e., assuming a homogeneous stress field, for the $\la 100\ra$ grains (8 active systems) and $\la 111\ra$ grains (6 active systems). However, for the $\la 110\ra$ grains, 4 semi-active slip systems can be identified, in addition to the 4 active systems predicted by the Reuss hypothesis. This is due to the multi-axial stress state, which becomes more pronounced with increasing anisotropy. Overall, we conclude that, depending on the degree of anisotropy and the specific grain orientation, the average resolved shear stress and the slip system activity may deviate notably from the classical values, obtained by only considering the (constant) mean stress. For future investigations, it would be of interest to check whether these deviations persist when the (shear) stress distribution over a whole polycrystal is considered instead of specific grain orientations. In this context, testing theoretical predictions based on a homogeneous stress field, e.g., for the average resolved shear stress and plastic strain \cite{le2020averaging}, would prove fruitful.

An analysis of the type of statistical distribution of the local shear stress distribution is performed using Q-Q plots and taking into account higher statistical moments (skewness and excess kurtosis).
It is shown that, even though the stress distribution gets smoother and more symmetric for larger ensembles, it does not converge to a normal distribution. Furthermore, the deviation from normal is found to get larger with increasing anisotropy. The log-normal distribution, as the natural choice for positive valued data, appears to be a better fit. However, a theoretical justification for this choice is currently outstanding.

In a previous investigation by Krause and Böhlke \cite{krause:20}, the MEM was unable to provide accurate predictions for single grain orientations when compared to computations on a single microstructure. Using the large data set of the present study, we confirm that the MEM results correspond to the representative distribution for an arbitrary neighborhood. In particular, the MEM results are found to be in excellent agreement to the full-field computations up to the second statistical moment.

In investigation on the spatial stress distribution, we observe that both stress maxima and minima occur at the grain boundary confirming the results by Rollett et al. \cite{rollett2010stress}. In general, we observe an asymmetry towards higher stresses, i.e., maximum stresses depart further from the mean value. Whereas stress maxima are usually associated with high stress gradients due to the mismatch of material properties at the grain boundaries, the occurrence of stress minima is suspected to be due to microstructural shielding mechanisms \cite{rollett2010stress}, which is yet to be investigated in more detail.

Last but not least, we observe that during the elastic-plastic transition plastification tends to start close to the grain boundary, in accordance to the results in the elastic regime. Subsequently the voxels at the inside of the grain plastifiy, whereas the last voxels to yield are again located close to the grain boundary. Concerning the elastic-plastic range, our observation is limited to small deformations (up to $0.5\%$ strain) as the behavior at higher deformations is strongly dependent on the specific material model in terms of flow rule and hardening law. A dedicated study on the impact of different phenomenological and physical hardening laws (see, for instance, Demir et al. \cite{demir2021investigation}, Lim et al. \cite{lim2019investigating}) on the stress distribution in the plastic regime would considerably expand upon the present work and aid to generalize its findings. 
In particular, physics-based models taking into account the evolution of dislocations \cite{KocksMecking2003, RotersOverview, sedighiani2022determination} and the underlying thermodynamics \cite{langer2010thermodynamic, LeNonUniform,langer2020scaling} appear promising for predicting the mechanical behavior of polycrystals over a wide range of deformations, temperatures and strain rates using the same physics-based parameter set.

\section*{Author contributions}
\textbf{Flavia Gehrig:} Methodology, Software, Formal analysis, Investigation, Data Curation, Visualization, Writing - Original Draft, Writing - Review \& Editing.
\textbf{Daniel Wicht:} Conceptualization, Methodology, Software, Formal analysis, Investigation, Writing - Review \& Editing.
\textbf{Maximilian Krause:} Software, Formal analysis, Writing - Review \& Editing.
\textbf{Thomas Böhlke:} Supervision, Project administration, Funding acquisition, Writing - Review \& Editing.

\bibliographystyle{ieeetr}
{\footnotesize
\bibliography{manuscript}

\begin{thebibliography}{10}

\bibitem{javaid2020local}
F.~Javaid, Y.~Xu, and K.~Durst, ``{Local analysis on dislocation structure and
  hardening during grain boundary pop-ins in tungsten},'' {\em Journal of
  Materials Science}, pp.~1--11, 2020.

\bibitem{blochwitz1996analysis}
C.~Blochwitz, J.~Brechb{\"u}hl, and W.~Tirschler, ``{Analysis of activated slip
  systems in fatigue nickel polycrystals using the EBSD-technique in the
  scanning electron microscope},'' {\em Materials Science and Engineering: A},
  vol.~210, no.~1-2, pp.~42--47, 1996.

\bibitem{clausen1999lattice}
B.~Clausen, T.~Lorentzen, M.~A. Bourke, and M.~R. Daymond, ``{Lattice strain
  evolution during uniaxial tensile loading of stainless steel},'' {\em
  Materials Science and Engineering: A}, vol.~259, no.~1, pp.~17--24, 1999.

\bibitem{pang2000generation}
J.~Pang, T.~Holden, J.~Wright, and T.~Mason, ``{The generation of intergranular
  strains in 309H stainless steel under uniaxial loading},'' {\em Acta
  Materialia}, vol.~48, no.~5, pp.~1131--1140, 2000.

\bibitem{tamura2003scanning}
N.~Tamura, A.~MacDowell, R.~Spolenak, B.~Valek, J.~Bravman, W.~Brown,
  R.~Celestre, H.~Padmore, B.~Batterman, and J.~Patel, ``{Scanning X-ray
  microdiffraction with submicrometer white beam for strain/stress and
  orientation mapping in thin films},'' {\em Journal of Synchrotron Radiation},
  vol.~10, no.~2, pp.~137--143, 2003.

\bibitem{lienert2004investigating}
U.~Lienert, T.-S. Han, J.~Almer, P.~Dawson, T.~Leffers, L.~Margulies, S.~F.
  Nielsen, H.~Poulsen, and S.~Schmidt, ``{Investigating the effect of grain
  interaction during plastic deformation of copper},'' {\em Acta Materialia},
  vol.~52, no.~15, pp.~4461--4467, 2004.

\bibitem{raabe2001micromechanical}
D.~Raabe, M.~Sachtleber, Z.~Zhao, F.~Roters, and S.~Zaefferer,
  ``{Micromechanical and macromechanical effects in grain scale polycrystal
  plasticity experimentation and simulation},'' {\em Acta Materialia}, vol.~49,
  no.~17, pp.~3433--3441, 2001.

\bibitem{tasan2014strain}
C.~C. Tasan, J.~P. Hoefnagels, M.~Diehl, D.~Yan, F.~Roters, and D.~Raabe,
  ``{Strain localization and damage in dual phase steels investigated by
  coupled in-situ deformation experiments and crystal plasticity
  simulations},'' {\em International Journal of Plasticity}, vol.~63,
  pp.~198--210, 2014.

\bibitem{berger2022experimental}
A.~Berger, J.~Witz, A.~El~Bartali, T.~Sadat, N.~Limodin, M.~Dubar, and
  D.~Najjar, ``{Experimental investigation of early strain heterogeneities and
  localizations in polycrystalline $\alpha$-Fe during monotonic loading},''
  {\em International Journal of Plasticity}, p.~103253, 2022.

\bibitem{demir2021investigation}
E.~Demir and I.~Gutierrez-Urrutia, ``{Investigation of strain hardening near
  grain boundaries of an aluminum oligocrystal: Experiments and crystal based
  finite element method},'' {\em International Journal of Plasticity},
  vol.~136, p.~102898, 2021.

\bibitem{voigt1928lehrbuch}
W.~Voigt, ``{Lehrbuch der Kristallphysik},'' {\em BG Teubner, Leipzig und
  Berlin}, 1928.

\bibitem{taylor1938plastic}
G.~I. Taylor, ``{Plastic strain in metals},'' {\em Journal of the Institute of
  Metals}, vol.~62, pp.~307--324, 1938.

\bibitem{bishop1951cxxviii}
J.~Bishop and R.~Hill, ``{CXXVIII. A theoretical derivation of the plastic
  properties of a polycrystalline face-centered metal},'' {\em The London,
  Edinburgh, and Dublin Philosophical Magazine and Journal of Science},
  vol.~42, no.~334, pp.~1298--1307, 1951.

\bibitem{bishop1951xlvi}
J.~Bishop and R.~Hill, ``{XLVI. A theory of the plastic distortion of a
  polycrystalline aggregate under combined stresses},'' {\em The London,
  Edinburgh, and Dublin Philosophical Magazine and Journal of Science},
  vol.~42, no.~327, pp.~414--427, 1951.

\bibitem{Reus1929}
A.~Reuss, ``{Berechnung der Flie{\ss}grenze von Mischkristallen auf Grund der
  Plastizit{\"a}tsbedingung f{\"u}r Einkristalle},'' {\em ZAMM-Journal of
  Applied Mathematics and Mechanics/Zeitschrift f{\"u}r Angewandte Mathematik
  und Mechanik}, vol.~9, no.~1, pp.~49--58, 1929.

\bibitem{sachs1928zietschrift}
G.~Sachs, ``{Zur Ableitung einer Flie{\ss}bedingung},'' {\em Zeitschrift Verein
  deutscher Ingenieure}, vol.~72, pp.~734--736, 1928.

\bibitem{hershey1954elasticity}
A.~Hershey, ``{The elasticity of an isotropic aggregate of anisotropic cubic
  crystals},'' {\em Journal of Applied Mechanics -Transactions of the ASME},
  vol.~21, no.~3, pp.~236--240, 1954.

\bibitem{kroner1958berechnung}
E.~Kr{\"o}ner, ``{Berechnung der elastischen Konstanten des Vielkristalls aus
  den Konstanten des Einkristalls},'' {\em Zeitschrift f{\"u}r Physik},
  vol.~151, no.~4, pp.~504--518, 1958.

\bibitem{hill1965self}
R.~Hill, ``{A self-consistent mechanics of composite materials},'' {\em Journal
  of the Mechanics and Physics of Solids}, vol.~13, no.~4, pp.~213--222, 1965.

\bibitem{hutchinson1976bounds}
J.~W. Hutchinson, ``{Bounds and self-consistent estimates for creep of
  polycrystalline materials},'' {\em Proceedings of the Royal Society of
  London. A. Mathematical and Physical Sciences}, vol.~348, no.~1652,
  pp.~101--127, 1976.

\bibitem{molinari1987self}
A.~Molinari, G.~Canova, and S.~Ahzi, ``{A self consistent approach of the large
  deformation polycrystal viscoplasticity},'' {\em Acta Metallurgica}, vol.~35,
  no.~12, pp.~2983--2994, 1987.

\bibitem{lebensohn1993self}
R.~A. Lebensohn and C.~Tom{\'e}, ``{A self-consistent anisotropic approach for
  the simulation of plastic deformation and texture development of
  polycrystals: Application to zirconium alloys},'' {\em Acta Metallurgica et
  Materialia}, vol.~41, no.~9, pp.~2611--2624, 1993.

\bibitem{LEBENSOHN20045347}
R.~A. Lebensohn, Y.~Liu, and P.~P. Casta{\~n}eda, ``{On the accuracy of the
  self-consistent approximation for polycrystals: Comparison with full-field
  numerical simulations},'' {\em Acta Materialia}, vol.~52, no.~18, pp.~5347 --
  5361, 2004.

\bibitem{BRENNER20093018}
R.~Brenner, R.~A. Lebensohn, and O.~Castelnau, ``{Elastic anisotropy and yield
  surface estimates of polycrystals},'' {\em International Journal of Solids
  and Structures}, vol.~46, no.~16, pp.~3018 -- 3026, 2009.

\bibitem{kreher:89}
W.~Kreher and W.~Pompe, {\em {Internal Stresses in Heterogeneous Solids}},
  vol.~9 of {\em Physical Research}.
\newblock Akademie-Verlag Berlin, 1989.

\bibitem{krause:20}
M.~Krause and T.~B\"{o}hlke, ``{Maximum-Entropy Based Estimates of Stress and
  Strain in Thermoelastic Random Heterogeneous Materials},'' {\em Journal of
  Elasticity}, vol.~141, pp.~321--348, 07 2020.

\bibitem{hashimoto1983role}
K.~Hashimoto and H.~Margolin, ``{The role of elastic interaction stresses on
  the onset of slip in polycrystalline alpha brass - I. Experimental
  determination of operating slip systems and qualitative analysis},'' {\em
  Acta Metallurgica}, vol.~31, no.~5, pp.~773--785, 1983.

\bibitem{kumar1996micro}
S.~Kumar, S.~K. Kurtz, and V.~K. Agarwala, ``{Micro-stress distribution within
  polycrystalline aggregate},'' {\em Acta Mechanica}, vol.~114, no.~1-4,
  pp.~203--216, 1996.

\bibitem{BARBE2001537}
F.~Barbe, S.~Forest, and G.~Cailletaud, ``{Intergranular and intragranular
  behavior of polycrystalline aggregates. Part 2: Results},'' {\em
  International Journal of Plasticity}, vol.~17, no.~4, pp.~537 -- 563, 2001.

\bibitem{sauzay2007cubic}
M.~Sauzay, ``{Cubic elasticity and stress distribution at the free surface of
  polycrystals},'' {\em Acta Materialia}, vol.~55, no.~4, pp.~1193--1202, 2007.

\bibitem{WONG20101658}
S.~Wong and P.~Dawson, ``{Influence of directional strength-to-stiffness on the
  elastic–plastic transition of fcc polycrystals under uniaxial tensile
  loading},'' {\em Acta Materialia}, vol.~58, no.~5, pp.~1658 -- 1678, 2010.

\bibitem{LAVERGNE2013387}
F.~Lavergne, R.~Brenner, and K.~Sab, ``{Effects of grain size distribution and
  stress heterogeneity on yield stress of polycrystals: A numerical
  approach},'' {\em Computational Materials Science}, vol.~77, pp.~387 -- 398,
  2013.

\bibitem{castelnau2006effect}
O.~Castelnau, R.~Brenner, and R.~A. Lebensohn, ``{The effect of strain
  heterogeneity on the work hardening of polycrystals predicted by mean-field
  approaches},'' {\em Acta Materialia}, vol.~54, no.~10, pp.~2745--2756, 2006.

\bibitem{moulinec1994fast}
H.~Moulinec and P.~Suquet, ``{A fast numerical method for computing the linear
  and nonlinear mechanical properties of composites},'' {\em Comptes rendus de
  l'Acad{\'e}mie des sciences. S{\'e}rie II, M{\'e}canique, physique, chimie,
  astronomie}, vol.~318, no.~11, pp.~1417--1423, 1994.

\bibitem{moulinec1998numerical}
H.~Moulinec and P.~Suquet, ``{A numerical method for computing the overall
  response of nonlinear composites with complex microstructure},'' {\em
  Computer Methods in Applied Mechanics and Engineering}, vol.~157, no.~1-2,
  pp.~69--94, 1998.

\bibitem{michel2001computational}
J.~Michel, H.~Moulinec, and P.~Suquet, ``{A computational scheme for linear and
  non-linear composites with arbitrary phase contrast},'' {\em International
  Journal for Numerical Methods in Engineering}, vol.~52, no.~1-2,
  pp.~139--160, 2001.

\bibitem{gelebart2013non}
L.~G{\'e}l{\'e}bart and R.~Mondon-Cancel, ``{Non-linear extension of FFT-based
  methods accelerated by conjugate gradients to evaluate the mechanical
  behavior of composite materials},'' {\em Computational Materials Science},
  vol.~77, pp.~430--439, 2013.

\bibitem{QuasiNewton2019}
D.~Wicht, M.~Schneider, and T.~Böhlke, ``{On Quasi-Newton methods in FFT-based
  micromechanics},'' {\em International Journal for Numerical Methods in
  Engineering}, vol.~121, no.~8, pp.~1665--1694, 2020.

\bibitem{kasemer2020numerical}
M.~Kasemer, G.~Falkinger, and F.~Roters, ``{A numerical study of the influence
  of crystal plasticity modeling parameters on the plastic anisotropy of rolled
  aluminum sheet},'' {\em Modelling and Simulation in Materials Science and
  Engineering}, vol.~28, no.~8, p.~085005, 2020.

\bibitem{BRETIN201936}
R.~Bretin, M.~Levesque, and P.~Bocher, ``{Neighborhood effect on the strain
  distribution in linearly elastic polycrystals: Part 1 - Finite element study
  of the interaction between grains},'' {\em International Journal of Solids
  and Structures}, vol.~176-177, pp.~36 -- 48, 2019.

\bibitem{GUILHEM20101748}
Y.~Guilhem, S.~Basseville, F.~Curtit, J.-M. St{\'e}phan, and G.~Cailletaud,
  ``{Investigation of the effect of grain clusters on fatigue crack initiation
  in polycrystals},'' {\em International Journal of Fatigue}, vol.~32, no.~11,
  pp.~1748 -- 1763, 2010.

\bibitem{castelnau2020multiscale}
O.~Castelnau, K.~Derrien, S.~Ritterbex, P.~Carrez, P.~Cordier, and H.~Moulinec,
  ``{Multiscale modeling of the effective viscoplastic behavior of
  Mg\textsubscript{2} SiO\textsubscript{4} wadsleyite: Bridging atomic and
  polycrystal scales},'' {\em Comptes Rendus. M{\'e}canique}, vol.~348,
  no.~10-11, pp.~827--846, 2020.

\bibitem{gelebart2021grain}
L.~G{\'e}l{\'e}bart, ``{Grain size effects and weakest link theory in 3D
  crystal plasticity simulations of polycrystals},'' {\em Comptes Rendus.
  Physique}, vol.~22, no.~S3, pp.~1--18, 2021.

\bibitem{bieler2009role}
T.~Bieler, P.~Eisenlohr, F.~Roters, D.~Kumar, D.~Mason, M.~Crimp, and D.~Raabe,
  ``{The role of heterogeneous deformation on damage nucleation at grain
  boundaries in single phase metals},'' {\em International Journal of
  Plasticity}, vol.~25, no.~9, pp.~1655--1683, 2009.

\bibitem{lebensohn2012elasto}
R.~A. Lebensohn, A.~K. Kanjarla, and P.~Eisenlohr, ``{An elasto-viscoplastic
  formulation based on fast Fourier transforms for the prediction of
  micromechanical fields in polycrystalline materials},'' {\em International
  Journal of Plasticity}, vol.~32, pp.~59--69, 2012.

\bibitem{kuhbach2020quantification}
M.~K{\"u}hbach and F.~Roters, ``{Quantification of 3D spatial correlations
  between state variables and distances to the grain boundary network in
  full-field crystal plasticity spectral method simulations},'' {\em Modelling
  and Simulation in Materials Science and Engineering}, vol.~28, no.~5,
  p.~055005, 2020.

\bibitem{cailletaud2003some}
G.~Cailletaud, S.~Forest, D.~Jeulin, F.~Feyel, I.~Galliet, V.~Mounoury, and
  S.~Quilici, ``{Some elements of microstructural mechanics},'' {\em
  Computational Materials Science}, vol.~27, no.~3, pp.~351--374, 2003.

\bibitem{rollett2010stress}
A.~D. Rollett, R.~A. Lebensohn, M.~Groeber, Y.~Choi, J.~Li, and G.~S. Rohrer,
  ``{Stress hot spots in viscoplastic deformation of polycrystals},'' {\em
  Modelling and Simulation in Materials Science and Engineering}, vol.~18,
  no.~7, p.~074005, 2010.

\bibitem{GONZALEZ201449}
D.~Gonzalez, I.~Simonovski, P.~Withers, and J.~Q. da~Fonseca, ``{Modelling the
  effect of elastic and plastic anisotropies on stresses at grain
  boundaries},'' {\em International Journal of Plasticity}, vol.~61, pp.~49 --
  63, 2014.

\bibitem{hure2016intergranular}
J.~Hure, S.~El~Shawish, L.~Cizelj, and B.~Tanguy, ``{Intergranular stress
  distributions in polycrystalline aggregates of irradiated stainless steel},''
  {\em Journal of Nuclear Materials}, vol.~476, pp.~231--242, 2016.

\bibitem{el2020full}
S.~El~Shawish, P.-G. Vincent, H.~Moulinec, L.~Cizelj, and L.~G{\'e}l{\'e}bart,
  ``{Full-field polycrystal plasticity simulations of neutron-irradiated
  austenitic stainless steel: A comparison between FE and FFT-based
  approaches},'' {\em Journal of Nuclear Materials}, vol.~529, p.~151927, 2020.

\bibitem{lim2019investigating}
H.~Lim, C.~C. Battaile, J.~E. Bishop, and J.~W. Foulk~III, ``{Investigating
  mesh sensitivity and polycrystalline RVEs in crystal plasticity finite
  element simulations},'' {\em International Journal of Plasticity}, vol.~121,
  pp.~101--115, 2019.

\bibitem{vincent2011stress}
L.~Vincent, L.~G{\'e}l{\'e}bart, R.~Dakhlaoui, and B.~Marini, ``{Stress
  localization in BCC polycrystals and its implications on the probability of
  brittle fracture},'' {\em Materials Science and Engineering: A}, vol.~528,
  no.~18, pp.~5861--5870, 2011.

\bibitem{Lemaitre1990}
J.~Lemaitre and J.-L. Chaboche, {\em {Mechanics of Solid Materials}}.
\newblock Cambridge University Press, 1990.

\bibitem{d1990suggestion}
R.~B. D'Agostino, A.~Belanger, and R.~B. D'Agostino~Jr, ``{A suggestion for
  using powerful and informative tests of normality},'' {\em The American
  Statistician}, vol.~44, no.~4, pp.~316--321, 1990.

\bibitem{von2005mean}
P.~T. Von~Hippel, ``{Mean, median, and skew: Correcting a textbook rule},''
  {\em Journal of statistics Education}, vol.~13, no.~2, 2005.

\bibitem{westfall2014kurtosis}
P.~H. Westfall, ``{Kurtosis as peakedness},'' {\em The American Statistician},
  vol.~68, no.~3, pp.~191--195, 2014.

\bibitem{huck2012reading}
S.~W. Huck, {\em {Reading Statistics and Research}}.
\newblock Pearson Education, 2012.

\bibitem{lei2005effect}
M.~Lei and R.~G. Lomax, ``{The effect of varying degrees of nonnormality in
  structural equation modeling},'' {\em Structural Equation Modeling}, vol.~12,
  no.~1, pp.~1--27, 2005.

\bibitem{kuhn2020fast}
J.~Kuhn, M.~Schneider, P.~Sonnweber-Ribic, and T.~B{\"o}hlke, ``{Fast methods
  for computing centroidal Laguerre tessellations for prescribed volume
  fractions with applications to microstructure generation of polycrystalline
  materials},'' {\em Computer Methods in Applied Mechanics and Engineering},
  vol.~369, p.~113175, 2020.

\bibitem{bourne2020laguerre}
D.~P. Bourne, P.~J. Kok, S.~M. Roper, and W.~D. Spanjer, ``{Laguerre
  tessellations and polycrystalline microstructures: A fast algorithm for
  generating grains of given volumes},'' {\em Philosophical Magazine},
  vol.~100, no.~21, pp.~2677--2707, 2020.

\bibitem{zeman2010accelerating}
J.~Zeman, J.~Vond{\v{r}}ejc, J.~Nov{\'a}k, and I.~Marek, ``{Accelerating a
  FFT-based solver for numerical homogenization of periodic media by conjugate
  gradients},'' {\em Journal of Computational Physics}, vol.~229, no.~21,
  pp.~8065--8071, 2010.

\bibitem{NewtonKrylov}
M.~Kabel, T.~B\"{o}hlke, and M.~Schneider, ``{Efficient fixed point and
  Newton–Krylov solvers for FFT-based homogenization of elasticity at large
  deformations},'' {\em Computational Mechanics}, vol.~54, no.~6,
  pp.~1497--1514, 2014.

\bibitem{wicht2020efficient}
D.~Wicht, M.~Schneider, and T.~B{\"o}hlke, ``{An efficient solution scheme for
  small-strain crystal-elasto-viscoplasticity in a dual framework},'' {\em
  Computer Methods in Applied Mechanics and Engineering}, vol.~358, p.~112611,
  2020.

\bibitem{StaggeredGrid}
M.~Schneider, F.~Ospald, and M.~Kabel, ``{Computational homogenization of
  elasticity on a staggered grid},'' {\em International Journal for Numerical
  Methods in Engineering}, vol.~105, no.~9, pp.~693--720, 2016.

\bibitem{jaynes:63}
E.~T. Jaynes, {\em {Statistical Physics}}, vol.~3 of {\em Brandeis Summer
  Institute Lectures in Theoretical Physics}.
\newblock W. A. Benjamin Inc., New York, 1963.

\bibitem{paufler1978physikalische}
P.~Paufler and G.~E. Schulze, {\em {Physikalische Grundlagen mechanischer
  Festk{\"o}rpereigenschaften II}}.
\newblock Braunschweig: Vieweg, 1978.

\bibitem{Simmons1971}
G.~Simmons and H.~Wang, ``{Single crystal elastic constants and calculated
  aggregate properties: A handbook},'' {\em Cambridge, USA: MIT Press}, 1971.

\bibitem{every1992second}
A.~Every and A.~McCurdy, ``{Second and higher order elastic constants},'' {\em
  Landolt-B{\"o}rnstein Numerical Data and Functional Relationships in Science
  and Technology New Series Group III: Crystal and Solid State Physics},
  vol.~29, p.~743, 1992.

\bibitem{castelnau2008micromechanical}
O.~Castelnau, D.~Blackman, R.~A. Lebensohn, and P.~Ponte~Casta{\~n}eda,
  ``{Micromechanical modeling of the viscoplastic behavior of olivine},'' {\em
  Journal of Geophysical Research: Solid Earth}, vol.~113, no.~B9, 2008.

\bibitem{bhattacharyya2021elastoplastic}
J.~J. Bhattacharyya, S.~Nair, D.~C. Pagan, V.~Tari, R.~A. Lebensohn, A.~D.
  Rollett, and S.~R. Agnew, ``{Elastoplastic transition in a metastable
  $\beta$-Titanium alloy, Timetal-18--An in-situ synchrotron X-ray diffraction
  study},'' {\em International Journal of Plasticity}, vol.~139, p.~102947,
  2021.

\bibitem{henning2005local}
M.~Henning and H.~Vehoff, ``{Local mechanical behavior and slip band formation
  within grains of thin sheets},'' {\em Acta materialia}, vol.~53, no.~5,
  pp.~1285--1292, 2005.

\bibitem{el2018intergranular}
S.~El~Shawish and J.~Hure, ``{Intergranular normal stress distributions in
  untextured polycrystalline aggregates},'' {\em European Journal of Mechanics
  - A/Solids}, vol.~72, pp.~354--373, 2018.

\bibitem{gallardo2021lath}
F.-J. Gallardo-Basile, Y.~Naunheim, F.~Roters, and M.~Diehl, ``{Lath Martensite
  Microstructure Modeling: A High-Resolution Crystal Plasticity Simulation
  Study},'' {\em Materials}, vol.~14, no.~3, p.~691, 2021.

\bibitem{thode2002testing}
H.~C. Thode, {\em {Testing for Normality}}, vol.~164.
\newblock CRC press, 2002.

\bibitem{das2016brief}
K.~R. Das and A.~Imon, ``{A brief review of tests for normality},'' {\em
  American Journal of Theoretical and Applied Statistics}, vol.~5, no.~1,
  pp.~5--12, 2016.

\bibitem{eghtesad2018openmp}
A.~Eghtesad, T.~J. Barrett, K.~Germaschewski, R.~A. Lebensohn, R.~J. McCabe,
  and M.~Knezevic, ``{OpenMP and MPI implementations of an elasto-viscoplastic
  fast Fourier transform-based micromechanical solver for fast crystal
  plasticity modeling},'' {\em Advances in Engineering Software}, vol.~126,
  pp.~46--60, 2018.

\bibitem{le2020averaging}
K.~C. Le, T.~H. Le, and T.~M. Tran, ``{Averaging in dislocation mediated
  plasticity},'' {\em International Journal of Engineering Science}, vol.~149,
  p.~103230, 2020.

\bibitem{KocksMecking2003}
U.~F. Kocks and H.~Mecking, ``{Physics and phenomenology of strain hardening:
  The FCC case},'' {\em Progress in Materials Science}, vol.~48, pp.~171--273,
  2003.

\bibitem{RotersOverview}
F.~Roters, P.~Eisenlohr, L.~Hantcherli, D.~Tjahjanto, T.~Bieler, and D.~Raabe,
  ``{Overview of constitutive laws, kinematics, homogenization and multiscale
  methods in crystal plasticity finite-element modeling: Theory, experiments,
  applications},'' {\em Acta Materialia}, vol.~58, no.~4, pp.~1152--1211, 2010.

\bibitem{sedighiani2022determination}
K.~Sedighiani, K.~Traka, F.~Roters, D.~Raabe, J.~Sietsma, and M.~Diehl,
  ``{Determination and analysis of the constitutive parameters of
  temperature-dependent dislocation-density-based crystal plasticity models},''
  {\em Mechanics of Materials}, vol.~164, p.~104117, 2022.

\bibitem{langer2010thermodynamic}
J.~Langer, E.~Bouchbinder, and T.~Lookman, ``{Thermodynamic theory of
  dislocation-mediated plasticity},'' {\em Acta Materialia}, vol.~58, no.~10,
  pp.~3718--3732, 2010.

\bibitem{LeNonUniform}
K.~Le, ``{Thermodynamic dislocation theory for non-uniform plastic
  deformations},'' {\em Journal of the Mechanics and Physics of Solids},
  vol.~111, pp.~157--169, 2018.

\bibitem{langer2020scaling}
J.~Langer and K.~Le, ``{Scaling confirmation of the thermodynamic dislocation
  theory},'' {\em Proceedings of the National Academy of Sciences}, vol.~117,
  no.~47, pp.~29431--29434, 2020.

\end{thebibliography}
}

\end{document}